\documentclass[useAMS,usenatbib]{mn2e}
\usepackage{color}
\usepackage{array} 
\usepackage{amsmath}
\usepackage{graphicx}
\usepackage{epstopdf}
\usepackage[labelfont=bf,labelsep=period,justification=raggedright,singlelinecheck=false,font=small]{caption}
\usepackage{wrapfig}
\usepackage{float}
\usepackage{amssymb}
\usepackage[normalem]{ulem}
\usepackage{undertilde}
\usepackage[T1]{fontenc}
\usepackage[utf8]{inputenc}
\usepackage{comment}
\usepackage[export]{adjustbox}
\usepackage{scalerel}
\usepackage{tikz}
\usetikzlibrary{svg.path}

\definecolor{orcidlogocol}{HTML}{A6CE39}
\tikzset{
  orcidlogo/.pic={
    \fill[orcidlogocol] svg{M256,128c0,70.7-57.3,128-128,128C57.3,256,0,198.7,0,128C0,57.3,57.3,0,128,0C198.7,0,256,57.3,256,128z};
    \fill[white] svg{M86.3,186.2H70.9V79.1h15.4v48.4V186.2z}
                 svg{M108.9,79.1h41.6c39.6,0,57,28.3,57,53.6c0,27.5-21.5,53.6-56.8,53.6h-41.8V79.1z M124.3,172.4h24.5c34.9,0,42.9-26.5,42.9-39.7c0-21.5-13.7-39.7-43.7-39.7h-23.7V172.4z}
                 svg{M88.7,56.8c0,5.5-4.5,10.1-10.1,10.1c-5.6,0-10.1-4.6-10.1-10.1c0-5.6,4.5-10.1,10.1-10.1C84.2,46.7,88.7,51.3,88.7,56.8z};
  }
}

\newcommand\orcidicon[1]{\href{https://orcid.org/#1}{\mbox{\scalerel*{
\begin{tikzpicture}[yscale=-1,transform shape]
\pic{orcidlogo};
\end{tikzpicture}
}{|}}}}

\usepackage{hyperref}

\def\apj{ApJ}
\def\aap{A\& A}
\def\apjl{ApJL}
\def\apjs{ApJS}
\def\mnras{MNRAS}

\def\pasa{PASA}
\def\pasj{PASJ}
\def\aj{AJ}

\def\nat{Nature}

\def\rmxaa{RMxAA}
\def\na{New Astron}
\def\jqsrt{J. Quant. Spec. Radiat. Transf.}

\newcommand\altaffiltext[1]{$^{#1}$}

\restylefloat{table}
\begin{document}

\title[Smoothed Particle Radiation Hydrodynamics]{Smoothed Particle Radiation Hydrodynamics: Two-Moment method with Local Eddington Tensor Closure}

\author[T. K. Chan et al.]
  {T. K. ~Chan\thanks{Email: (TKC)tsang.k.chan@durham.ac.uk}$^{\orcidicon{0000-0003-2544-054X}}$, Tom Theuns $^{\orcidicon{0000-0002-3790-9520}}$, Richard Bower and Carlos Frenk\\
  \altaffiltext{}{ Institute for Computational Cosmology, Department of Physics, Durham University, South Road, Durham DH1 3LE, UK}\\ {}\\
}

\maketitle

\begin{abstract}
 We present a new radiative transfer method ({\small SPH-M1RT}) that is coupled dynamically with smoothed particle hydrodynamics ({\small SPH}). We implement it in the (task-based parallel) {\small SWIFT} galaxy simulation code but it can be straightforwardly implemented in other {\small SPH} codes. Our moment-based method simultaneously solves the radiation energy and flux equations in {\small SPH}, making it adaptive in space and time. We modify the {\small M1} closure relation to stabilize radiation fronts in the optically thin limit. We also introduce anisotropic artificial viscosity and high-order artificial diffusion schemes, which allow the code to handle radiation transport accurately in both the optically thin and optically thick regimes.  Non-equilibrium thermo-chemistry is solved using a semi-implicit sub-cycling technique. The computational cost of our method is independent of the number of sources and can be lowered further by using the reduced speed of light approximation. We demonstrate the robustness of our method by applying it to a set of standard tests from the cosmological radiative transfer comparison project of Iliev et al. The {\small SPH-M1RT} scheme is well-suited for modelling situations in which numerous sources emit ionising radiation, such as cosmological simulations of galaxy formation or simulations of the interstellar medium.
\end{abstract}


\label{firstpage}

\section{Introduction}
\label{sec:Introduction}
Almost everything we know about galaxies and most of what we know about stars comes from studying their radiation. However, radiation is not just a messenger informing us about the sources and sinks of radiation, but may impact gas directly, for example through photo-heating or the suppression of cooling, or by affecting its chemistry. Radiation pressure on gas and dust can also affect the dynamics of the gas directly. Unfortunately, including the effects of radiation in numerical models is challenging: the equation that accounts for the change of intensity of a light ray resulting from emission and absorption is 7 dimensional. To make matters worse, radiation travels at the speed of light, requiring dramatically shorter time-steps than those required to solve the associated hydrodynamics equations.

Progress has been made by concentrating on particular aspects of the impact of radiation. We briefly mention some of these aspects and the codes in which they are implemented, without aiming to be exhaustive. The {\sc cloudy} code, last described by \cite{Ferland17Cloudy}, implements in great detail the interaction between radiation and matter in simple geometries assuming equilibrium conditions. {\sc cloudy} has been instrumental in interpreting the spectra of galaxies. Accounting for absorption and re-emission of light by dust in more complex geometries has been implemented using Monte-Carlo radiative transfer in for example the {\sc skirt} \citep{Baes15Skirt}, {\sc sunrise} \citep{Jonsson06Sunrise}, {\sc CMacIonize} \citep{Vand18MCRT}, and {\small AREPO-MCRT} \citep{Smit20AREPOMCRT} codes. The resonant scattering of Lyman-$\alpha$ has been implemented by, for example
\cite{Zheng02Lya, Cantalupo05Lya, Verhamme06Lya,Smit15COLT} and others. Radiation can also regulate star formation through radiative feedback. The infrared radiation on the interstellar medium (ISM) is modelled in, e.g., \cite{Turn01FLD,Davi12RTSC}. Radiative feedback is also important in the formation of the first stars and galaxies \citep[e.g][]{Brom091ststar, Kim2019}.

In this paper we concentrate on the propagation of (hydrogen) ionizing
photons. Radiative transfer (henceforth RT) of ionizing photons is
important in the context of galaxies, governing the evolution of HII
regions in the interstellar medium (ISM), and in the physics of the
intergalactic medium (IGM), which is highly ionized
  \citep{Gunn65} by radiation from active galactic nuclei (AGN,
  \citealt{Sargent80}) and massive stars in galaxies \citep{Shapiro87,
    Madau94}. In both situations, the following considerations are
relevant to the design of a successful RT implementation: (1) there is
no useful symmetry to be exploited; (2) radiation is emitted by
numerous sources; and (3) gas and radiation interact under
non-equilibrium conditions. In addition, even without including RT,
simulating the ISM and the IGM is computationally demanding requiring
the inclusion of many other physical processes. These considerations
motivate us to build RT on top of an existing hydrodynamics code, and
implement a method that is independent of the number of sources.

 Smooth Particle Hydrodynamics (SPH) \citep{Lucy77SPH,Ging77SPH} is a Lagrangian hydrodynamics scheme that has been applied to a large variety of astrophysical problems (from planet to star to galaxy formation simulations) as well as non-astrophysical problems. In this scheme, the hydrodynamic properties of a fluid are carried by a set of discrete {\it particles} that move with the fluid and are used to interpolate physical quantities such as density using a smooth function called the \lq kernel\rq. The method is computationally efficient, highly adaptive in space and time, and can easily be coupled to gravity. Many current state-of-the art astrophysical hydrodynamics codes are SPH based \citep[e.g.][]{Spri05Gadget2,FIRE,Scha15EAGLE,Wads17GASOLINE2,Pric18phantom,Spri20Gadget4}.

We briefly discuss available options for including RT in hydrodynamical codes, especially for transporting ionizing photons. Conceptually most intuitive is direct ray tracing (also called \lq long characteristics\rq), where each source casts a number of rays and the equation for RT is solved along all rays simultaneously. With a computational cost scaling as $N_{\rm source}\times N^2_{\rm sink}$, where $N_{\rm source}$ and $N_{\rm sink}$ are the number of sources and sinks, this method may be accurate but it is also computationally extremely demanding. Approximations to full ray-tracing are possible though, for example using short characteristic \citep{Miha78shortchar,Mell06c2ray}, hybrid characteristic \citep{Rijk06Hybridchar}, or adaptive ray-tracing \citep{Abel02ART,Wise11ART,Kim17ART}. While the short characteristic method is faster than the long characteristic methods, its angular resolution is lower \citep{Finl09VET}; it is difficult to handle bright point sources \citep[e.g][]{Davi12RTSC};  and it has not yet been implemented directly on top of irregular meshes or particle-based codes such as SPH (though see \citealt{Finl09VET}). Adaptive ray-tracing is fast and can be applied to irregular meshes (and particle methods), so it remains a viable option for RT in SPH, although its computational cost still scales with the number of sources. In cases where the radiation field is largely known, reverse ray-tracing \citep{Kess00reverseRT,Alta13revrt} has been used to calculate the attenuation of ionizing radiation in high density regions. A variation of reverse ray-tracing has proved to be efficiently parallelizable in SPH \citep[e.g.][]{Susa06RSPH,Hase10START}.

An alternative to ray-tracing is to discretize radiation directions in a finite number of cones \citep{Pawl08TRAPHIC,Petk11coneRT}. The scaling of this implementation is independent of the number of sources provided a \lq cone merging scheme\rq\ is implemented. The method has been applied in reionization simulations \citep{Pawl17Aurora}. However, the method is still relatively expensive, given that a high number of cones is required to avoid excessive loss of angular resolution. It also requires substantial modifications to the hydrodynamics code, e.g. virtual particles and rotation of cones to improve the angular sampling and avoid artificial radiation spikes. Another strategy, the Monte Carlo method (e.g. \citealt{Alta08SPHRAY,Baek09MCRTSPH, Graziani13}), is even more expensive requiring a large number of photon packets to reduce shot noise to acceptable levels.

A different starting point for an RT algorithm is to compute angular and spectral moments of the radiative transfer equation and integrate the resulting \lq moment\rq\ equations numerically. It is the radiation equivalent of integrating the fluid equations rather than the full Boltzmann equation. In both cases, doing so leads to a dramatic reduction in the dimensionality of the problem.  Just as in the case of the fluid equations, there is an infinite hierarchy of moment equations which needs to be 
truncated by a \lq closure relation\rq. The closure relation is not unique and obtaining a good closure relation is challenging, because it needs to be able to handle the very different nature of the transport of optically thick and optically thin radiation.

RT moment methods vary in terms of the order of the moments used and in the choice of closure relation. 
Ideally, the closure relation uses only local properties of the gas and the radiation: this makes the computational cost independent of the number of sources and makes the implementation easily parellelisable. Moment methods do not require fine angular discretization - unlike cone-based or short characteristic methods - so the computational cost per cell or particle can be lower.

The \lq Flux Limited Diffusion\rq\ method (FLD, \citealt{Leve81FLD}) solves only for the zeroth moment of the RT equation, which is a diffusion equation provided the time derivative of the first moment is neglected. The speed with which a radiation front propagates is not limited by the speed of light but can be infinite;  however, it is possible to impose a \lq flux limiter\rq\ to enforce causality. FLD has been used in many astrophysical simulations \citep[e.g.][]{Turn01FLD,Reyn09FLD,Comm11FLD,Krum12FLD}, some of which use  SPH \citep{Whit04SPHFLD}. \cite{Gned01OTVET} developed the {\sc otvet} method, which also evolves a diffusion equation of radiation energy density, but with a closure relation applicable to optically thin radiation. FLD and {\sc otvet} are fast with a compute time that is largely independent of the number of sources.

However, the relatively diffusive nature of transport in FLD makes it hard to preserve the propagation direction of radiation accurately. As a consequence, neither standard FLD nor {\sc otvet} cast sharp shadows behind optically thick regions, albeit for slightly different reasons \citep{Haye03RT,Gned01OTVET}. The time-step for propagating radiation in these methods is very restrictive as a consequence of the infinite propagation speed of information; thus, it may be more efficient to use an implicit integration scheme. Unfortunately, an implicit method is computationally inefficient in a scheme like SPH for a large neighbour number (\citealt[e.g][]{Whit04SPHFLD,Petk09OTVET}; see also the discussion about the efficiency of the implicit FLD scheme in \citealt{Skin13M1}). 

The \lq Two Moment\rq\ method solves the zeroth and first order moments of the RT equation simultaneously.
A popular closure relation for this method was introduced by \cite{Leve84} to which we will refer  as the \lq M1\rq\ closure relation\footnote{The \lq M\rq\ in \lq M1\rq\ refers to Minerbo, who introduced the maximum entropy closure in \cite{Mine79closure}.}. The M1 method was first used in astrophysics by \cite{Gonz07M1}, and has also been implemented in other hydrodynamics codes,  e.g. grid-based (\citealt{Aube08M1}, \citealt{Skin13M1}, \citealt{Rosd13ramsert}, and \citealt{Kann19AREPORT}) and hybrid schemes (\citealt{Hopk20rhd}). 

This computational scheme is accurate up to order $v/c$ (the fluid velocity divided by the speed of light; \citealt{Buch83rtff}) for a single source in the optically thin or thick limits \citep{Leve84}. In the optically thick case, it captures the minimum entropy (production) principle in the presence of one preferred direction \citep{Leve96keclosure,Dubr99M1}. In the optically thin case, it preserves the radiation's direction - and hence it can cast shadows - with radiation fronts moving at the speed of light. It may be surprising at first, but this second-order method is generally {\em faster} than FLD or {\sc otvet} if solved explicitly. This is a consequence of the hyperbolic nature of the equations which result in a much less restrictive time-step \citep{Thom98pde}.  The speed and accuracy of the method makes it a promising scheme for including RT in astrophysical hydrodynamics calculations.

While the M1 method works well on structured and unstructured meshes, to date, it has not been implemented in SPH. One reason is that SPH has zeroth-order errors under irregular particle distributions meaning that the SPH estimate does not converge to the true value in the limit of vanishing smoothing length \citep{Lucy77SPH,Ravi85SPHinconsistent,Lans08bettergrad, Dehnen12}. Secondly, devising a good artificial dissipation scheme in SPH is not trivial. However, such artificial dissipation is necessary in order to suppress numerical oscillations around discontinuities. As we will demonstrate, the usual scheme, e.g. \cite{Pric08artcon}, for implementing artificial dissipation fails when applied to the M1 scheme. Finally, the original M1 closure relation artificially amplifies noise in the optically-thin regime which requires changes to the closure relation.

Despite these difficulties, implementing the M1 RT method in SPH would be highly desirable: SPH is highly adaptive and ideal for problems that are characterised by a very large dynamic range, whereas the M1 method is efficient and accurate in both the optically thick and thin limits\footnote{But it has issues in handling multiple sources in the optically thin region; see \S\ref{sec:discussions}.}. Furthermore, the M1 method can be straightforwardly implemented on top of any SPH code, since the structures of the hydrodynamics equations and radiation moment equations are quite similar. This results in an accurate and fast code that can handle a very large number of sources in a computationally efficient way. As such, the method described in this paper goes some way towards enabling the inclusion of RT in simulations of galaxy formation as a matter of course.

In this paper, we describe how to incorporate the M1 method into SPH and examine its performance through standard RT problems. We begin in \S\ref{sec:method} by briefly illuminating the analogy between taking moments of the Boltzmann equation to derive the fluid equations, and taking moments of the RT equation to derive the two-moment method. We then discuss closure relations and discuss our modification to the M1 closure. Next, we show how the SPH equations can be dicretized to yield the more accurate gradients required for implementing the two-moment method and discuss ways of capturing discontinuities in the radiation field. We finish \S\ref{sec:method} by discussing the coupling of radiation to the thermodynamics and chemistry of the gas, explain and discuss the advantages and drawbacks of the \lq reduced speed-of-light\rq\ approximation, and discuss how we inject radiation. In \S \ref{sec:results}, we present the results of tests with a known solution and compare more realistic tests without a known solution
to those in the RT code comparison project \citep{Ilie06RTcom,Ilie09RTcom}. In \S \ref{sec:discussions}, we comment on the strengths and weaknesses of our scheme and compare with other radiative transfer methods. In \S \ref{sec:conclusions}, we briefly summarise our findings and foresee possible improvements in the future.

\section{Methods}
\label{sec:method}
In the following, we first describe the two moment method including modifications to the M1 closure relation. We continue by discussing
the implementation in SPH as well as the thermo-chemistry solver we employ. The equations contain numerous variables which we have collated for easy reference in Table \ref{table:variables}.

\begin{table*}[]
\centering
\begin{tabular}{lp{23mm}p{10mm}|lp{23mm}p{10mm}|lp{23mm}p{10mm}}
\hline
\hline
 $D/Dt$ & Lagrangian Derivative &\ref{eq:masscon}
 &$\rho$ &  gas density& \ref{eq:masscon}
 &${\bf v}$ & gas velocity& \ref{eq:masscon}
 \\$p$ & gas pressure& \ref{eq:momcon}
 &$\chi$ & opacity& \ref{eq:momcon}
 &$\tilde{c}$&reduced speed of light& \ref{eq:dfrad}
 \\${\bf f}$&radiation flux / gas density& \ref{eq:durad}
 &${\bf F}$&radiation flux &\ref{eq:durad}
 &$u$& gas internal energy& \ref{eq:intcon}
 \\$\xi$& radiation energy density / gas density&\ref{eq:durad}
 &$\mathbb{P}$& radiation stress tensor& \ref{eq:durad}
 &$\mathbb{F}$& Eddington tensor& \ref{eq:stressrad}
 \\$e$& gas thermal energy &\ref{eq:intcon}
 &$E$& radiation energy density& \ref{eq:durad}
 &$f_{\rm Edd}$& Eddington factor &\ref{eq:stressrad}
 \\$\hat{\bf n}$& radiation direction& \ref{eq:eddclose}
 &$\nu$ & photon frequency &\ref{eq:ngammahv}
 &$\Omega$ & solid angle &\ref{eq:intIE}
 \\$I$& specific intensity &\ref{eq:RT}
 &$\hbar$& reduced Planck constant &\ref{eq:ngammahv}
 &$2\pi\hbar\bar{\nu}$& spectral mean photon energy  &\ref{eq:ngammahv}
 \\$\varepsilon$& optical thickness estimator &\ref{eq:feddest}
 &$h$& SPH particle size &\ref{eq:sphxi}
 &$m$& SPH particle mass &\ref{eq:sphxi}
 \\$D$& diffusion coefficient &\ref{eq:xidiss}
&$W$& SPH kernel &\ref{eq:sphxi}
&$v_{\rm sig}$& SPH signal speed &\ref{eq:signalkappa}
\\$\phi$& slope limiter &\ref{eq:highorderdiss}
&$n_\gamma$&photon number density & \ref{eq:ngammahv}
&${\bf f}_\gamma$& photon number flux  & \ref{eq:ngammahv}
\\$x$,$x_{\rm HI}$& neutral hydrogen fraction &\ref{eq:dngammadt}
&$\alpha_{\bf f},\alpha_\xi$& artificial dissipation factor & \ref{eq:alphaf}
&$\Lambda$& combined heating and cooling rate &  \ref{eq:intcon}
\\$S$& injection source rate&  \ref{eq:intcon}
&$c$&physical speed of light&
&$\epsilon_\gamma$& photon-ionization heating per ionization&  \ref{eq:photoheatperion} \& \ref{eq:dedt}
\\$\sigma_\gamma$& photon-ionization cross-section&  \ref{eq:dfgammadt} \& \ref{eq:sigmagamma}
&$\Omega_i$ &correction for variable smoothing length &\ref{eq:omegai}
&&& 
\\\hline
\hline
\end{tabular}
\caption{A list of variables, including their symbols, names, and the equations they are first used. }
\label{table:variables}
\end{table*}

\subsection{The radiative transfer equation}
The radiative transfer equation expresses the constancy of the specific intensity ($I$; in ${\rm erg\; cm^{-2}s^{-1} Hz^{-1} sr^{-1}}$) of a beam of light in the absence of sources or sinks and fluid motion \cite[e.g.][]{Pomraning73}
\begin{equation}
    \frac{1}{c}\frac{\partial}{\partial t}I+\hat{\bf n}\cdot \frac{\partial}{\partial{\bf x}}I=\left(\frac{DI}{Dt}\right)_{SS}\,.
    \label{eq:RT}
\end{equation}
Here, $I$ is a function of position (${\bf x}$), direction ($\hat{\bf n}$), frequency ($\nu$) and time ($t$). The right-hand side term represent sources, sinks, and/or the scattering of photons. In the astronomical literature, $I$ is usually called the \lq surface brightness\rq, and just as surface brightness suffers from redshifting, but this is not included in Eq. \ref{eq:RT}. We refer the reader to \cite{Buch83rtff} for the derivation of the complete RT equation. 

{\em Moment methods} drastically simplify the solution of this equation by multiplying Eq.~(\ref{eq:RT}) with some function of direction and integrating the resulting equation over solid angle. This yields an infinite number of moment equations, with the hierarchy closed after a finite number of moments by 
a \lq closure relation\rq. Solving the resulting RT moment equations is the radiation equivalent of solving the fluid equations rather than the collisional Boltzmann equation. We point the interested reader to a sketch of the derivation of these moment equations and the relation to fluid equations in Appendix \ref{sect:appendix-moments}. It is worth recalling that fluid equations, being differential equations, do not properly describe the behaviour of a set of particles in case of discontinuities such as shocks or contact discontinuities. Their numerical integration requires the addition of extra terms (such as \lq artificial viscosity\rq\ or \lq artificial conduction\rq). The same is true for moments of the RT equation, and we describe the discontinuity capturing scheme below, after we introduce the full moment equations for fluid and radiation combined in the next section.

\subsection{Radiation Moments}
We convert the specific intensity $I$ to angular moments by integrating over the solid angle:
\begin{align}
E_\nu&=\frac{1}{\tilde{c}}\int I{\rm d} \Omega,\nonumber\\
{\bf F}^i_\nu&=\int \hat{\bf n}^i\,I{\rm d} \Omega,\nonumber\\
{\mathbb P}^{ij}_\nu&=\frac{1}{\tilde{c}}\int \hat{\bf n}^i\,\hat{\bf n}^j\,I\,{\rm d}\Omega\,,
\label{eq:intIE}
\end{align} 

When additionally integrated over frequency (in the \lq grey\rq\ approximation, e.g. \citealt[][]{Turn01FLD}), those angular moments become radiation energy density $E$, radiation flux ${\bf F}$ and radiation stress tensor ${\mathbb P},$ respectively. Integrating over small frequency intervals, in order to mimic multi-frequency RT, is challenging when Doppler shifts or redshifts are large. This case is not considered here.

The relations between the photon number density, $n_\gamma$, and the radiation energy density, $E$, and between the photon flux, ${\bf F}_\gamma$ and the radiation flux, ${\bf F}$, are
\begin{align}
n_\gamma = \frac{E}{2\pi\hbar\bar{\nu}};\;{\bf F}_\gamma \approx  \frac{\bf F}{2\pi\hbar\bar{\nu}},
\label{eq:ngammahv}
\end{align}
where the mean photon energies $2\pi\hbar\bar{\nu}$ are
\begin{align}
&\hbar\bar{\nu} = \left[\int I{\rm d} \Omega{\rm d} \nu\right] \left[\int (I/\hbar\nu){\rm d} \Omega{\rm d} \nu\right]^{-1};\nonumber\\
&\hbar\bar{\nu} \approx \left[\left|\int \hat{\bf n}I{\rm d} \Omega{\rm d} \nu\right|\right] \left[\left|\int \hat{\bf n}(I/\hbar\nu){\rm d} \Omega{\rm d} \nu\right|\right]^{-1}\,.
\label{eq:meanhnu}
\end{align}
The second relation is a good approximation when the radiation is either isotropic or optically thin.
For reference, the mean photon energy $2\pi\hbar\bar{\nu}$ of ionizing radiation is 29.6eV for a black-body spectrum at $T=10^5{\rm K}$; $\hbar$ is Planck's constant divided by $2\pi$ (see Appendix \ref{sec:heatcoolparam} for details).

We further defined the ratio of the radiation energy density over the fluid's density, $\xi\equiv E/\rho$, and the ratio of radiation flux over the fluid's density as ${\bf f}\equiv{\bf F}/\rho$.

\subsection{Two-moment equations}
\label{sec:two-moment}
The moment equations describing the interaction of gas with radiation are \citep[e.g][]{Buch83rtff,Miha84rhd}:
\begin{align}
\frac{D\rho}{Dt}+\rho{\bf \nabla\cdot} {\bf v}=0,
\label{eq:masscon}
\end{align}
\begin{align}
\frac{D{\bf v}}{Dt}=-\frac{\nabla p}{\rho}-\nabla\phi+\frac{\chi\rho}{\tilde{c}}{\bf f} + {\bf S}_{\bf v},
\label{eq:momcon}
\end{align}
\begin{align}
\frac{Du}{Dt}=-\frac{p}{\rho}{\bf \nabla\cdot\bf v}+\Lambda_{u}+S_{u},
\label{eq:intcon}
\end{align}
\begin{align}
\frac{D\xi}{Dt}=-\frac{1}{\rho}\nabla\cdot(\rho{\bf f}) -\frac{{\bf\nabla v:} \mathbb{P}}{\rho}+\Lambda_{\xi}+S_{\xi},
\label{eq:durad}
\end{align}
\begin{align}
\frac{1}{\tilde{c}^2}\frac{D}{Dt}{\bf f}=-\frac{{\bf \nabla\cdot \mathbb{P}}}{\rho}-\frac{\chi\rho}{\tilde{c}}{\bf f}+{\bf S}_{\bf f},
\label{eq:dfrad}
\end{align}
\begin{align}
\mathbb{P}=\mathbb{F}E=\mathbb{F}\rho\xi\,.
\label{eq:stressrad}
\end{align}
These equations are series expansions including all terms up to $v/c$, in which properties of the radiation field are measured in the local fluid frame. As such, they (partially) account for 
changes in radiation energy density due to fluid velocities \citep{Buch83rtff}. A list of variable descriptions is given in Table~\ref{table:variables}.

Eqs. (\ref{eq:masscon}-\ref{eq:intcon}) express the local conservation of mass, momentum, and energy respectively. The fluid variables are mass density ($\rho$), velocity (${\bf v}$), pressure ($p$), and thermal energy per unit mass ($u$); $\nabla\phi$ is the gravitational acceleration. $D/Dt$ is the Lagrangian time derivative. The terms ${\bf S}_{\bf v}$ and $S_u$ are sources or sinks for the injection of momentum and energy respectively, e.g. due to feedback from stars. The term $\Lambda_u$ is the combined heating and cooling rate.
The case of photo-heating and radiative cooling will be discussed in detail in \S\ref{sec:thermochemisty}.  Finally, the term $({\chi\rho}/{\tilde{c}}){\bf f}$ represents radiation pressure\footnote{Currently, we apply radiation pressure inferred from the quantities averaged over the volume of each particle. However, \cite{Hopk18M1} demonstrated that it is more accurate to apply radiation pressure to the interface between particles, an improvement we intend to implement in future. In the case of ionizing radiation propagating through a low resolution simulation - for example when simulating cosmic reionization - the resulting differences are expected to be small because the radiation imparts little momentum.  However, 
a more accurate treatment of radiation pressure may be required
in high-resolution simulations to capture radiation pressure
from massive stars or active galactic nuclei.}. Here, $\tilde{c}$ is the reduced speed of light (see \S\ref{sec:RSL}) and $\chi$ is the opacity related to the optical depth per unit length as $d\tau/dr=\chi\,\rho$.

Eqs. (\ref{eq:durad}-\ref{eq:dfrad}) express the local conservation of radiation energy and momentum respectively. The radiation variables are radiation energy per unit mass ($\xi$), radiation flux per unit mass (${\bf f}$), and the \lq radiation stress tensor\rq\ (${\mathbb P}$). ${\bf\nabla v:} \mathbb{P}$ is short hand for the contraction $\mathbb{P}^{ij}v_{i,j}$. In Eq.~(\ref{eq:stressrad}), the tensor $\mathbb{F}$ is the Eddington tensor, which we will discuss in \S\ref{sec:momentclosure}. 

Some further source/sink terms appear on
the right-hand sides of Eqs.~(\ref{eq:durad}-\ref{eq:dfrad}). $\Lambda_\xi$ is the rate at which the radiation density changes due to heating and cooling, discussed in more detail in \S \ref{sec:thermochemisty}. $S_{\rm \xi}$ and ${\bf S}_{\bf f}$ are the source terms for radiation energy and flux, respectively. The injection of radiation will be described in more detail in \S\ref{sec:injection}.

In this paper we propagate radiation at the speed $\tilde{c}<c$, which is a \lq reduced\rq\ speed of light. The motivation, validity and limitations of this approximation are discussed in \S\ref{sec:RSL}.

In the two-moment method, the time derivatives of the radiation density and radiation flux are kept, unlike in the case of flux limited diffusion (FLD, \citealt{Leve81FLD}). There are some advantages in keeping this term. Firstly, \cite{Buch83rtff} showed that the time derivative of ${\bf f}$ may be significant in the optically thin (free streaming) regime, making the two-moment method more accurate than FLD. Secondly, because of this time derivative, M1 can maintain the direction of the radiation, whereas in FLD the radiation follows the gradient in energy density and hence incorrectly goes around corners in the optically thin limit.

Finally, including the time derivative yields hyperbolic equations rather than the parabolic equation of FLD. Solving a parabolic differential equation explicitly requires 
a more restrictive timestep, $\Delta t\propto  (\Delta x)^2/\tilde{c}$, compared to the hyperbolic case where
$\Delta t\propto  \Delta x/\tilde{c}$; where $\Delta x$ is the spatial resolution. Combined with using a reduced speed of light approximation ($\tilde c$ rather than $c$) improves the efficiency of the RT implementation compared to FLD\footnote{The M1 method can be faster even if FLD is solved implicitly because the inversion step in the implicit solver is expensive (see, e.g. \citealt{Skin13M1}).}.

\subsection{Closure relation}
\label{sec:momentclosure}
Taking successive angular moments of the RT relation leads to an infinite set of coupled moment equations \citep{Miha84rhd}. A \lq closure relation\rq\, which relates higher order moments to lower-order ones, is required to break this hierarchy. Unfortunately, the closure relation is not unique and depends on the problem at hand. \cite{Leve84} derived a closure relation as follows. We
consider the RT equations assuming ${\bf v}=0$ and additionally neglecting the $\Lambda$ and $S$ terms. Then, Eqs.~(\ref{eq:durad}) and (\ref{eq:dfrad}) simplify to
the following two moment equations:
\begin{align}
\frac{\partial E}{\partial t}=-\nabla\cdot{\bf F}\,, 
\label{eq:duradS}
\end{align}
\begin{align}
\frac{1}{\tilde{c}^2}\frac{\partial }{\partial t}{\bf F}=-{\bf \nabla\cdot \mathbb{P}}-\frac{\chi\rho}{\tilde{c}}{\bf F}\,.
\label{eq:dfradS}
\end{align}
 Provided that the radiation field is symmetric around a given direction $\hat{\bf n}$, \cite{Leve84} demonstrated that the second moment can be written as
\begin{align}
&\mathbb{P}=\mathbb{F}E=\frac{E}{2}(1-f_{\rm Edd})\mathbb{I}+\frac{E}{2}(3f_{\rm Edd}-1){\bf \hat{n}}{\bf \hat{n}},
\label{eq:eddclose}
\end{align}
where $f_{\rm Edd}$ is called the \lq Eddington factor\rq. 

When the radiation field is {\em almost} isotropic, $\mathbb{P}_{ij}\approx (E/3)\delta_{ij}$ which corresponds to $f_{\rm Edd}=1/3$. Combining the two moment equations with this relation yields
\begin{align}
\frac{\partial E}{\partial t} = \nabla\cdot\left(\frac{\tilde c}{3\chi\rho}\nabla E\right)-\nabla\cdot\left( \frac{1}{\tilde{c}\chi\rho}\frac{\partial }{\partial t}{\bf F}\right)\,.
\label{eq:optthickdelP}
\end{align}
This describes isotropic diffusion of the energy density,  $E$, in case the rate of change of the flux (the last term on the right hand side) can be neglected. Of course, if the radiation field were exactly isotropic everywhere it has to be uniform as well - but this diffusion approximation can be used, provided $E$ varies sufficiently slowly in space and time \citep{Leve84}. This case corresponds to the classical \lq Eddington\rq\ approximation for the propagation of radiation in the isotropic case, and we will refer to as the \lq optically-thick\rq\ solution.

In contrast, the value $f_{\rm Edd}=1$ leads to anisotropic radiation propagation with
\begin{align}
{\bf \nabla\cdot \mathbb{P}}= \hat{\bf n}\left(\hat{\bf n}\cdot\nabla E\right)\,.
\label{eq:optthindelP}
\end{align}
In this \lq optically-thin\rq\ case, 
\begin{align}
    \frac{\partial^2E}{\partial t^2}+\tilde c^2
    (\hat{\bf n}\cdot\nabla)^2 E=-{\tilde c}\chi\rho\frac{\partial E}{\partial t}\,,
    \label{eq:streaming}
\end{align}
and radiation \lq streams\rq\ in direction $\hat{\bf n}$ with speed $\tilde c$, with its intensity decreasing due to absorption as quantified by the right hand side of the equation.

The closure relation of Eq.~(\ref{eq:eddclose}) therefore 
captures the propagation correctly in the two limiting cases
of (1) high-optical depth, with the solution describing isotropic diffusion, and (2) the optically-thin regime of negligible optical depth, where the solution describes free propagation at the speed
${\tilde c}$ in the characteristic direction ${\bf \hat{n}}$.
The expectation is then that Eq.~(\ref{eq:eddclose}) also 
provides a good approximation to any intermediate case \citep{Leve81FLD}.

One disadvantage of the scheme is that radiation behaves as a \lq collisional\rq\ fluid: beams of light with different propagation directions $\hat{\bf n}$ that intersect will collide. This is because
the local Eddington tensor closure relation of Eq. (\ref{eq:eddclose}) can only handle one direction $\hat{\bf n}$ at a time (in addition to an isotropic component). We will discuss this issue in more details in \S\ref{sec:radcoll}.

\subsubsection{Choice of Eddington factor}
Next we turn to the choice of $f_{\rm Edd}$. As shown by \cite{Leve84}, given that ${\bf F}/(\tilde{c}E)$ and $\mathbb{P}/E$ are the first and second moments of a non-negative unit density variable requires that
\begin{align}
\varepsilon^2= \left|\frac{\bf F}{\tilde{c}E}\right|^2=\left|\frac{\bf f}{\tilde{c}\xi}\right|^2\leq f_{\rm Edd} \leq 1,
\label{eq:feddcon}
\end{align}
which we term the  `original' closure.

Of course, even if we demanded that the Eddington factor should only depend 
on the local values of ${\bf F}$ and $E$, then this would not specify $f_{\rm Edd}$ uniquely (see \citealt{Leve84} for a summary of reasonable choices). One particular choice is the `M1' closure, which \cite{Leve84} derived by assuming that there exist inertial frames in which the radiation density is isotropic (not necessarily isotropic in the lab or fluid frame). This original
M1 relation is
\begin{align}
f_{\rm Edd} = \frac{3+4\varepsilon^2}{5+2\sqrt{4-3\varepsilon^2}}.
\label{eq:feddest}
\end{align}
\cite{Dubr99M1} showed that this corresponds to the simplest moment closure that maximizes the entropy and is anisotropic\footnote{Note that in the mathematics community, the entropy has the opposite sign compared to that in the physics community.}.

The evaluation of this expression for M1 is computational efficient as well as highly parallelisable, as compared to e.g. ray-tracing or Monte Carlo methods, because $f_{\rm Edd}$ depends only on local quantities. Because of this, several astrophysical RT implementations use this M1 closure relation, e.g. \cite{Gonz07M1,Aube08M1,Skin13M1,Rosd13ramsert,Kann19AREPORT,Skin19FORNAX}.
However, this choice is not without its problems (as are
other variants based on local variables). Firstly, consider the case
of two otherwise identical beams of radiation propagating in opposite directions. Where the beams hit the net flux is zero, $|{\bf f}/(\tilde{c}\xi) |\sim 0$ so that $\varepsilon=0$ and $f_{\rm Edd}=1/3$: this corresponds to the optically-thick solution, even in the system were optically thin. It is as if the beams of radiation collide with each other \citep[see also][]{Rosd13ramsert}. Clearly, this behaviour is incorrect.

This choice of closure relation also results in artificial dispersion, since radiation does not move at the same speed when $|{\bf f}/(\tilde{c}\xi) |$ varies: radiation propagates with speed between
$\tilde{c}/3$ and $\tilde{c}$, when Eq.~(\ref{eq:optthickdelP})
or Eq.~(\ref{eq:streaming}) applies, respectively.

An improved closure relation can be derived from the following considerations. In the optically thick limit, we desire that $f_{\rm Edd}=1/3$, since the corresponding isotropic diffusion captures the
random walks of photons through the medium as a consequence of numerous independent scattering events. In the opposite limit of 
an optically thin medium, we desire that $f_{\rm Edd}=1$, since that correctly describes streaming of radiation at the speed of light. Note that in this strategy, we set $f_{\rm Edd}=1$ only according to the optical depth ($\tau$) and independent of $|{\bf f}/(\tilde{c}\xi) |$, since the latter can be small even in the optically thin regime, e.g. head-on collision. Finally, we require that $f_{\rm Edd}\le 1$. 

%{\color{red} Hence, we only require that $f_{\rm Edd}\sim 1$ (as well as $\varepsilon\sim 1$) in the optically thin limit, whereas $f_{\rm Edd}\sim 1/3$ in the optically thick case (as in the original M1 scheme).}

Our proposed `modified' M1 closure relation is
\begin{align}
\varepsilon=\max\left[\exp(-\tau),|{\bf f}/(\tilde{c}\xi)|\right],
\label{eq:codeest}
\end{align}
where $\tau\equiv\chi\rho h$ is the local optical depth across the extent $h$ of a resolution element. This choice satisfies $\varepsilon\le 1$, and has the correct limiting behaviour. In the optically thin limit ($\tau\rightarrow 0$), $\varepsilon\rightarrow 1$, while in the optically thick case ($\tau\rightarrow\infty$), $\varepsilon\rightarrow |{\bf f}/(\tilde{c}\xi)|\rightarrow 0$, since the flux $|{\bf f}|$ is small. In case $|{\bf f}|$ is small due to the \lq collision\rq\ of two beams of radiation, $\varepsilon$ can still be of order 1 and describe radiation streaming rather than diffusion provided the optical depth is small. We will demonstrate in \S\ref{sec:radcoll} that our modified M1 closure (Eq.\ref{eq:codeest}) can handle head-on beam collisions and more generally, 1D RT problems.

We choose to modify M1 by the factor $\exp(-\chi\rho h)$ to mimic the diffusion of radiation when the optical depth is large.
The choice is also motivated by a desire to help numerical convergence: the combined contributions of two resolution elements, for example two SPH particles with extents $h_i$ and $h_j$, 
is approximately $\exp(-\chi\rho h_i)\times\exp(-\chi\rho h_j)=\exp[-\chi\rho (h_i + h_j)]$, which corresponds to the approximate effect of a lower resolution SPH particle with size $(h_i + h_j)$.

We will also show in Fig.\ref{fig:radstream2d_difdiss} that our scheme is more stable than the original M1 closure in optically thin regions when simulated with SPH. However, the scheme does 
not solve the problem of the artificial collision of radiation beams in case they are not head on (\S\ref{sec:radcoll}), since the radiation directions will still merge locally according to Eq.~(\ref{eq:eddclose}). Fortunately, even in this case, our closure (Eq. \ref{eq:codeest}) will still prevent the numerical diffusion in the optically thin limit.

Finally, we justify the use of physical quantities other than radiation energy and flux in the Eddington factor. The Eddington tensor should be derived from the RT equation, which contains information about the gas, e.g. density, velocity, and the opacity (through the collisional term). Thus, the Eddington tensor should be also a function of these gas properties. In fact, in the absence of the collision term ($\chi=0$), the radiation should always be free streaming at the the speed of light regardless of the value of $\xi$ and ${\bf f}$. 

\subsection{SPH forms for the two moment method and the numerical solution to the propagation equation}
\label{sec:SPHform}
In the standard formulation of SPH \citep[e.g][]{Mona02sphtur}, the density, $\rho_i$, at the location of particle $i$ is calculated through interpolating over \lq neighbouring\rq\ particles in a gather approach,
\begin{align}
\rho_i =\sum_jm_jW_{ij}(h_i),
\end{align}
where the kernel $W_{ij}(h_i)=W(|{\bf r}_i-{\bf r}_j|,h_i)$ is a function with compact support (by default the $M_4$ cubic B-spline function), $h_i$ is the smoothing length and $m_i$ the mass of particle $i$. We follow the variable smoothing length treatment similar to that in \cite{Spri02esph} such that the number of neighbour particles that contribute to the sum is $N_{\rm ngb}$(=48 in 3D) (see the {\small GADGET-2} SPH section in \citealt{Scha15EAGLEhydro} for more details, including the SPH formulation of the hydrodynamics in {\small SWIFT}).

In the radiation hydrodynamics tests presented below, we do not use the standard SPH formulation but rather the modifications
introduced by \cite{Borr20SPHENIX} called {\small SPHENIX}, which uses the density and energy hydrodynamic variables, rather than pressure and energy. {\small SPHENIX} applies the \cite{Cull10} shock detector to minimize artificial viscosity away from shocks and the artificial diffusion term to capture fluid mixing described by \cite{Pric08artcon}.

One of the main hurdles to overcome for implementing a moment method in SPH is that such a higher-order method requires the calculation of derivatives, and these tend to be noisy when the particle distribution is irregular. For example, there are several ways to estimate the divergence of a vector field ${\bf X}$, which include \citep[e.g.][]{Tric12divBclean} the
`{\it symmetric}' estimate:
\begin{align}
\left.\left ( \nabla\cdot{\bf X} \right )_i\right|_{\rm sym} = \rho_i\sum_jm_j&\left[\frac{{\bf X}_i}{\Omega_i\rho_i^2}\cdot\nabla_iW_{ij}(h_i)\right.\nonumber\\&\left.+ \frac{{\bf X}_j}{\Omega_j\rho_j^2}\cdot\nabla_iW_{ij}(h_j)\right ],
\label{eq:sym}
\end{align}
and the `{\it difference}' estimate:
\begin{align}
\left.\left ( \nabla\cdot{\bf X} \right )_i\right|_{\rm diff} = -\sum_j\frac{m_j}{\Omega_i\rho_i}\left ( {\bf X}_i-{\bf X}_j \right )\cdot\nabla_iW_{ij}(h_i),
\label{eq:dif}
\end{align}
where ${\bf X}$ is an arbitrary vector or tensor associated with each particle, and
\begin{align}
\Omega_i = 1+\frac{h_i}{3\rho_i}\sum_jm_j\frac{\partial W_{ij}(h_i)}{\partial h_i},
\label{eq:omegai}
\end{align}
is a correction factor introduced by \cite{Spri02esph} to account for spatial variations in the value of the smoothing length, $h$.

We use the difference form to evaluate
Eqs.(\ref{eq:durad}) and (\ref{eq:dfrad}):

\begin{align}
&\left.\left(\frac{D\xi_i}{Dt}\right)\right|_{\rm diff}=\left.\left[-\frac{1}{\rho}\nabla\cdot(\rho{\bf f})\right]\right|_{i}\nonumber\\
&=-\sum_j \frac{m_j}{\Omega_i\rho_i^2}(\rho_i{\bf f}_i-\rho_j{\bf f}_j)\cdot\nabla_i W_{ij}(h_i),
\label{eq:sphxi}
\end{align}
and 
\begin{align}
&\left.\left(\frac{1}{\tilde{c}^2}\frac{D {\bf f}_i}{Dt}\right)\right|_{\rm diff}=\left.\left[-\frac{{\bf \nabla\cdot \mathbb{P}}}{\rho}-\frac{\chi\rho}{\tilde{c}}{\bf f}\right]\right|_{i}\nonumber\\
&=- \sum_{j}\frac{m_j}{\Omega_i\rho_i^2}\left ( \rho_i\xi_i\mathbb{F}_i -\rho_j\xi_j\mathbb{F}_j\right )\cdot\nabla_i W_{ij}(h_i)-\frac{\chi_i\rho_i}{\tilde{c}}{\bf f}_i.
\label{eq:SPHdfrad}
\end{align}
The difference formulation subtracts the zeroth-order errors
that occur in SPH explicitly, yielding first-order accuracy regardless of the underlying particle distribution. This results in superior numerical estimates of the divergence particularly near steep gradients. However, the difference estimate does not manifestly conserve flux, unlike the `symmetric' estimate\footnote{The `symmetric' SPH form can also help to regularize the particle distribution in hydrodynamics calculations,
albeit by introducing purely numerical forces \citep{Pric12SPHMHD}. This is less important when propagating ionising radiation which
does not usually exert strong forces on the gas particles.}. Fortunately, we find that the level of non-conservation of flux is
small in our experiments (Typically less than one percent, 
better accuracy could be reached by increasing the order of the scheme, if required.). There is no known formulation that simultaneously avoids zeroth-order errors and is manifestly conservative in SPH \citep[see the discussion in][]{Pric12SPHMHD}. 

We add the term $-\chi_i\rho_i{\bf f}_i/\tilde{c}$ to Eq.~(\ref{eq:dfrad}) using operator splitting,
\begin{align}
{\bf f}_i(t+\Delta t)=\exp(-\chi_i\rho_i \tilde{c}\Delta t)\times{\bf f}_i(t).
\end{align}
Though this scheme in unconditionally stable, it nevertheless 
yields the wrong answer when the time-step, $\Delta t$, is too long. This could be avoided by limiting the time-step to $\Delta t\le 1/(\chi\rho \tilde{c})$, but that would result in unacceptably
short time-steps in regions of high optical depth. Since in such regions the impact of radiation may be small anyway, we will limit the time-step by the usual\footnote{In the time-step determination, we will use the smallest $h$ of all neighbouring SPH particles and of the particle itself, to ensure stability and conservation.} Courant–Friedrichs–Lewy (CFL) condition, $\Delta t\le 0.1h/\tilde{c}$. To ensure that our results are physically meaningful, numerically stable and satisfy causality, we apply the following additional limiters at the beginning of each time-step: ({\em i})
$|{\bf f}|\le {\tilde c}\xi$, ({\em ii}) $\xi\ge 0$, and ({\em iii})
we zero unused components of ${\bf f}$ in 1D or 2D simulations.
The latter limiter corrects for any numerical scatter of radiation into unused dimensions, as may happen if the Eddington tensor is non-zero but $|{\bf f}|$ is small.

\subsubsection{Optically thin radiation}
\label{sec:opticalthindir}
In general, we set the propagation direction, $\hat{\bf n}$, to be that of the local flux, $\hat{\bf n}=\hat{\bf f}$. However, 
radiation propagates in a constant direction in the optically thin case. Since the flux is computed numerically, round-off errors or
numerical noise may rotate the flux vector so that imposing $\hat{\bf n}=\hat{\bf f}$ does not guarantee that radiation travels in a straight line, even in the optically thin case.

In some special cases, for example of light emanating from a single point source or the propagation of a plane parallel radiation front, the direction $\hat{\bf n}$ is known a priori, and we can therefore choose to simply impose the propagation direction, and use that direction to compute the optically thin Eddington tensor, Eq.~(\ref{eq:eddclose}).

Not surprisingly, imposing the direction of radiation propagation yields spherical ionization regions around a point source (Fig.\ref{fig:stromgren3d_tm}) and casts sharp shadows behind optically thick absorbers even at low resolution (Fig.\ref{fig:radshadow3d}). Of course in general, the propagation direction $\hat{\bf n}$ is not generally known, for example, there may be several sources or an additional isotropic background, which require improvements of our scheme.

\subsection{Discontinuity-capturing dissipation terms}
\label{sec:artdiss}
The fluid equations encoded in SPH are differential equations and hence need to be supplemented with extra terms in order to correctly capture discontinuities such as shocks and contact discontinuities.
These terms broaden discontinuities by introducing numerical
dissipation so that they can be resolved by the interpolation scheme \citep[see e.g][]{Mona97SPHRiem, Ager07SPH, Pric08artcon}.

The SPH implementation of the moment method needs to be extended with similar dissipation terms to handle discontinuities in the radiation field, and we base these on the artificial diffusion and artificial viscosity terms of the SPH fluid equations. The energy dissipation term is similar to the {\it artificial diffusion} term in SPH hydrodynamics,
\begin{align}
\left.\left(\frac{{\rm D}\xi_i}{{\rm D} t}\right)\right|_{\rm diss}=\sum^N_{j=1}D_{\xi,ij}\frac{m_j}{\bar{\rho}^2}(\tilde{\rho_i\xi_i}-\tilde{\rho_j\xi_j})\frac{{\bf \hat{r}}_{ij}\cdot\overline{\nabla_i W_{ij}}}{r_{ij}},
\label{eq:xidiss}
\end{align}
where $\bar{\rho}=\sqrt{\rho_i\rho_j}$ is the geometric mean of
the densities of the pair of interacting particles $i$ and $j$. If the density contrast is larger than 10, we found the scheme to be more stable with the choice $\bar{\rho}=\min(\rho_i,\rho_j)$, but this choice is not used in the tests in this paper.

Flux dissipation can modelled with the {\it artificial viscosity} term in SPH hydrodynamics\footnote{This is not our default choice,  see Eq.~(\ref{eq:fdissaniso}) below.},
\begin{align}
\left.\left(\frac{{\rm D}{\bf f}_i}{{\rm D} t}\right)\right|_{\rm diss}=\left\{\begin{matrix}
\sum^N_{j=1} D_{{\bf f},ij}\frac{m_j}{\bar{\rho}^2}(\rho_i{\bf f}_i-\rho_j{\bf f}_j)\cdot{\bf \hat{r}}_{ij}\frac{\overline{\nabla_i W_{ij}}}{r_{ij}},\\{\rm if}\; (\rho_i{\bf f}_i-\rho_j{\bf f}_j)\cdot{\bf \hat{r}}_{ij}<0,\\ \\   
0, \;{\rm otherwise}.
\end{matrix}\right.
\label{eq:fdiss}
\end{align}
In these expression, $D_{\xi, ij}$ and $D_{{\bf f}, ij}$ are the \lq artificial dissipation\rq\ coefficients, they have the units of
a diffusion constant and we write them as
\begin{align}
D_{\xi,ij} = \alpha_{\xi,i} v_{{\rm sig},i}h_{i} +\alpha_{\xi,j} v_{{\rm sig},j}h_{j},\nonumber\\
D_{{\bf f},ij} = \alpha_{{\bf f},i} v_{{\rm sig},i}h_{i} +\alpha_{{\bf f},j} v_{{\rm sig},j}h_{j}\,.\nonumber\\
\label{eq:artDalpha}
\end{align}
Here, $v_{\rm sig}$ is the signal speed,

\begin{align}
v_{\rm sig} = |{\bf \hat{f}}\cdot{\bf \hat{r}}_{ij}| \tilde{c},,
\label{eq:signalkappa}
\end{align}
and $(\alpha_{\bf f},\alpha_\xi)\leq 1$ are dimensionless numbers 
that quantify the strength of the numerical dissipation. The forms of Eqs.~(\ref{eq:xidiss}) and (\ref{eq:fdiss}) are consistent with the Riemann solver across the boundary of two SPH particles \citep{Mona97SPHRiem}. The kernel averaged over smoothing length is $\overline{\nabla_i W_{ij}} = 0.5[\nabla_iW_{ij}(h_j)+\nabla_iW_{ij}(h_j)]$.

Eqs.~(\ref{eq:xidiss}) and (\ref{eq:fdiss}) are 
diffusion equations \citep[see e.g.][]{Jube04thcon,Pric08artcon}.
The maximum value of the diffusion speed is $h\tilde{c}$, where $h$ is the particle size and $\tilde{c}$ is the propagation speed of the radiation. A numerical diffusion coefficient larger than this maximum value will result in numerical instabilities if the time-step is set by the CFL condition, $\Delta t\lesssim h/\tilde{c}$. 

\cite{Pric08artcon} set $\tilde{\rho_i\xi_i}=\rho_i\xi_i$, the values associated with individual SPH particles, and minimized the amount of numerical dissipation by choosing how the signal speed depends on local quantities. However, another way to minimize artificial dissipation is by reconstructing fluid quantities at the interface between particles \citep[see e.g.][]{Fron17CRKSPH,Ross19MAGMA2}.
To do so, we reconstruct the radiation energy density at the interface using a Taylor series expansion,
\begin{align}
&\tilde{\rho_i\xi_i}-\tilde{\rho_j\xi_j}  = |\hat{\bf f}\cdot\hat{\bf r}_{ij}|\{(\rho_i\xi_i-\rho_j\xi_j) \nonumber\\
&+\phi [\frac{h_i}{h_i+h_j}{\bf r}_{ji}\cdot\nabla(\rho_i\xi_i)-\frac{h_j}{h_i+h_j}{\bf r}_{ij}\cdot\nabla(\rho_j\xi_j)]\},
\label{eq:highorderdiss}
\end{align}
where $\phi$ is the slope limiter (implemented using the {\em minmod} function, {\em minmod}($x$)={\em max}(0,{\em min}($x$,1))
to minimize spurious oscillations. The term $\hat{\bf f}\cdot\hat{\bf r}_{ij}$ limits unwanted dissipation perpendicular to direction of the flux. We find that for $\alpha_\xi=1$, discontinuity-capturing is good while dissipation is small in smooth regimes. A discontinuity \lq detector\rq\\ for artificial diffusion is therefore not required.

\cite{Chow97SRSPH} suggested turning off artificial dissipation when $(\rho_i{\bf f}_i-\rho_j{\bf f}_j)\cdot{\bf \hat{r}}_{ij}>0$, in order to reduce unnecessary diffusion, e.g. behind a discontinuity. However in our experiments we found that such a switch makes the scheme unstable, in particular in cases
where radiation beams collide in the optically thin regime.
The instability results in significant non-conservation of energy. We therefore do not use the \cite{Chow97SRSPH} switch, but instead
apply a discontinuity detector to minimize the artificial viscosity as described in \S\ref{sec:dissswitch}.

The artificial flux dissipation term of Eq.~(\ref{eq:fdiss}) causes numerical dissipation of the flux in directions perpendicular to the flux. This is problematic, in particular in the optically thin case\footnote{In the optically thick regime, $\chi$ already provides the necessary dissipation. This is one of the reason why the flux-limited diffusion does not require artificial dissipation. Another reason is that there are no artificial oscillations when solving a diffusion equation).} where it leads to the destruction of a packet of radiation as shown in Fig.\ref{fig:radstream2d_difdiss}. To avoid this, it requires
that any artificial flux dissipation should be in the direction of the flux itself. Simply multiplying the right hand side of Eq.~(\ref{eq:fdiss}) by $\hat{\bf f}\cdot\hat{\bf r}_{ij}$ does not work: any component of numerical flux perpendicular to the actual flux, for example due to numerical noise, will still lead to the artificial destruction of an optically thin radiation packet\footnote{It is possible to use Eq. \ref{eq:fdiss} without disrupting radiation directions if the optically thin direction is imposed, as in \S\ref{sec:opticalthindir}. In this case, we will only consider the dissipation component (in Eq. \ref{eq:fdiss}) along the optically thin direction, and only consider the flux difference in that direction.}. 

A better solution is to implement the dissipation scheme
as as anisotropic diffusion\footnote{`Anisotropic artificial viscosity' was also considered by \citealt{Owen04artvisaniso}, but our SPH form is simpler and different from theirs.}.

Here we outline our {\it default} choice of {\it artificial flux dissipation}. We begin by rewriting Eq.~(\ref{eq:fdiss}) in the form of an anisotropic diffusion equation,
\begin{align}
\left.\left(\frac{{\rm D}{\bf f}}{{\rm D} t}\right)\right|_{\rm diss} &= \frac{1}{\rho} \nabla\cdot\left[ \mathbb{D}^{\bf f}\nabla\cdot(\rho{\bf f})\right],\nonumber\\
&\equiv\frac{1}{\rho} \nabla\cdot\left[ \rho\mathbb{D}^{\bf f}\,\psi\right]\,,
\label{eq:fdissaniso}
\end{align}
where the tensor $\mathbb{D}^{\bf f}$ and the scalar $\psi$ are given by
\begin{align}
\mathbb{D}^{\bf f} &=\alpha_{\bf f} v_{\rm sig} h \hat{{\bf n}}\hat{{\bf n}}\nonumber\\
\psi &\equiv \rho^{-1}\,\nabla\cdot(\rho{\bf f})\,.
\label{eq:fdissaniso2}
\end{align}
We implement the diffusion equation in SPH as
\begin{align}
\psi_i &=-\sum_j \frac{m_j}{\Omega_i\rho_i^2}(\rho_i{\bf f}_i-\rho_j{\bf f}_j)\cdot\nabla_i W_{ij}(h_i)\,,
\label{eq:psianiso}\\
\left.\left(\frac{{\rm D}{\bf f}}{{\rm D} t}\right)\right|_{i,{\rm diss}}
&=-\sum_{j} \frac{m_j}{\Omega_i\rho_i^2}\left (\rho_i\psi_i\mathbb{D}^{\bf f}_i - \rho_j\psi_j\mathbb{D}^{\bf f}_j\right )\cdot\nabla_i W_{ij}(h_i).
\end{align}

This formulation of anisotropic viscosity in SPH is novel and we suggest that it may be applicable to other situations as well,
for example when implementing magneto-hydrodynamics or cosmic ray propagation.

\subsubsection{A switch for applying flux dissipation}
\label{sec:dissswitch}
Clearly it would be advantageous to activate flux dissipation only near discontinuities in the radiation, which requires
efficient detection of such discontinuities. Such switches are also regularly used to activate dissipation in the SPH equations for hydrodynamics itself. 

\cite{Morr97artvisswitch} proposed to use the divergence of the velocity as a measure of how discontinuous the fluid flow evolves, but this cannot distinguish compression - which conserves entropy - from true discontinuities. In addition, flux dissipation may be activated unnecessarily in the case of wave-like disturbances. \cite{Ross19entropyswitch} suggested to use changes in entropy as a discontinuity  detector, however
it is not clear how to apply this in the case of radiation.
\citep{Cull10} suggested to track the time derivative
of the velocity divergence, $\nabla\cdot {\bf v}$, so that the diffusion coefficient is of the form $h^2\,|\dot\nabla{\bf v}|/v_{\rm sig}^2$, where $v_{\rm sig}$ is the signal velocity. This effectively corresponds to a switch that is based on the second time-derivative of the density and hence can distinguish between gas in the pre- and post-shock regions. Such a switch is implemented in {\small SPHENIX}.

Inspired by the \cite{Cull10} switch and after experimenting with various forms of how their expression can be applied to the case of radiation, we settled on the following target value for the diffusion coefficient,
\begin{align}
\alpha_{{\bf f},{\rm aim}}={\rm min}\left({\rm max}\left\{-\frac{Ah^2}{\rho\xi \tilde{c}^2}\frac{{\rm D}[\nabla\cdot(\rho{\bf f})]}{{\rm D}t},0\right\},1\right)\,.
\label{eq:alphaf}
\end{align}
The denominator is $\rho\xi \tilde{c}^2$ rather than $\rho|{\bf f}|\tilde{c}$ because $\rho c\xi\gg|\rho{\bf f}|$ in the optically thick limit where artificial dissipation is not needed. $A(=200)$ is a constant multiplication factor to compensate for the large $\tilde{c}^2$ in the equation.

Upstream from a discontinuity, we require that the diffusion coefficient be large enough so that the discontinuity can be captured by the interpolation scheme. Downstream from the discontinuity, a smaller level of diffusion is still required to suppress any numerical oscillations. We follow \cite{Morr97artvisswitch} and implement this by making the diffusion coefficient time-dependent, as follows: (1) when $\alpha\le \alpha_{\rm aim}$, we set $\alpha=\alpha_{\rm aim}$; (2) when $\alpha\ge \alpha_{\rm aim}$, we evolve $\alpha$ back to $\alpha_{\rm aim}$ by solving
\begin{align}
\frac{{\rm D}}{{\rm D}t}(\alpha-\alpha_{\rm aim})=-\frac{1}{\tau_{\rm relax}}=-(\tilde{c}/h +\tilde{c}\chi\rho),
\end{align}
where $\tau_{\rm relax}$ is the relaxation time scale.
The term $\tilde{c}\chi\rho$ ensures that a large value for the $\alpha$ quickly relaxes back to the target value in the optically thick yet smooth regime.  Finally, for gas particles in which we inject radiation we set $\alpha=1$ to better capture any discontinuities associated with radiation injection.

Before ending this section, we comment on the required number of SPH neighbour loops associated with our RT scheme. If the radiation moment and hydrodynamics equations (Eqs. \ref{eq:masscon}-\ref{eq:dfrad}) are solved simultaneously, then at least three neighbour loops are required to compute the
right-hand-size of the anisotropic diffusion equation,
Eq.~(\ref{eq:fdissaniso}): the variable $\psi$ in Eq.~(\ref{eq:fdissaniso2}) requires (1) a loop to compute $\rho$ and (2) a second loop to compute the gradient; and finally the scheme requires (3) a third loop to compute the gradient of $\psi$ Eq.~(\ref{eq:fdissaniso}). Similarly, the dissipation 
switch of Eq.~(\ref{eq:alphaf}) requires three loops.
The scheme may be optimized by solving the radiation transport equation on a shorter time-scale than used to update the hydrodynamics. During such sub-cycling, the density is kept a constant, in which case the radiation transport only requires two SPH neighbour loops. We will report on such improvements elsewhere.

\subsection{Tests of the Eddington tensor closure and artificial dissipation schemes}
\label{sec:testartdiss}
\begin{figure}
 \includegraphics[width=0.48\textwidth]{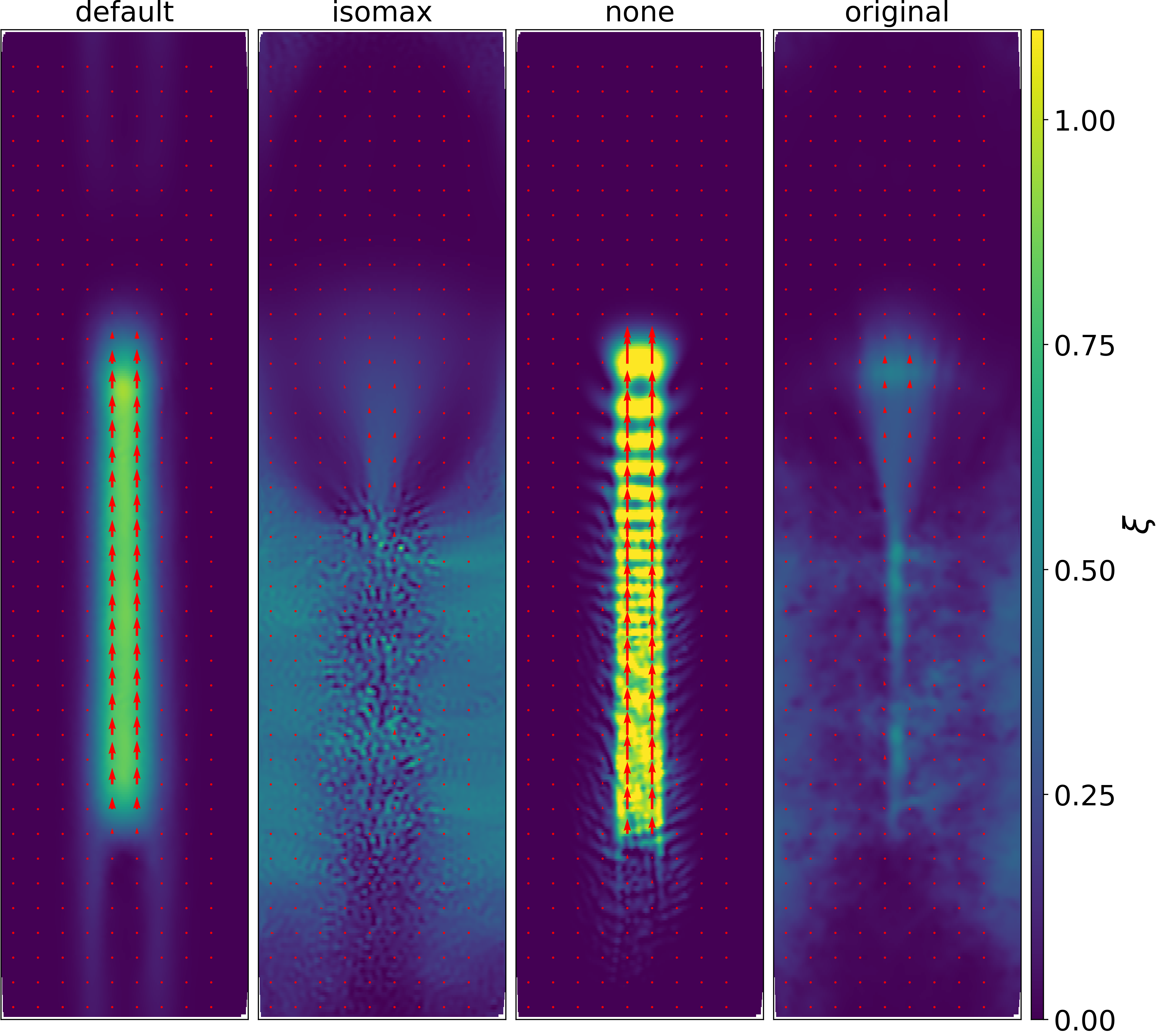}
\caption{A packet of radiation propagating upwards (from $x=0$ to $x=2$)
over a 2D glass-like distribution of $48\times192$ particles shown at time $t=0.4$; the reduced speed of light is $\tilde{c}=1$ and the extent of the simulation volume is $\Delta x=2$ in the vertical direction and $\Delta y=0.5$ in the horizontal direction. Colours represent the radiation energy density $\xi$ and small red-arrows the radiation flux density, ${\bf f}$. Panels from left to right illustrate the {\it default}, {\it isomax}, {\it none} and {\it original} choices for the artificial dissipation and closure scheme (see text). The optical depth of the medium is zero and the packet should propagate freely at the speed of light while retaining its square form. With isotropic artificial dissipation ({\it isomax}) or with the {\it original} closure scheme, the radiation packet incorrectly dissolves quickly. The case without artificial dissipation ({\it none}) results in strong oscillations which leads to significant non-conservation of energy. The {\it default} choice correctly maintains the morphology of the beam while suppressing artificial numerical oscillations.}
\label{fig:radstream2d_difdiss}
\end{figure}
A test of the artificial dissipation scheme and the choice of the Eddington tensor closure by propagating a single short beam of light is illustrated in Fig.~\ref{fig:radstream2d_difdiss}. The underlying $48\times192$
particle distribution is glass-like. The figure shows four choices
of the artificial dissipation and closure relation labelled {\it default}, {\it isomax}, {\it none}, and {\it original}:
\begin{itemize}
    \item {\it default}: uses the {\it default} artificial dissipation
    of Eqs.~\ref{eq:xidiss}, \ref{eq:highorderdiss}, and \ref{eq:fdissaniso} described in \S\ref{sec:dissswitch}, and the {\it default} modified closure M1 scheme of Eq.~(\ref{eq:codeest});
    \item {\it isomax}: uses isotropic artificial dissipation of Eqs.~ \ref{eq:xidiss} and \ref{eq:fdiss}, setting $\alpha_{\bf f}=\alpha_\xi=1$, see Eq.~(\ref{eq:artDalpha}), and the {\it default} modified closure scheme;
    \item {\it none}: does not use any artificial dissipation (i.e. $\alpha_{\bf f}=\alpha_\xi=0$ in Eq.~\ref{eq:artDalpha}) and the {\it default} closure scheme;
    \item {\it original}: uses the {\it default} artificial dissipation but the `{\it original}' closure Eq.~(\ref{eq:feddcon}).
\end{itemize}
Figure~\ref{fig:radstream2d_difdiss} demonstrates that artificial diffusion is needed to suppress the artificial oscillations seen in panel {\it none}:
oscillations lead to non-physical negative values of $\xi$, which, if zeroed, lead to a catastrophic artificial increase in radiation energy. However, such diffusion should be anisotropic to avoid that the beam artificially diffuses perpendicular to the propagation direction as seen in panel {\it isomax}. The {\it original} scheme fails to preserve the beam's shape: it does not handle well non-uniform particle distributions.
Fortunately, our {\it default} scheme preserves the beam shape,  suppresses artificial oscillations and conserves energy.

\subsection{Thermo-chemical processes}
\label{sec:thermochemisty}
In this section we briefly describe how we implement the interaction between matter and radiation, limiting the discussion to the particular case of ionizing radiation in a pure hydrogen gas.

\subsubsection{Pure Hydrogen Gas thermo-chemistry}
The processes of collisional ionization and photo-ionization, photo-heating, and collisional and radiative cooling in a hydrogen gas are
\citep[e.g.][]{Aube08M1}:
\begin{align}
\frac{\partial n_\gamma}{\partial t}&=\frac{\rho(\Lambda_\xi+S_\xi)}{2\pi\hbar\bar{\nu}}\nonumber\\
&=-n_{{\rm HI}}\tilde{c}\sigma_\gamma n_\gamma + n_en_{{\rm HII}}(\alpha_A-\alpha_B)+S_\gamma\,,
\label{eq:dngammadt}
\end{align}
\begin{align}
\frac{\partial {\bf f}_\gamma}{\partial t}&=-\chi\rho\tilde{c}{\bf f}_\gamma+\frac{\rho{\bf S}_{\bf f}}{2\pi\hbar\bar{\nu}}=-n_{{\rm HI}}\tilde{c}\sigma_\gamma {\bf f}_\gamma + {\bf S}_\gamma\,,
\label{eq:dfgammadt}
\end{align}
\begin{align}
\frac{\partial n_{{\rm HI}}}{\partial t}&=-n_{{\rm HI}}\tilde{c}\sigma_\gamma n_\gamma +n_en_{{\rm HII}}\alpha_A-n_e n_{{\rm HI}}\beta\,,
\label{eq:dnH0dt}
\end{align}
\begin{align}
\frac{\partial e_{\rm tot}}{\partial t} &= \rho\frac{\partial u}{\partial t} =\rho(S_u+\Lambda_u)\nonumber\\  &=\epsilon_\gamma n_{{\rm HI}}\tilde{c}\sigma_\gamma n_\gamma - n_{{\rm HI}}n_e\Gamma_{e{\rm HI}}-n_{{\rm HII}}n_e\Gamma_{e{\rm HII}}\,.
\label{eq:dedt}
\end{align}

\begin{comment}
In addition to the variables described in section \ref{sec:two-moment},
$n_{\rm X}$ is the number density of element species $X$ ({\em i.e} neutral and ionised hydrogen, and electrons), $e_{\rm tot} \equiv \rho u = \frac{3}{2}n_{\rm tot}k_{\rm B}T$ is the thermal energy density of the fluid, $n_{\rm tot}=\rho/(\mu m_{\rm H})=n_{\rm HI}+2\times n_{\rm HII}$ is the total number density of all particles, $\mu=1/(2-x)$ is
the mean molecular weight and $x\equiv n_{\rm HI}/n_{\rm H}$ is the neutral fraction {\color{red}(sometimes we refer the neutral fraction as $x_{\rm HI}$)}. Furthermore, $\alpha_A$ and $\alpha_B$ are the \lq case A\rq\ and \lq case B\rq\ recombination coefficients, respectively, $\beta$ is the collisional ionization coefficient, $\sigma_\gamma$ is the photoionization cross section, $\Gamma_{e{\rm HI}}$ is the collisional cooling rate per unit volume (combining collisional ionization and collisional excitation),
$\Gamma_{e{\rm HII}}$ represents the sum of recombination cooling
and thermal Bremsstrahlung, and $\epsilon_\gamma$ is the photoionization heating energy per ionization. The values of these coefficients, together with their dependence on photon frequency, $\nu$, and/or gas temperature, $T$, are listed in Appendix \ref{sec:heatcoolparam}. Finally, $S_\xi$ and $S_\gamma$ are external sources of photons.
\end{comment}

Eq.~(\ref{eq:dngammadt}) 
accounts for changes in the photon density due to the sink term  $\Lambda$ and the source term $S$
(see \S\ref{sec:two-moment}). 
In the second line, we specialize the sink term to photo-ionization ($\sigma_\gamma$ is the photo-ionization cross-section and $n_{\rm HI}$ is the neutral hydrogen number density) and add recombination as a source term ($\alpha_A$ and $\alpha_B$ are the \lq case A\rq\ and \lq case B\rq~ recombination coefficients, respectively). The final term $S_\gamma$ represents any other source of photons.  Eq.~(\ref{eq:dfgammadt}) is the corresponding
equation for the photon flux ${\bf f}_{\gamma}$, which includes a photo-ionization term and a source term. 

Eq.~(\ref{eq:dnH0dt}) accounts for the corresponding changes in the density of neutral hydrogen, $n_{\rm HI}$. The terms on the right hand side are the photo-ionization, recombination, and collisional ionization rates respectively ($n_{\rm HII}$ is the density of ionized hydrogen,  $n_e$ is the electron density and $\beta$ is the collisional ionization coefficient).

Eq.~(\ref{eq:dedt}) is the corresponding
thermal energy equation ($e_{\rm tot}$ is the internal energy per unit volume and $u$ is the internal energy per unit mass). In the second line, 
terms from left to right are, respectively, photo-ionization heating ($\epsilon_\gamma$ is the excess thermal energy per ionization) and gas cooling (quantified by the coefficients $\Gamma$). The values of the various constants and coefficients, together with any dependence on photon frequency, $\nu$, and/or gas temperature, $T$, are summarized in Appendix \ref{sec:heatcoolparam}.

The above set of differential equations is in general numerically stiff, meaning that the numerical solution is unstable unless the equations are integrated in time using a very short time-step, $\Delta t$. The reason is that the coefficients in these equations are large in some situations, e.g. $n_{\rm HI}\tilde{c}\sigma_\gamma n_{\gamma}$ is large in the neutral region near radiation sources. The usual remedy is to use an implicit scheme
because this is stable, however its solution may not be sufficiently accurate. Our strategy described below is to combine explicit and implicit methods. 

\subsubsection{Solving the thermo-chemistry equations with a semi-implicit scheme combined with sub-cycling}
\label{sec:semiimplicit}
To illustrate the solution method we make the \lq on-the-spot\rq\ approximation by assuming that recombinations directly to the ground state produce an ionizing photon that is absorbed close to where it was emitted ({\em i.e.} \lq on the spot\rq). In this approximation we set $\alpha_A=\alpha_B$, resulting in the following set of three coupled differential equations,
\begin{align}
\frac{\partial n_\gamma}{\partial t}&=-x n_{{\rm H}}\tilde{c}\sigma_\gamma n_\gamma,
\label{eq:OTSngamma}
\end{align}
\begin{align}
\rho\frac{\partial u}{\partial t}= \epsilon_\gamma x n_{\rm H} \tilde{c}\sigma_\gamma n_\gamma - n^2_{\rm H}x(1-x)\Gamma_{e{\rm HI}}-n^2_{\rm H}(1-x)^2\Gamma_{e{\rm HII}},
\label{eq:thermaleq}
\end{align}
\begin{align}
\frac{\partial x}{\partial t}&=-x \tilde{c}\sigma_\gamma n_\gamma +(1-x)^2n_{{\rm H}}\alpha_B-x(1-x)n_{\rm H}\beta,
\label{eq:chemHeq}
\end{align}
where $x=n_{\rm HI}/n_{\rm H}$ is the neutral hydrogen fraction. Note that we denote neutral fraction as $x_{\rm HI}$ in the figures for clarity and as $x$ in text for simplicity.

The partial time derivatives refer to changes due to interaction between radiation and gas only. There may be additional terms, for example, due to other photon sources or sinks, and heating and cooling due to adiabatic processes or shocks. Here, we restrict ourselves to solving these radiative equations, treating any other source/sink terms in operator split fashion.

We integrate these equations following the approach of \cite{Petk09OTVET}: solve the first two equations explicitly and use that solution to solve
the third equation (the chemistry equation) implicitly. However, we additionally perform {\it sub-cycling}, requiring that $n_\gamma$ and $u$ do not change by more than 10\% in each sub-cycle.

\begin{figure}
\begin{center}
 \includegraphics[width=0.6\textwidth]{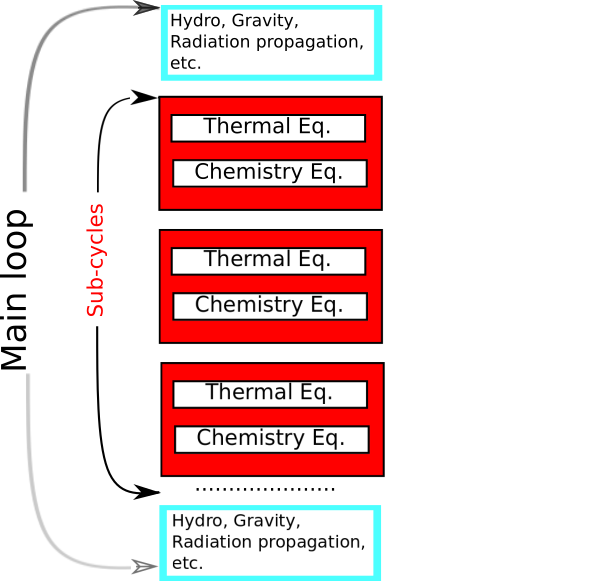}
\caption{A schematic diagram illustrating the sub-cycling method for solving non-equilibrium thermo-chemistry equations. In the main loop, we advance the radiation, hydrodynamics, gravity, and radiation injection equations, and any other equations not related to radiation ({\em e.g.} sub-grid schemes). In between we sub-cycle the thermo-chemistry equations (Eqs.~\ref{eq:OTSngamma}, \ref{eq:thermaleq} \& \ref{eq:chemHeq}), with a smaller time-step. Note that the whole set of the thermo-chemistry equations is solved sequentially within one sub-cycle, accurately accounting for any rapid changes of ionization state or photon density.}
\label{fig:subcyclethermal}
\end{center}
\end{figure}

We do so by requiring that $\Delta t\le 0.1\,{\rm min}(1/(x n_{{\rm H}}c\sigma_\gamma), u/|\partial u/\partial t|)$. While the implicit solver for the neutral fraction $x$ is unconditionally stable, it can be inaccurate if the time-step is too large. Therefore, we further limit the sub-cycle time-step to $\Delta t\le \mathit{C}x/|\partial x/\partial t|$, where $\mathit{C}$ is a parameter we choose to be 0.1 (but can be larger depending on the tolerance). So in summary, we take the sub-cycling step to be
\begin{align}
\Delta t=0.1\,{\rm min}(\frac{n_\gamma}{|\partial n_\gamma/\partial t|}, \frac{u}{|\partial u/\partial t|}, \frac{x}{|\partial x/\partial t|})\,.
\label{eq:chemHeqDt}
\end{align}
Fig. \ref{fig:subcyclethermal} illustrates the semi-implicit sub-cycle scheme.

We update $n_\gamma$ using the analytic solution of Eq.~(\ref{eq:OTSngamma}) in case $x$ and $n_{\rm H}$ are held constant,
\begin{align}
n_\gamma(t^{n+1})=n_\gamma(t^{n})\exp(-\Delta t \sigma_\gamma c n_{\rm H}x)\,.
\end{align}
Doing so guarantees that $n_\gamma$ is always positive, as it should be, and that the solution is asymptotically correct when recombinations are negligible. We update $u$ using the corresponding analytical solution of Eq.~(\ref{eq:thermaleq}).

The implicit solution to the chemistry equation uses the updated values for the radiation and internal energies,
\begin{align}
&\frac{x(t^{n+1})-x(t^n)}{\Delta t} = - \tilde{c} \sigma_\gamma n_\gamma (t^{n+1}) x(t^{n+1})\nonumber\\
&+n_{\rm H}\alpha_B(t^{n+1})[1-x(t^{n+1})]^2- n_{\rm H}\beta(t^{n+1}) x(t^{n+1})[1-x(t^{n+1})]\,.
\end{align}
which is a quadratic equation for $x(t^{n+1})$.

This method can be generalised to the case of more elements by adopting the approach of \cite{Anni97multchem}, {\em i.e.} updating each element implicitly one by one in order of increasing timescale.
 A possible alternative scheme uses the {\small CVODE} library \citep{Cohe96CVODE} to solve these stiff equation, {\em e.g.} \cite{Kann19AREPORT}.

For now, we have restricted this discussion to the on-the-spot approximation.
This limitation can be relaxed by adding recombination radiation
as a source terms to each gas particle. Given that the computing time of our RT scheme is independent of the number of sources, this could be feasibly implemented in the future.

Our semi-implicit sub-cycling scheme is accurate, as we show below, as well as computationally efficient, as shown in Appendix \ref{sec:thermochemsub}. Sub-cycles are initiated only when the system is out of equilibrium, for example when a source of photons is suddenly switched on. But even in such situations, we find that only a few dozen sub-cycles occur. Sub-cycling should also help with load balancing the computation. Without sub-cycling, most of the computing time will be spent in the vicinity of ionization fronts, where the thermo-chemistry is highly out of equilibrium. With sub-cycling enabled, the time-step of the main loop is instead limited by the overall CFL time-step. We proceed by showing some tests of the thermo-chemistry implementation.

\subsubsection{Thermo-chemistry test I: ionizing a single gas parcel}
\label{sec:singlegasparcel}
This test is a variation of Test 0 in \cite{Ilie06RTcom}:
an initially neutral parcel of pure hydrogen gas at low temperature is suddenly ionized and heated by a source of ionizing radiation with a specified spectrum of ionizing photons. After a specified time, the ionizing source is switched off. The total density of the gas parcel is kept constant. The radiation is assumed to be optically thin at all times. The test involves following the evolution of the neutral fraction, $x$, and of the temperature of the gas, $T$. To enable a fair comparison between codes, it is, of course, important to make sure that the physical constants used - such as, for example, the frequency dependence of the ionization cross section - are the same.

As the source is switched on and the hydrogen gas gets ionized, the temperature increases to a value that depends on the shape of the ionizing spectrum. The gas is then in ionization equilibrium. However, it takes longer for the gas to be in thermal equilibrium - where photo-heating balances radiative cooling. When the source is switched off, the gas starts to recombine and cool. We do not include molecule formation in the calculation, and hence the cooling rate drops with decreasing $T$. 

\noindent $\bullet$ {\bf Analytical description} The evolution can be understood by writing Eq.~(\ref{eq:chemHeq}) as a Ricatti equation,
\begin{align}
\frac{dx}{dt}&=-\frac{x}{\tau_i}+\frac{(1-x)^2}{\tau_r}-\frac{x(1-x)}{\tau_e}\equiv\frac{1}{\tau}(x-x_1)(x-x_2)\,,
\end{align}
where we defined three characteristic time-scales,
\begin{align}
\tau_i&\equiv \frac{1}{c\sigma_\gamma n_\gamma}\,;\quad \tau_e\equiv\frac{1}{n_H\beta}\,;\quad \tau_r\equiv \frac{1}{n_H\alpha_B}\,,
\end{align}
and
\begin{align}
\frac{1}{\tau}&=\frac{1}{\tau_e}+\frac{1}{\tau_r}\,;\quad 
x_1+x_2=2+\frac{\tau_r}{\tau_i}\,;\quad x_1x_2=\frac{\tau}{\tau_r}\,.
\end{align}
Choosing initial condition $x=x_0$ at $t=0$, the general solution in case all $\tau$'s are constant, is
\begin{align}
x(t) &= \frac{x_2(x_0-x_1)-x_1(x_0-x_2)f(t)}
{(x_0-x_1)-(x_0-x_2)f(t)}\nonumber\\
f(t)&=\exp[-(x_1-x_2)t/\tau]\,.
\label{eq:xt}
\end{align}

In the special case where collisional ionizations are neglected, $\tau_e\to\infty$ and when $\tau_i\ll \tau_r$, this solution simplifies to approximately $x(t)=x_0 \exp(-t/\tau_i)+\tau_i/\tau_r$: the neutral fraction approaches its {\it ionization equilibrium} exponentially on the ionization timescale, $\tau_i$.

This approximate description assumes that $\tau_r$ - and hence the temperature of the gas - remains a constant. An estimate of the change in temperature following rapid ionization, $\tau_i\ll \tau_r$, follows from neglecting radiative cooling in the short time it takes to ionize the gas, so that
\begin{align}
    \rho\frac{du}{dt}&\approx -\epsilon_\gamma \frac{dn_{\rm HI}}{dt}=-\epsilon_\gamma\,n_{\rm H}\frac{dx}{dt}\,,
\end{align}
where $\epsilon_\gamma$ is the mean energy injected into the gas per photo-ionization, see \S A. Writing the thermal energy per unit mass, $u$, in terms of the neutral fraction, $x$, as $u=k_{\rm B}T/(\mu m_{\rm H})$ with $\mu=(2-x)^{-1}$, yields the following relation between the initial temperature $T=T_0$ at $t=0$, and the temperature $T$ when the neutral fraction is $x$:
\begin{align}
    k_{\rm B}T&=\frac{2-x_0}{2-x}k_{\rm B}T_0+\frac{2\epsilon_\gamma}{3}\frac{x_0-x}{2-x}\,.
\end{align}

In the test described below, the gas is initially neutral, $x_0=1$, and at low temperature, $k_{\rm B}T_0\ll \epsilon_\gamma$, and the photo-ionization rate is high so that in photo-ionization equilibrium, $x\ll 1$. In this case, the photo-ionization equilibrium temperature is $T_1\approx \epsilon_\gamma/(3k_{\rm B})$, which depends only on the ionization energy per photon, regardless of radiation intensity or gas density.

On a longer timescale, the parcel of gas will reach thermal equilibrium (temperature $T_2$), where photo-heating balances radiative cooling. Provided $T_2>T_1$, the timescale to reach this equilibrium can be estimated by simply neglecting cooling and noting that the rate at which the gas is heated is approximately
the product of the ionization rate, $x/\tau_i$, times the energy injected per photo-ionization per hydrogen atom, $\epsilon_\gamma/m_{\rm H}$, hence
\begin{align}
    \frac{{\rm d}u}{{\rm d}t}=\frac{\epsilon_\gamma}{m_{\rm H}}\frac{x}{\tau_i}\approx \frac{\epsilon_\gamma}{m_{\rm H}\,\tau_r}\,.
\end{align}
Therefore it takes approximately a recombination time to reach thermal equilibrium (see also \citealt{Pawl11multifRT}).

When the source is suddenly switched off, gas will start to recombine. This is still described by a Ricatti equation of the form of Eq.~(\ref{eq:xt}), except that now
\begin{align}
x(t) &= \frac{x_2(x_{\rm eq}-x_1)-x_1(x_{\rm eq}-x_2)f(t)}
{(x_{\rm eq}-x_1)-(x_{\rm eq}-x_2)f(t)}\nonumber\\
f(t)&=\exp[-(x_1-x_2)(t-t_{\rm eq}))/\tau]\,,
\label{eq:xt2}
\end{align}
where $x_{\rm eq}$ is the neutral fraction in thermal equilibrium, $t_{\rm eq}$ is the time that the ionizing source is switched off, 
and $x_1+1/x_1=1+\tau/\tau_r$ with $x_2=1/x_1$. If the gas is no longer heated, it will of course simply keep on cooling and there is no further equilibrium state.

\noindent $\bullet$ {\bf Numerical solution}
For the numerical values for this test, we take\footnote{Note that the value of ${\tilde c}$ is irrelevant to the solution since the pure thermo-chemistry equation is independent of ${\tilde c}$ if $F_{\rm photon}$ is given.} ${\tilde c}=c$, $n_{\rm H}=1~{\rm cm}^{-3}$, $x_0=1$ and $T_0=100~{\rm K}$. From time $t=0$, the parcel of gas is being irradiated with a black body spectrum of temperature $T=10^5{\rm K}$ (for which $\epsilon_\gamma=6.33\,{\rm eV}$ in the optically thin limit) and photon flux $F_{\rm photon} = 10^{12} \;{\rm photons}\;{\rm s}^{-1}\;{\rm cm}^{-2}$. The source is switched off after a time $t_s=5\times 10^7 {\rm yr}$ and we follow the evolution until time $t_e=10^8 {\rm yr}$. 

For a reference temperature of $T=10^4~{\rm K}$, the three timescales are $\tau_i\approx 10^{-2.3}{\rm yr}$, $\tau_r\approx 10^{5.1}{\rm yr}$ and $\tau_e\approx 10^{9.3}{\rm yr}$, so that $\tau_i\ll \tau_r\approx\tau\ll t_e<\tau_e$. The temperature in photo-ionization equilibrium is $T=6.33~{\rm eV}/(3k_{\rm B})\approx 10^{4.39}{\rm K}$, and the temperature in thermal equilibrium is about twice that. The timescale for the gas to reach thermal equilibrium can be estimated as follows. The heating rate of the gas, when in photo-ionization equilibrium, is $du/dt\approx \epsilon_\gamma/\tau_r\approx 8.3\times 10^{-11}{\rm erg}~{\rm Myr}^{-1}$. Therefore the timescale to reach thermal equilibrium is approximately
\begin{align}
    t_{\rm eq}\approx \frac{\Delta u}{du/dt}=\frac{u_{\rm eq}-u_i}{du/dt}\approx 
    \frac{u_i}{du/dt}\approx 10^{5.1}~{\rm yr}\,,
\end{align}
where we have used the fact that for the parameters of this test, $u_{\rm eq}\approx 2 u_i$. In summary: the gas should reach its photo-ionization equilibrium temperature by a time $\tau_i$, reach its thermal equilibrium temperature by the time $t_{\rm eq}$, and start to cool and recombine after time $t_s$.

We want to verify that the combination of explicit sub-cycling and implicitly solving the chemistry equations yields the correct solution, independently of a globally imposed time-step. To demonstrate the accuracy of the integration scheme, we also want to compare to a run in which we integrate the equations with a short, fixed time-step. However, the ionization timescale $\tau_i$ is much smaller than the evolution timescale $t_e$, and it is impractical to simulate the whole time evolution with a time-step much shorter than $\tau_i$. Here we follow \cite{Pawl11multifRT} and perform a dozen simulations with different (fixed) time-steps, from $\Delta t\ll \tau_i$ to $\Delta t\gg \tau_i$.

\begin{figure}
 \includegraphics[width=0.45\textwidth]{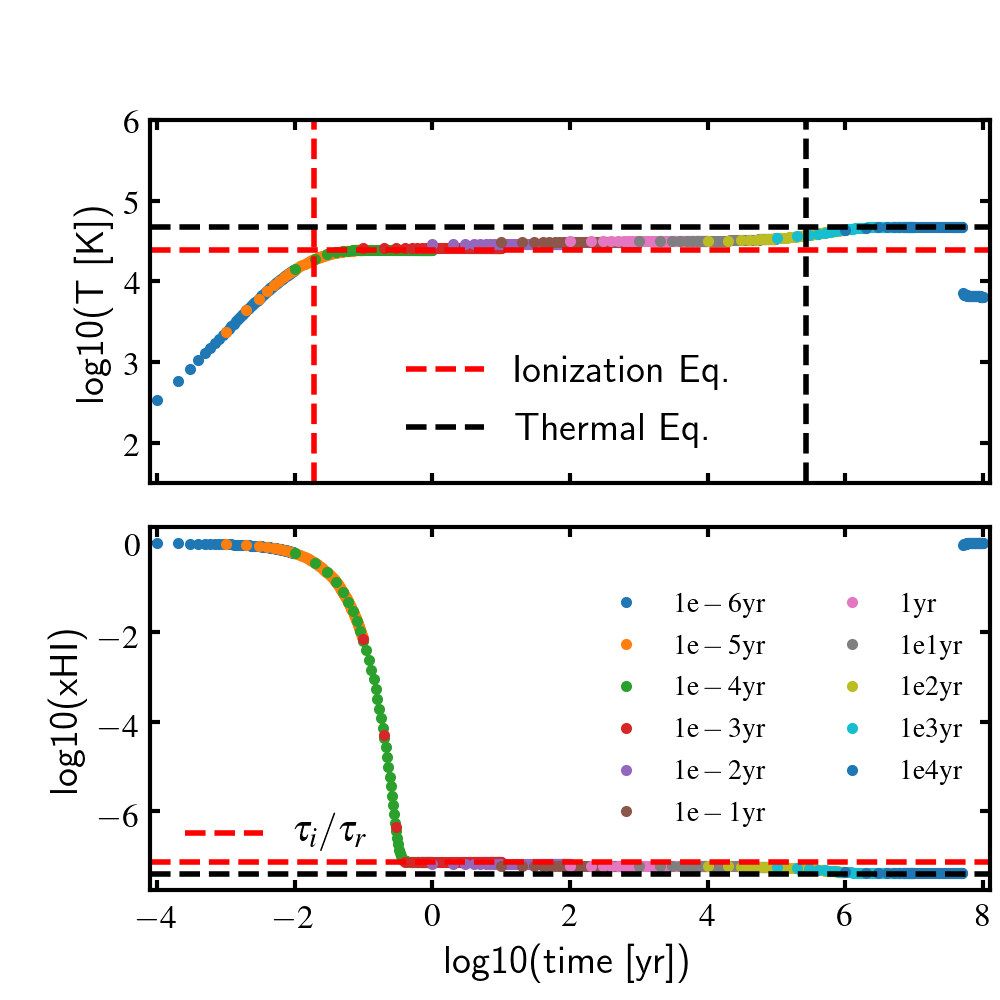}
\caption{A variation of \protect\cite{Ilie06RTcom} Test 0: photo-heating of a single gas parcel irradiated with black body radiation of temperature $T=10^5~{\rm K}$. The source is switched off after time $t_s=5\times 10^7{\rm yr}$. Vertical dashed lines indicate the ionization timescale $\tau_i$ ({\em red}) and the time to reach thermal equilibrium, $t_{\rm eq}$ ({\em black}), horizontal dashed lines indicate the expected photo-ionization equilibrium temperature (or neutral fraction) ({\em red}) and the thermal equilibrium temperature ({\em black}).  {\em Points with different colours} show the evolution computed with different, fixed, global time-steps, as per the legend. Simulations with different $\Delta t$ step sizes yield the same evolution, which also matches the analytical estimate a well as the curves in \protect\cite{Ilie06RTcom} and \protect\cite{Pawl11multifRT} (not shown here).}
\label{fig:SCTrealc}
\end{figure}

Results are shown in Fig. \ref{fig:SCTrealc}, where differently coloured curves show the evolution for different values of the global time-step, $\Delta t$. All curves follow the analytical expectation: gas heats and gets almost fully ionized on a timescale $\tau_i$ (vertical dashed red line), reaching its photo-ionization equilibrium temperature (horizontal dashed line), continues to be heated on a timescale $\tau_r$ (vertical dashed black line) to reach thermal equilibrium (horizontal dashed black line), and finally starts to recombine and cool when the source is switched off. All simulation runs fall on top of each other, demonstrating that the numerical solution is independent of the global fixed value of $\Delta t$.

After the radiation is switched off, gas recombines and cools rapidly to $T\sim 10^4{\rm K}$, below which the cooling rate drops rapidly as we only include cooling by neutral hydrogen.

There are two major takeaways from Fig. \ref{fig:SCTrealc}. Firstly, the simulated evolution follows the analytical expectation as well as the results from other simulation codes in \cite{Ilie06RTcom} (see their Fig. 5), therefore the scheme is accurate. Secondly, the evolution is independent of the global time-step, demonstrating the convergence of the sub-cycle thermo-chemistry solver. With this solver, the (main) time-step of the simulation is only limited by the Courant condition in solving the moment equation (\S\ref{sec:two-moment}).

\subsection{Radiation injection}
\label{sec:injection}
In our implementation, radiation is injected by \lq star particles\rq\, which, in our SPH implementation, have a smoothing length $h$, which is calculated in the same way as that of gas particles ({\em i.e.} by requiring that each star particle interacts with the desired number of kernel-weighted gas neighbours). 

A star particle $i$ with time-step $\Delta t_i$ and energy injection rate $\dot{e}_{i,{\rm rad}}$ distributes a total amount of radiation $\dot{e}_{i,{\rm rad}}\Delta t_i$ into all of its neighbouring gas particles. Each individual neighbouring gas particle $j$ receives an amount of energy equal to
\begin{align}
\Delta e_{ij} = m_j\Delta \xi_j = \frac{m_j}{N_{\rm nor}\rho_j r_{ij}^2}\dot{e}_{i,{\rm rad}}\Delta t_i\,.
\end{align}
This kernel-weighted energy transfer is normalized by $N_{\rm nor}$, computed such that $\sum_j \Delta e_{ij}=\,\dot{e}_{i,{\rm rad}}\Delta t_i$, where the sum is performed over all of $i$'s gas neighbours $j$.

We inject the corresponding isotropic radiation flux, $m_j\Delta {\bf f}_j$, as if the surrounding medium were optically thin:
\begin{align}
\Delta {\bf f}_j = \tilde{c}\hat{\bf r}_{ji}\Delta \xi_j\, .
\end{align}

Because the distribution of gas neighbours around any star particle is generally relatively disordered, the resulting radiation field may not be very isotropic unless energy is injected over a sufficiently large number of gas particles. To avoid that the source of photons is unacceptably anisotropic,  we increase the smoothing lengths of star particles to be a few times the smoothing length of gas particles, e.g. $h_{\rm star}=2h_{\rm gas}$ (see Fig. \ref{fig:stromgren3dhs}).

An alternative way of ensuring isotropic radiation around sources is to impose the radiation direction in the optically thin limit (\S\ref{sec:opticalthindir}). In some tests, e.g. tests of Str\"omgren spheres, we calculate the total radiation energy within the injection region, and then reset the radiation distribution according to the optically thin expectation\footnote{It is possible to inject radiation energy only - without updating the radiation flux - provided we apply the original M1 closure (Eq.\ref{eq:feddcon}), since the moment equations will generate an isotropic radiation field in the absence of initial flux. But this is not possible with the modified M1 closure in the optically thin environment for which the (initial) direction of the Eddington tensor needs specifying.}.

\subsection{Implementation details}
\label{sec:code}
Our RT scheme is implemented in the public version of the {\small SWIFT} ({\bf S}PH {\bf w}ith {\bf i}nterdependent {\bf f}ine-grained {\bf t}asking) code \citep{Scha16SWIFT}\footnote{\url{http://swift.dur.ac.uk/}}, which has been applied in galaxy formation and planetary giant impact simulations \citep{Kege19planetimpact}. The target application of {\small SWIFT} are
zoomed cosmological simulations and simulations in representative volumes, with subgrid physics modules similar to {\sc eagle} \citep{Scha15EAGLE}.

{\small SWIFT} is an SPH code that solves cosmological or non-cosmological hydrodynamic equations, including self-gravity, and is designed to work on hybrid shared/distributed memory computer architectures. Load balance is optimised using task-based parallelism, with tasks assigned by a graph-based domain decomposition, and using dynamic, asynchronous communication. For hydrodynamics, \cite{Borr18SWIFT} found {\small SWIFT} to have good weak scaling from 1 to 4096 codes (losing only 25\% performance) in low redshift cosmological galaxy simulations (with {\small EAGLE} physics from \citealt{Scha15EAGLE}).

The time-stepping of the RT scheme follows the \cite{Hern89TREESPH} factor-of-two time-step hierarchy implemented in 
{\small SWIFT}: a particle with time-step $\Delta t$  is assigned to the time-step bin $N$ such that $2^{\rm N} \le \Delta t/t_{\rm min}
< 2^{\rm N+1} $, where $t_{\rm min}$ is some small minimum time-step.  At each step in time, the radiation field in particles in all bins $N$ with   $N \le M$ are updated with a forward Euler method, where $2^M t_{\rm min}$ is the time-step of the active particles with the largest time step (see \citealt{Borr18SWIFT} for the time-stepping strategy in {\small SWIFT} in the absence of RT).

When hydrodynamics and other processes are included, the time-step of each individual particle is the minimum time-step required by all these processes combined, although typically the radiation time-step ($\Delta t_{\rm rad}\sim 0.1 h/\tilde{c}$) is the most limiting. We do not (yet) sub-cycle the radiative transport step, therefore all processes (including gravity and hydrodynamics) are integrated using the smallest time-step. This is an avenue for future optimization. However, we {\em do} sub-cycle the thermo-chemistry differential equations, as described in \S\ref{sec:thermochemisty}. This leads to significant saving in computation time, since the time-step associated with these chemistry equations can be orders of magnitude shorter than the RT time step. 
\subsection{The Reduced Speed of Light approximation}
\label{sec:RSL}
When radiation travels at the speed of light, the time-step to advance a radiation front correctly is of order $\Delta t_c\sim h/c$, for a smoothing length $h$ of an SPH particle. This is, of course, much shorter than the CFL step, which is of order $\Delta t_s\sim h/v_s$, where $v_s$ is the sound speed. However, ionising radiation with flux $F$ moves at the speed of the ionization front, $v_I\sim F/(2\pi n_{\rm H}\hbar\nu)$, through neutral gas with density $n_{\rm H}$. When $v_I\ll c$, the code can be sped up by a large factor by reducing the speed of light, from $c$ to $\tilde c$. As long as $\tilde c>v_I$, the speed of an ionization front can still be correct for a given $F$ (see e.g. the discussion in \citealt{Rosd13ramsert}).

This \lq reduced speed of light\rq\ (RSL) approximation was introduced by \citet{Gned01OTVET} to simulate radiative transfer efficiently and has been applied to other radiative transfer simulations, e.g. by \cite{Aube08M1}. They demonstrate that RSL performs well in problems involving ionization, photo-heating, and expansion of HII regions. 

However, there is no unique way to implement RSL. \cite{Skin13M1} 
implemented the RSL approximation in simulating RT in the interstellar medium, e.g. modeling radiation reprocessed by dust. However, their approach does not conserve total radiation plus matter energy and momentum, and the non-equilibrium solution might not be correct.

\cite{Ocvi19DSL} examined the \lq dual speed of light\rq\ (DSL) approximation, where $c\to\tilde c$ in the propagation equation
but not in the thermo-chemistry equations. Unfortunately, DSL fails to reproduce the correct equilibrium gas properties. This is because when $c$ is reduced to $\tilde c$ in the propagation equation, the photon-matter interaction rate does not change accordingly in DSL. This can be seen by considering the analytical solution of the Str\"omgren sphere (Appendix \ref{sec:anaStromgren}), 
\begin{align}
n_{{\rm HI}}{c}\sigma_\gamma n_\gamma&=
\left(\frac{c}{\tilde c}\right)
\frac{n_{{\rm HI}}\sigma_\gamma \dot{N}_\gamma}{4\pi r^2}\exp\left(-\int^r_0n_{\rm HI}\sigma_\gamma {\rm d}r\right)\nonumber\\
&=\,n_en_{{\rm HII}}\alpha_B\,,
\label{eq:rslstromgren}
\end{align}
where the factor ${\tilde c}$ arises from the propagation equation in the optically thin limit, and the factor $c$ comes from the thermo-chemistry equation. The equilibrium neutral fraction will deviate from the correct solution due to the ${c}/{\tilde c}$ factor.

Thus, our {\it default} treatment is to replace $c\to\tilde c$ in all equations \citep{Aube08M1,Rosd13ramsert}, including the propagation and thermo-chemistry equations. For a fixed photon flux, $F$, (or photon injection rate), this choice reduces the interaction strength between light and matter (e.g. $\sigma_\gamma \tilde{c}$) to compensate for higher photon density (due to the slower photon propagation speed). As a result, the photo-ionization rate will be independent of $\tilde{c}$ (as long as $\tilde{c}$ is larger than other speeds). Furthermore, the choice of $\tilde{c}$ will {\it not} affect the equilibrium gas properties, as demonstrated in Eq. \ref{eq:rslstromgren} (and see the tests in next sections).

However, there are limitations to RSL. First, $\tilde{c}$ should exceed the speed $v_I$ of any ionization front. For example, \cite{Baus15Hreion} showed that using $\tilde{c}=c/10$ affects the timing of reionization. Another issue of RSL is that using ${\tilde c}<c$ increases the momentum term $\nabla(\mathbb{F}E)$, and if this is not corrected for
then the radiation pressure will be too large \citep[see also][]{Jian12VET,Jian18}. This may be problematic in cases
where radiation pressure is crucial, for example when modelling
radiation pressure from AGN. In the case of reionization simulations, the photon-density is low and radiation pressure is mostly neglected anyway.

\section{Validation}
\label{sec:results}
This section contains an extensive series of tests to validate the numerical scheme and its implementation in the {\sc swift} code. The tests combine the {\it default} scheme for radiation (\S\ref{sec:testartdiss}) with the {\small SPHENIX} SPH formulation for hydrodynamics (\S\ref{sec:SPHform}), unless explicitly stated otherwise. Some test impose the optically thin direction in the Eddington tensor (\S\ref{sec:opticalthindir}). Otherwise, the flux propagates in the direction $\hat{\bf n}=\hat{\bf f}$, as computed for each gas particle. In all except the shadowing test (\S\ref{sec:statictest}), we apply periodic boundary conditions. We will make the on-the-spot approximation in all of the tests (\S\ref{sec:semiimplicit}). We do not use the reduced speed of light (RSL) approximation in \S\ref{sec:purepro} in which we aim to compute the
radiation distribution, but we {\em do} use the RSL approximation in \S\ref{sec:statictest}-\ref{sec:rhdtest}, which focuses on properties of the gas (see \S\ref{sec:RSL}).

\subsection{Optically Thin Propagation tests}
\label{sec:purepro}

\begin{figure}
 \includegraphics[width=0.5\textwidth]{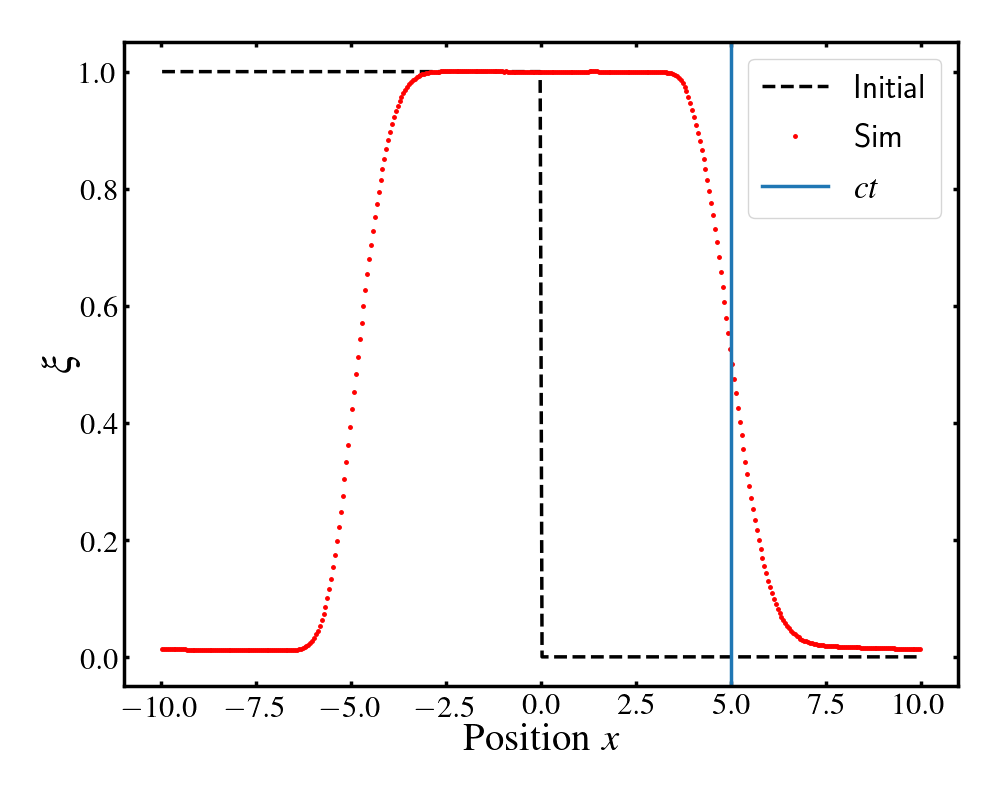}
 \includegraphics[width=0.5\textwidth]{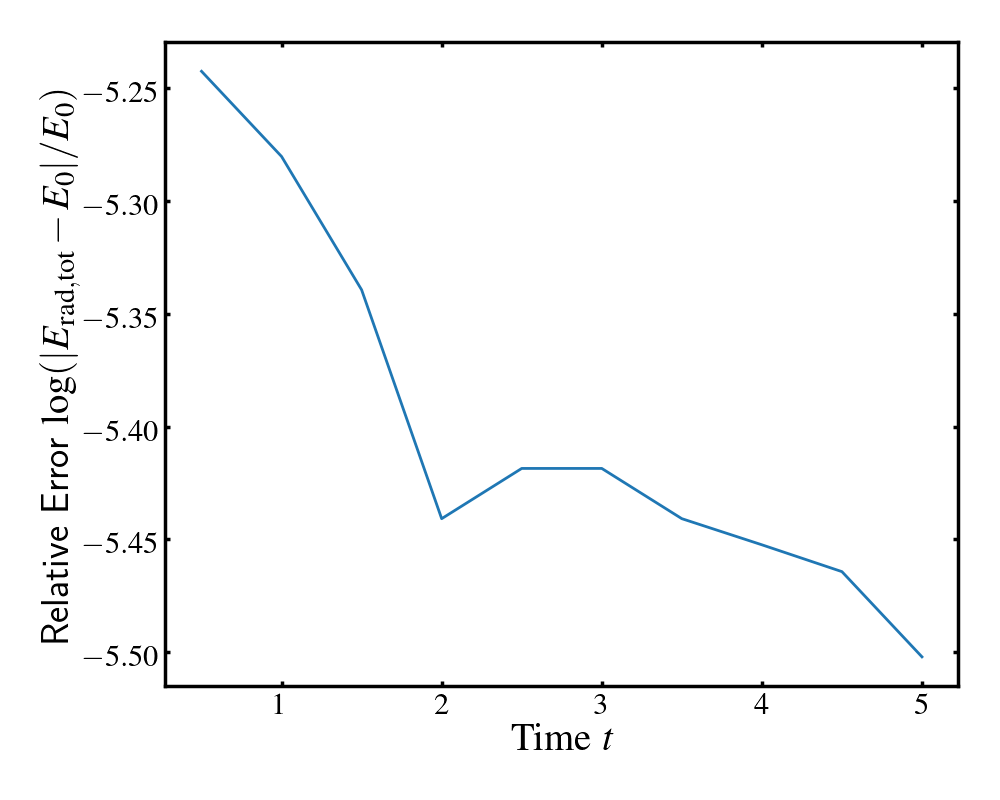}
\caption{Propagation in 1D: {\it Upper panel:} radiation energy density, $\xi(x)$
of a package of radiation propagating to the right at the speed of light ($c=1$ in these units). The {\em black dashed line} is the initial profile in units of the initial value of $E$, the {\em blue} line is the edge of the initial profile shifted to the right by $\Delta x=5$, the {\em red points} are the simulated values of $E$ for individual SPH particles at time $t=50$. The simulation uses 400 SPH particles located on the vertices of a regular grid. The test shows that the front moves at the correct speed. {\it Lower panel:} the relative error in total radiation energy as a function of time, where $E_0$ and $E_{\rm rad,tot}$ are the total radiation energy at the beginning and at time $t$, respectively. The deviation from energy conservation is less than 0.01\% at all times.}
\label{fig:radfront1d_tm}
\end{figure}

\begin{figure}
 \includegraphics[width=0.5\textwidth,left]{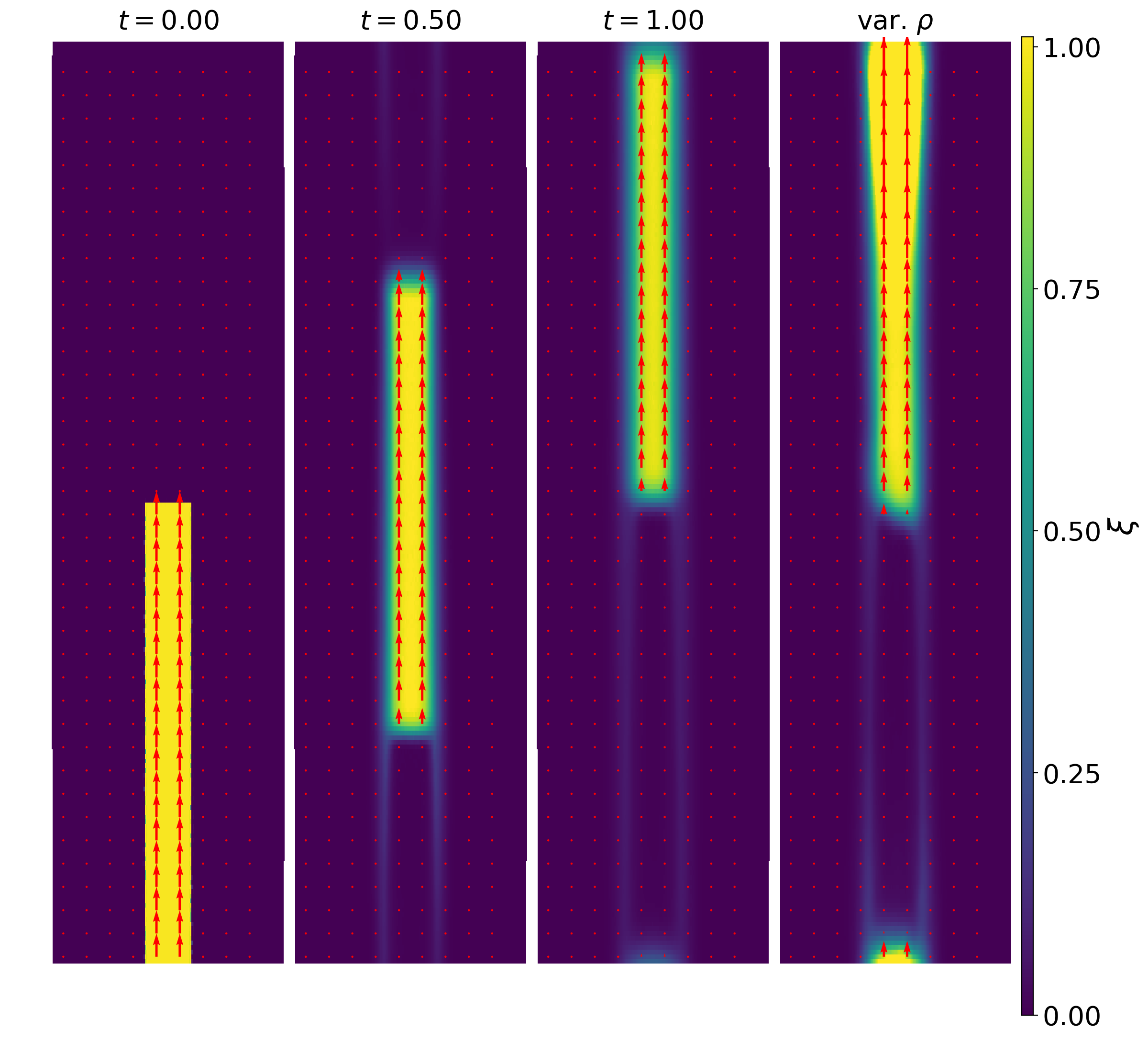}
 \includegraphics[width=0.48\textwidth,left]{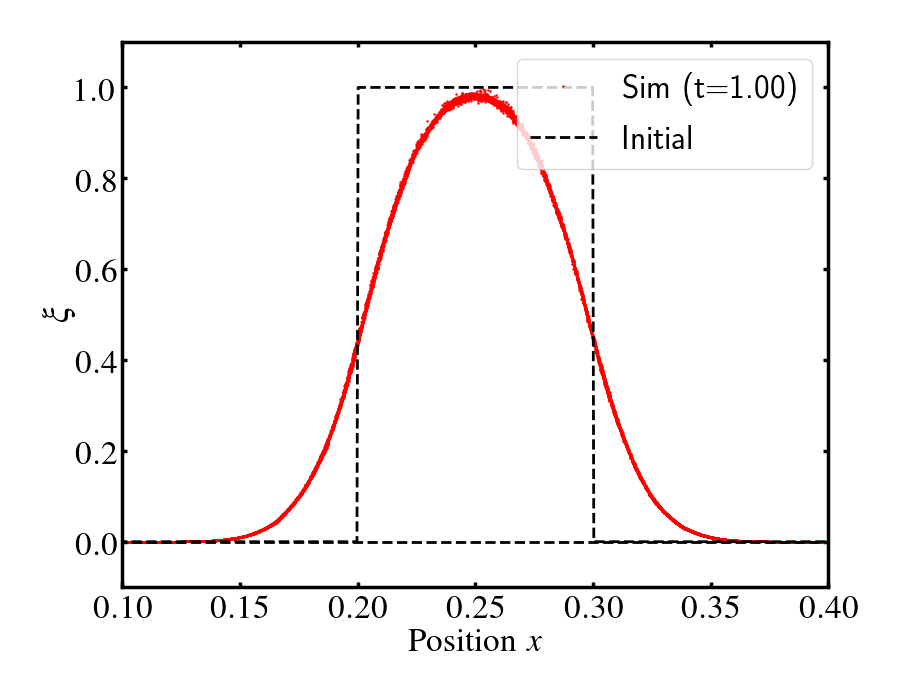}
\caption{Propagation in 2D: propagation of a packet of radiation in two dimension at the speed of light, $c=1$.
{\it Upper panels}: Radiation energy density of the packet in units of its initial value. The {\em left panel} shows the initial state, where the packet has length $\delta y=1$. The central panels show the system at different times at a constant particle density. The rightmost panel labelled `var. $\rho$' shows the radiation beam in case of an underlying density gradient (see main text) at t=1. Colors represent the radiation energy density and {\em red arrows} show the radiation fluxes. {\it Lower panel}: radiation energy density, $\xi(x,y)$, of SPH particles with $1.4\le y\le 1.6$ at time $t=1.0$ ({\em red line}), the {\em dashed-black line} is the initial shape of the packet.}
\label{fig:radstream2d_defaultdiss}
\end{figure}

\begin{figure*}
 \includegraphics[width=0.95\textwidth]{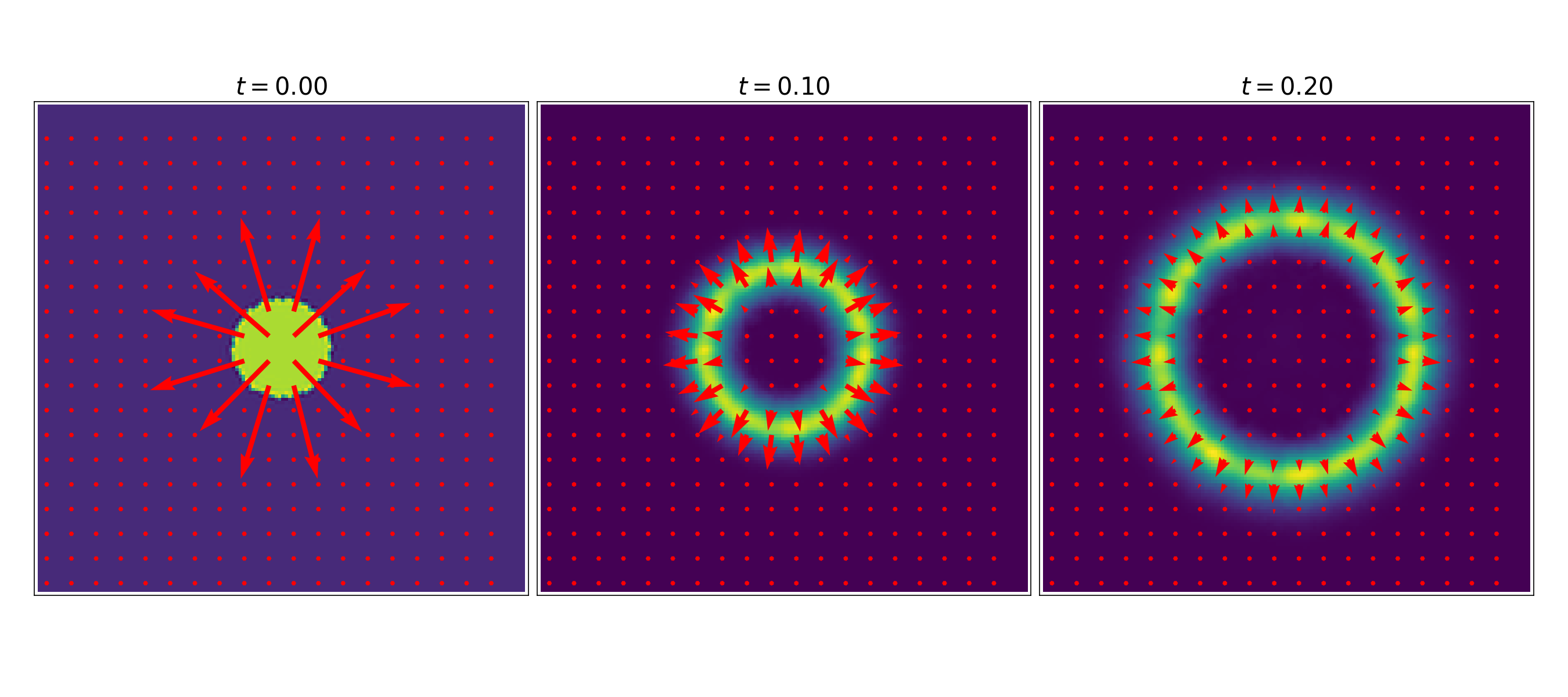}
\includegraphics[width=0.95\textwidth]{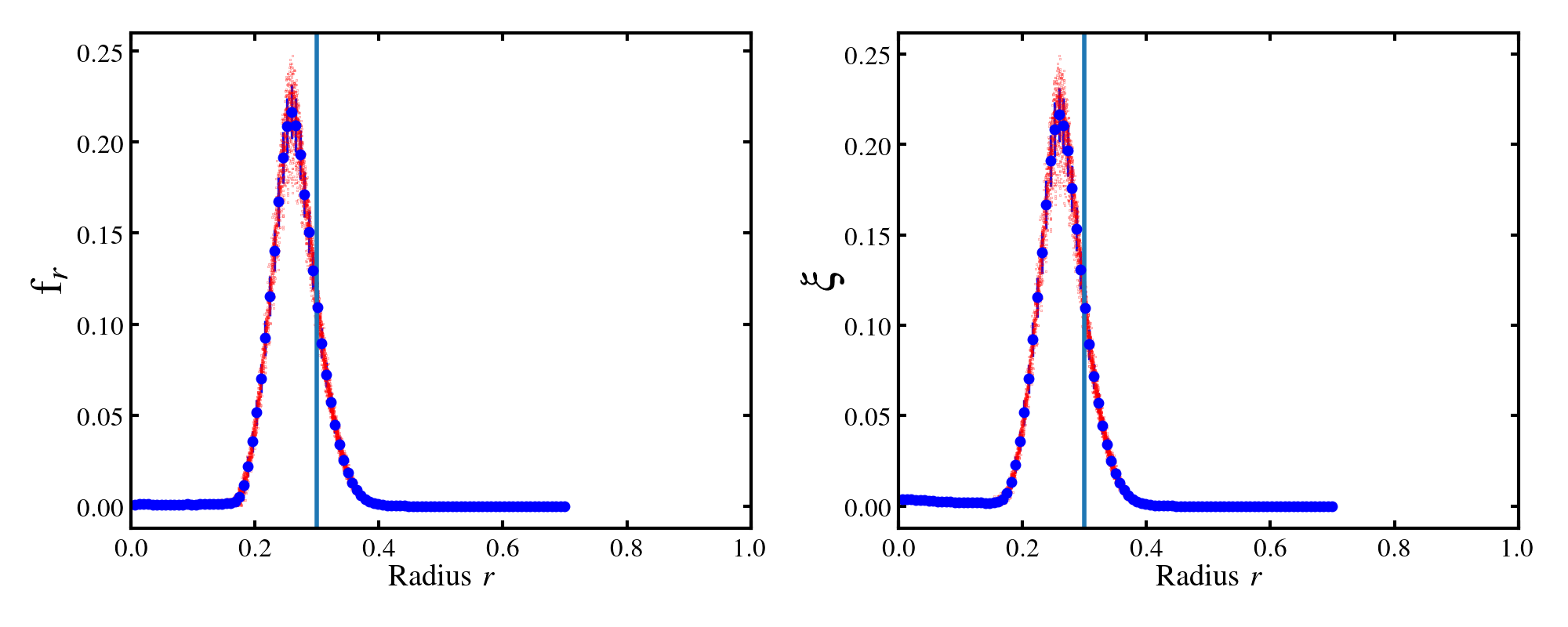}
\caption{Propagation in 2D: propagation of a shell of light away from a source at the speed of light, $c=1$. {\it Upper panel}: radiation energy density of the shell of light in units of its initial value. The {\em left panel} shows the initial state, where radiation is filled uniformly within a circle with radius $r=0.1$ and points out radially. Panels to the right show the system at times $t=0.1$ and $t=0.2$. The SPH particle distribution is glass-like, with $128\times 128$ particles filling the computational volume of horizontal extent $\Delta x=2$ and vertical extent $\Delta y=2$. Colours represent the radiation energy density and {\em red arrows} show the radiation fluxes; the colour scale is not the same in each panel to bring out the smoothness of the radiation density as the shell moves out. {\it Lower panel}: radiation flux ({\em left panel}) and radiation density ({\em right panel}) as a function of radius at time $t=0.2$. The {\em red points} represent values of all individual SPH particles, {\em blue points} show binned values. The {\em blue vertical lines} at $r=0.3$ show the location of the radiation front at time $t=0.2$.}
\label{fig:radfrontsph2d_tm}
\end{figure*}

\begin{figure*}
\includegraphics[width=0.95\textwidth]{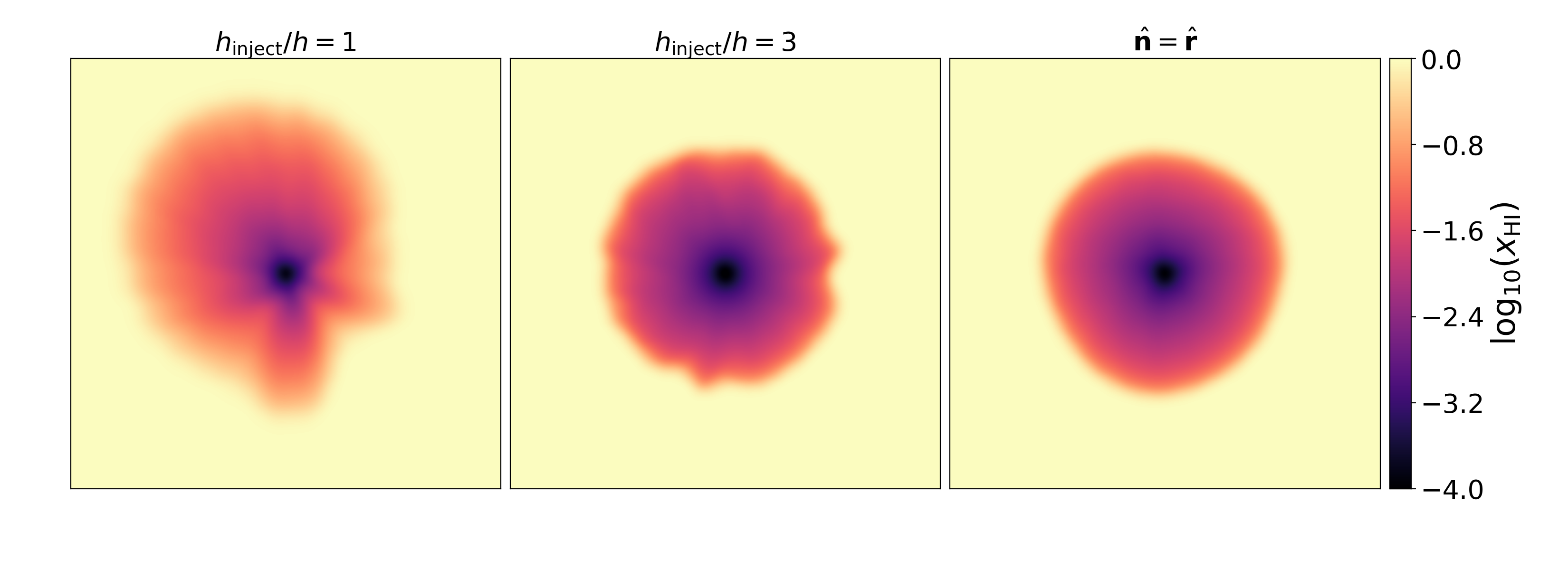}
\caption{Isothermal Str\"omgren sphere from \protect\cite{Ilie06RTcom} Test 1: a source of radiation, located at the centre of the panels, photoionizes hydrogen gas, kept at constant density and constant temperature. Panels show the neutral fraction, ${n_{\rm HI}}/n_{\rm H}$, at time $t=500~{\rm Myr}$ in a slice through the centre of the three-dimensional volume. {\it Left panel}: radiation is injected in all gas particles within one smoothing length from the source, see \S\ref{sec:injection} for the method of injection. {\it Central panel:} radiation is injected in all gas particles within two smoothing lengths from the source.
{\em Right panel:} as in left panel, but the direction in which radiation travels is set to  ${\hat{\bf n}}={\hat{\bf r}}$, see \S\ref{sec:opticalthindir} for details. The ionization front becomes more spherical if the injection region is enlarged, or if the radiation is forced to propagate radially.}
\label{fig:stromgren3dhs}
\end{figure*}

\begin{figure}
\includegraphics[width=0.48\textwidth]{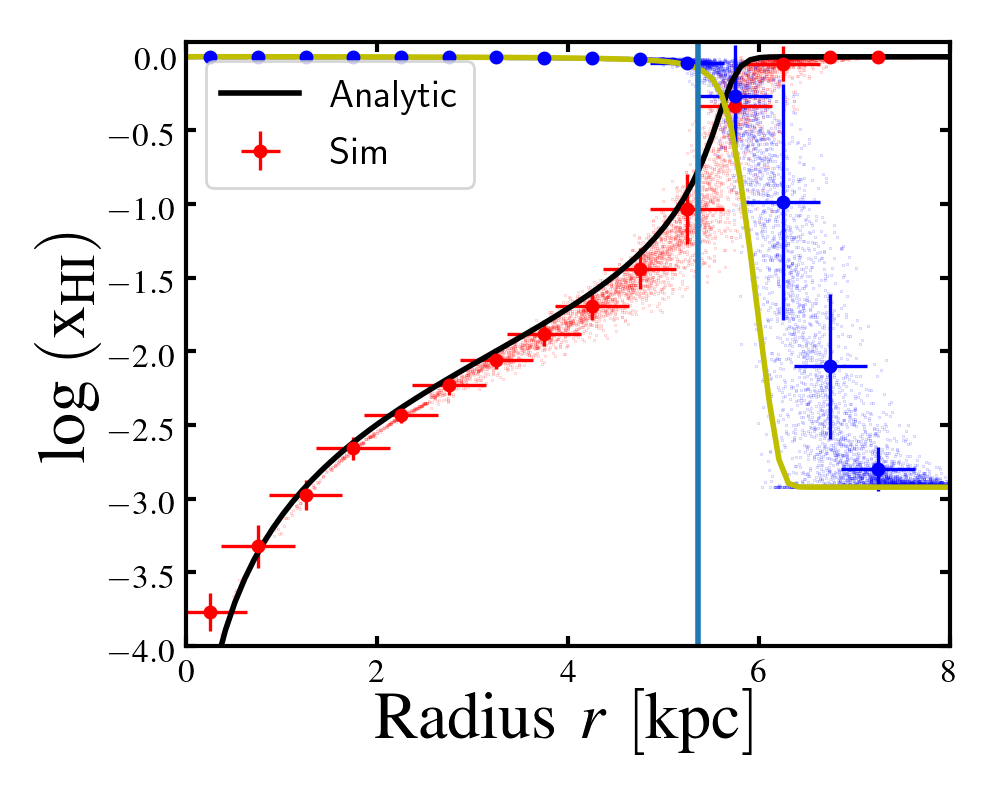}
\caption{Isothermal Str\"omgren sphere from \protect\cite{Ilie06RTcom} Test 1:
a source of radiation, located at radius $r=0$, photoionizes hydrogen gas, kept at constant density and constant temperature. The system is shown a time $t=500~{\rm Myrs}$ after the source is switched on. {\em Red points} are the neutral hydrogen fraction, $x_{\rm HI}=n_{\rm HI}/n_{\rm H}$, of individual gas particles, with the {\em thick blue points} showing binned values, with horizontal error bars indicating the bin width, and vertical error bars the standard deviation; the {\em black line} is the analytical solution derived in Appendix \ref{sec:anaStromgren}. {\em Small blue points}, and {\em thick black points} with black error bars show the corresponding ionized fraction, $1-x_{\rm HI}$, with the 
{\em yellow line} the analytical solution. The {\em vertical blue line} is the approximate analytic location of the Str\"omgren radius from the balance between injection and recombination
(Eq.~\ref{eq:anars}). The mean interparticle separation of the SPH particles is 0.625~kpc. Here we inject radiation over two smoothing lengths.}
\label{fig:stromgren3d_tm}
\end{figure}

\begin{figure}
\includegraphics[width=0.48\textwidth]{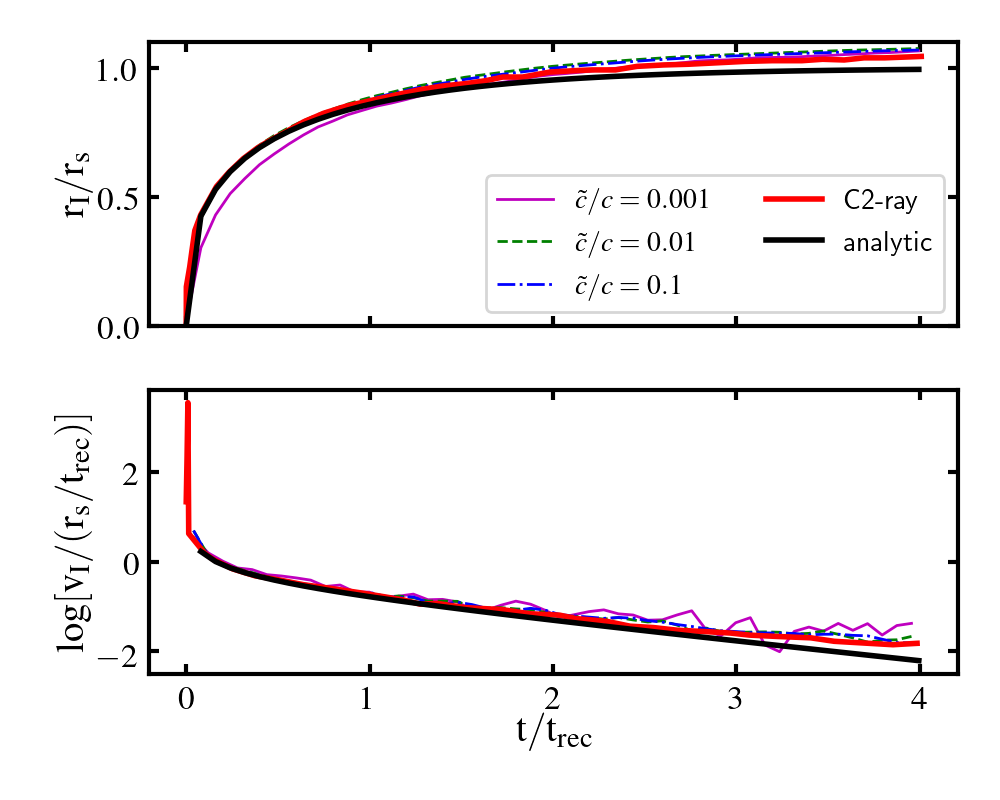}
\caption{Test 1 \protect\citep{Ilie06RTcom}: pure hydrogren isothermal HII region expansion from time $t = 0$ to $500~{\rm Myr}$. {\it Top}: the evolution of the radius of the ionization front, defined by where 50\% hydrogen in the spherical shell is ionized, divided by the Str\"omgren radius (Eq.~\ref{eq:Stromgrenradius}). {\it Bottom}: the evolution of the velocity of the ionization front, divided by the Str\"omgren radius over the recombination time $t_{\rm rec}=(n_{\rm H}\alpha_B)^{-1}$. The $c/\tilde{c}$ lines represent our simulation results with different reduced speeds of light and the fixed optically thin direction (\S\ref{sec:opticalthindir}). The black lines show the analytic solutions (Eq.~\ref{eq:anars} and its derivative; note that the analytic solution is only approximately correct). Finally, the `{\small C2-ray}' lines represent the {\small C2-ray} \protect\citep{Mell06c2ray} result in \protect\cite{Ilie06RTcom}.}
\label{fig:stromgren3devo}
\end{figure}

\begin{figure*}
  \includegraphics[width=0.48\textwidth]{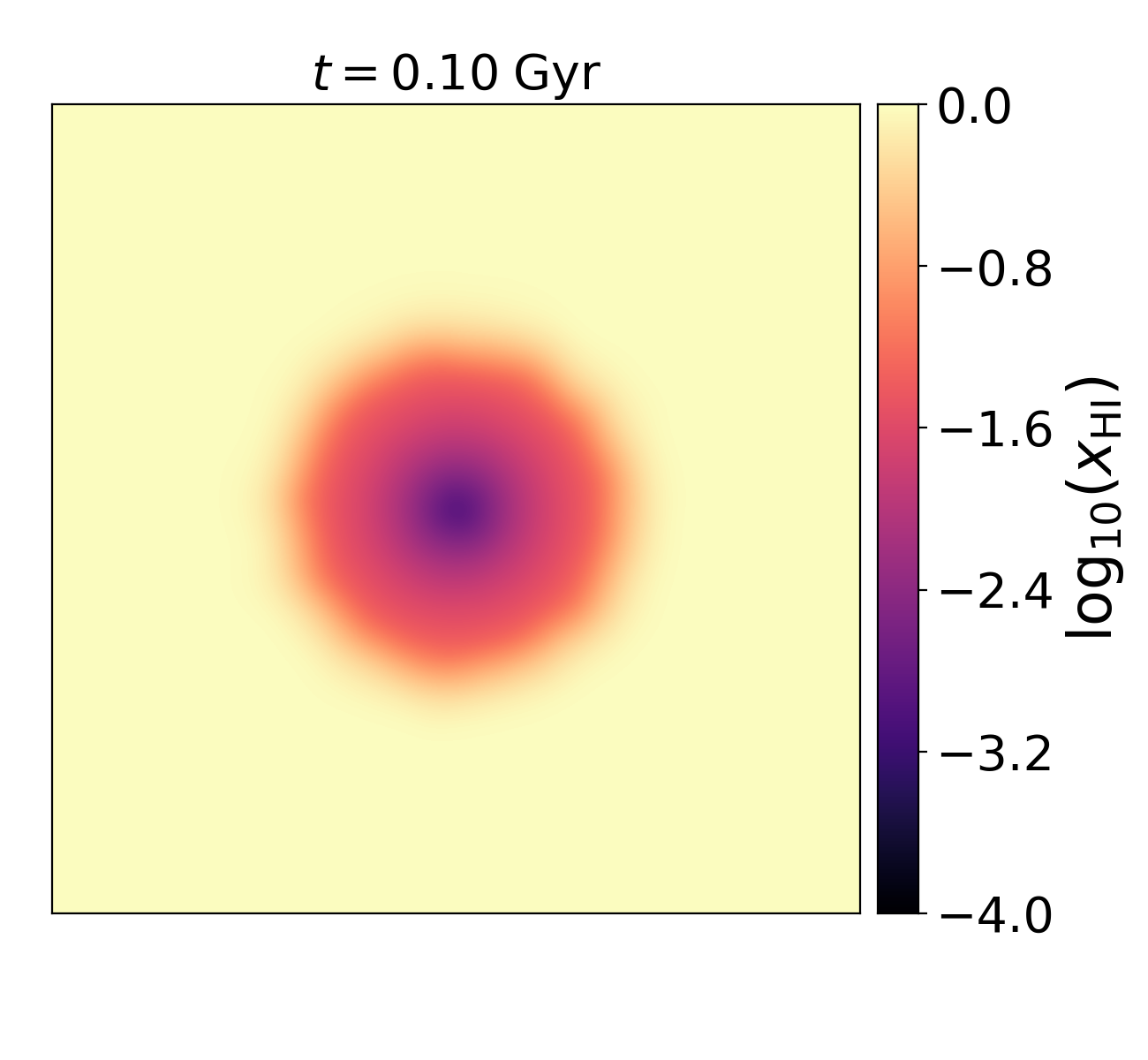}
 \includegraphics[width=0.48\textwidth]{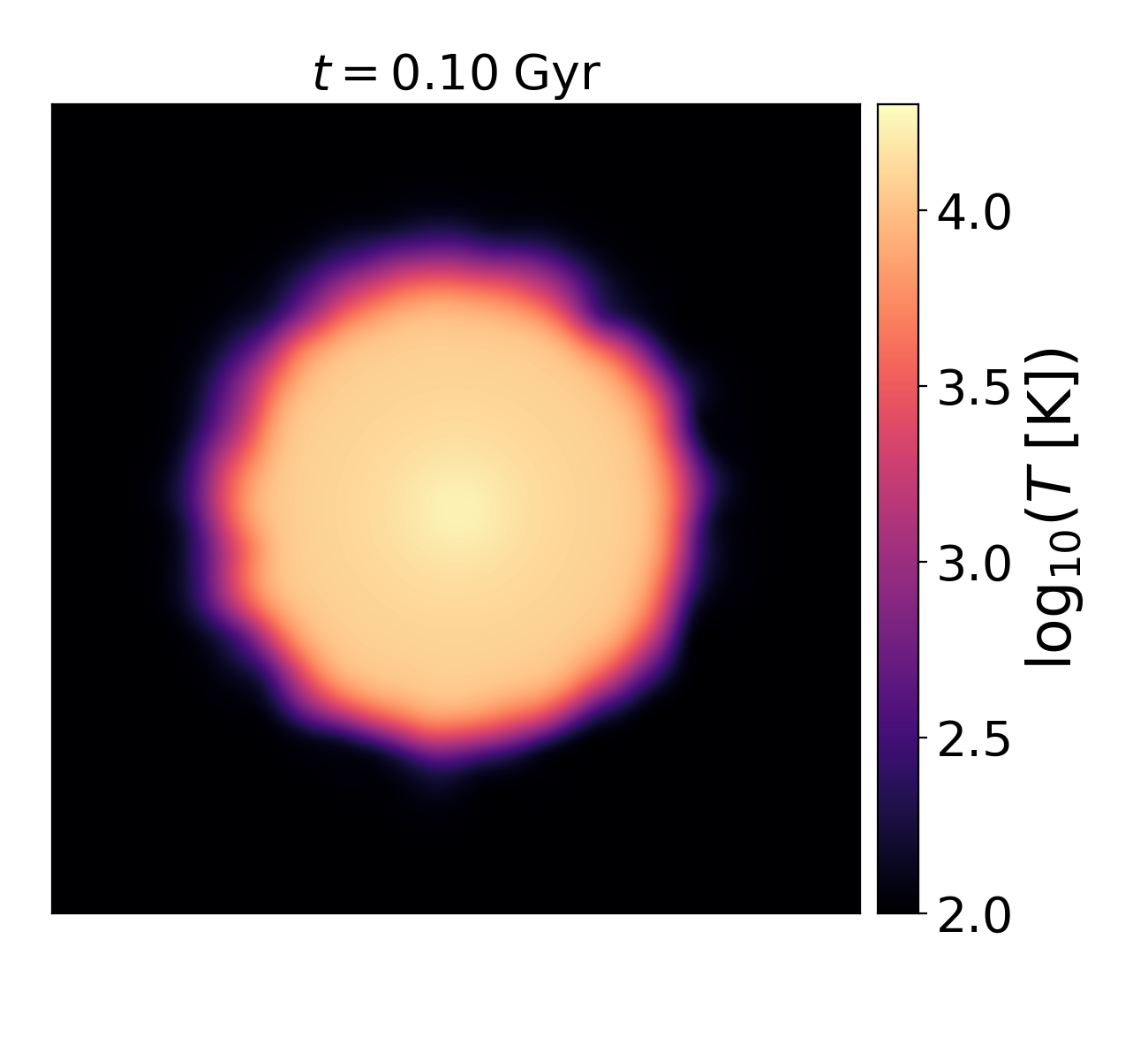}
 \includegraphics[width=0.95\textwidth]{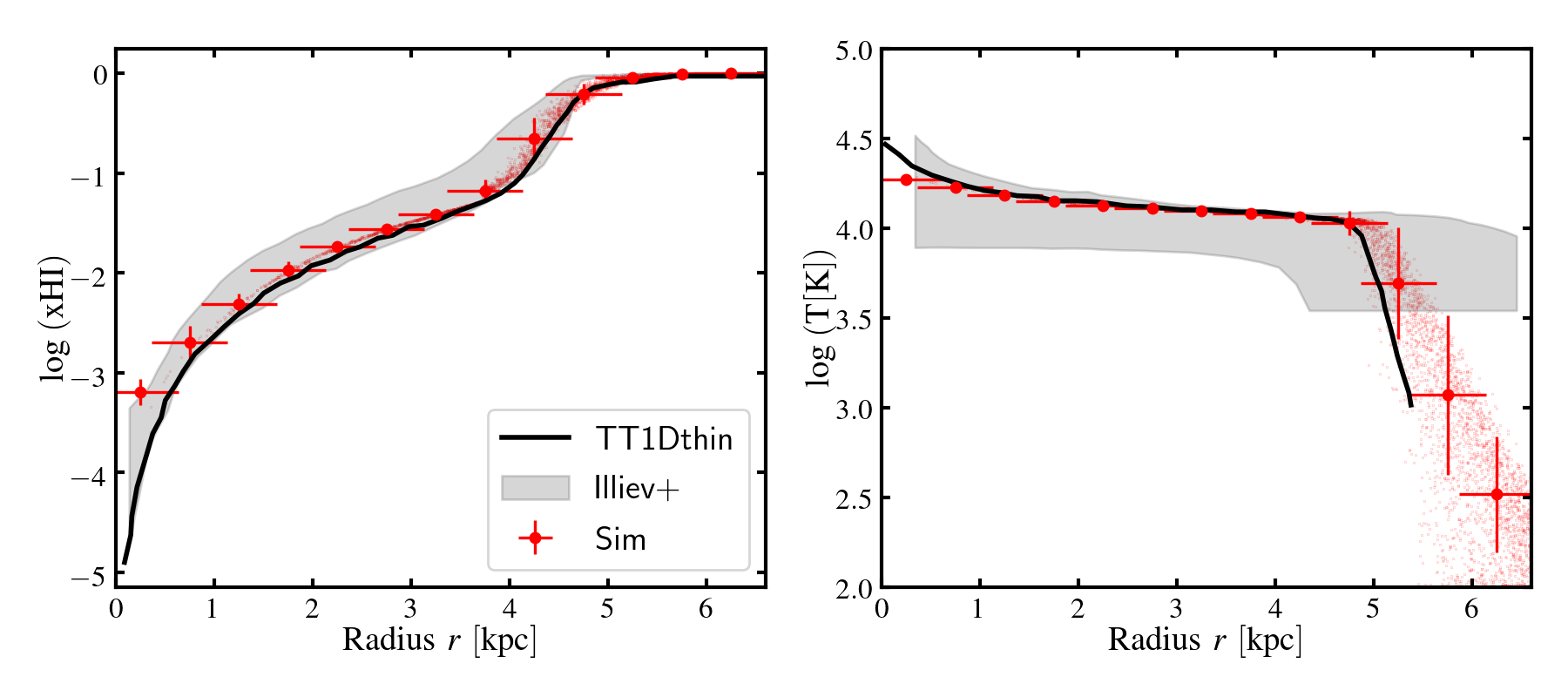}
\caption{\protect\cite{Ilie06RTcom} Test2: HII region expansion in a uniform gas with varying temperature and static gas particles at time $t=100~{\rm Myr}$. {\it Top left}: a slice of the hydrogen neutral fraction through the center; {\it Top right}: a slice of the gas temperature through the center; {\it Bottom left}: hydrogen neutral fraction as a function of radius; {\it Bottom right}: temperature as a function of radius. The red points represent the values of individual particles. The vertical error bars show the mean and standard deviation. The horizontal error bars show the smoothing length. The solid black line is the `{\small TT1D thin}' result \protect\citep{Pawl11multifRT} and the shaded region is the upper and lower bounds of the \protect\cite{Ilie06RTcom} results. Here we inject radiation over three smoothing lengths.}
\label{fig:stromvartemp}
\end{figure*}

\begin{figure*}
  \includegraphics[width=0.48\textwidth]{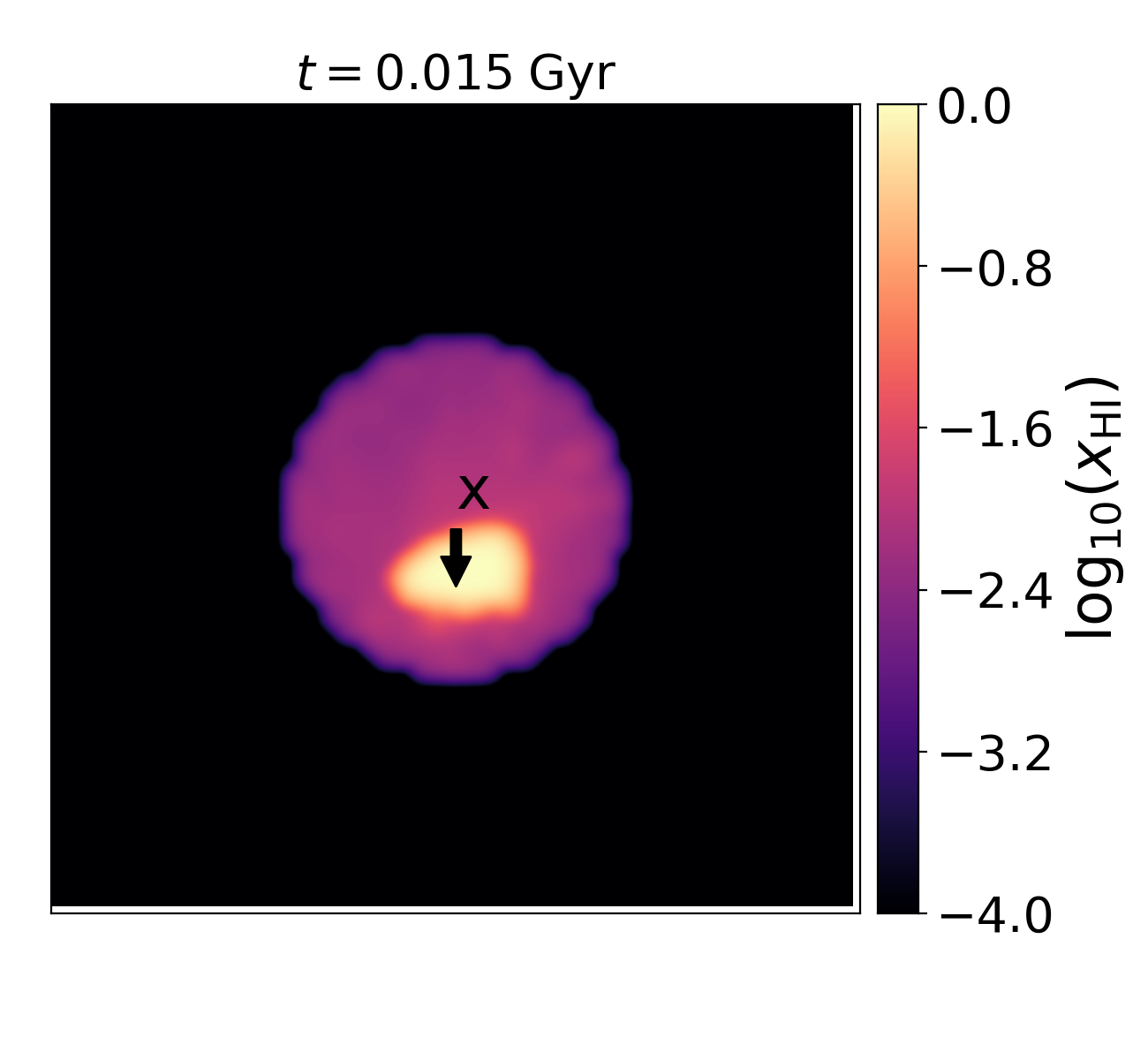}
 \includegraphics[width=0.48\textwidth]{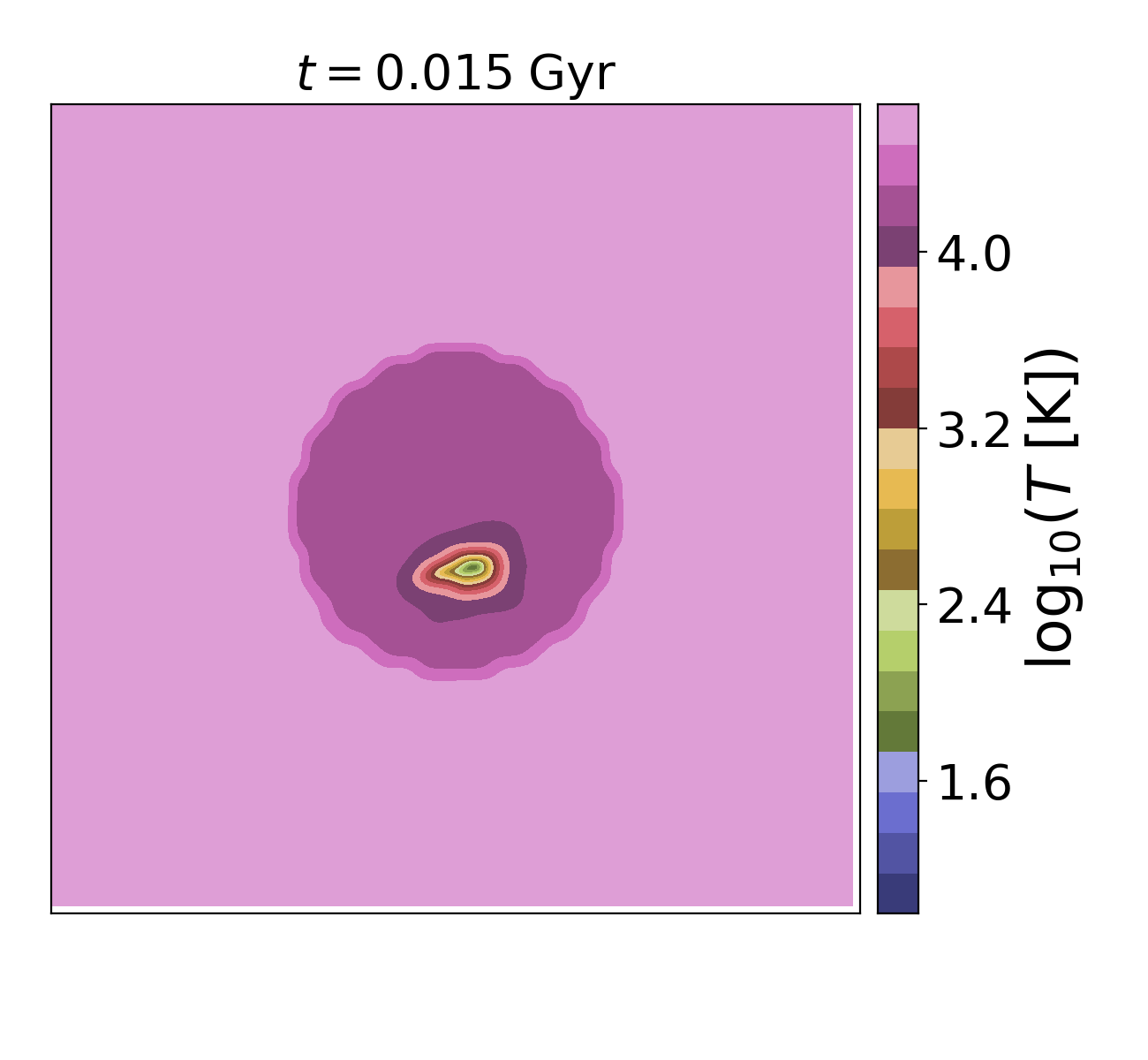}
  \includegraphics[width=0.48\textwidth]{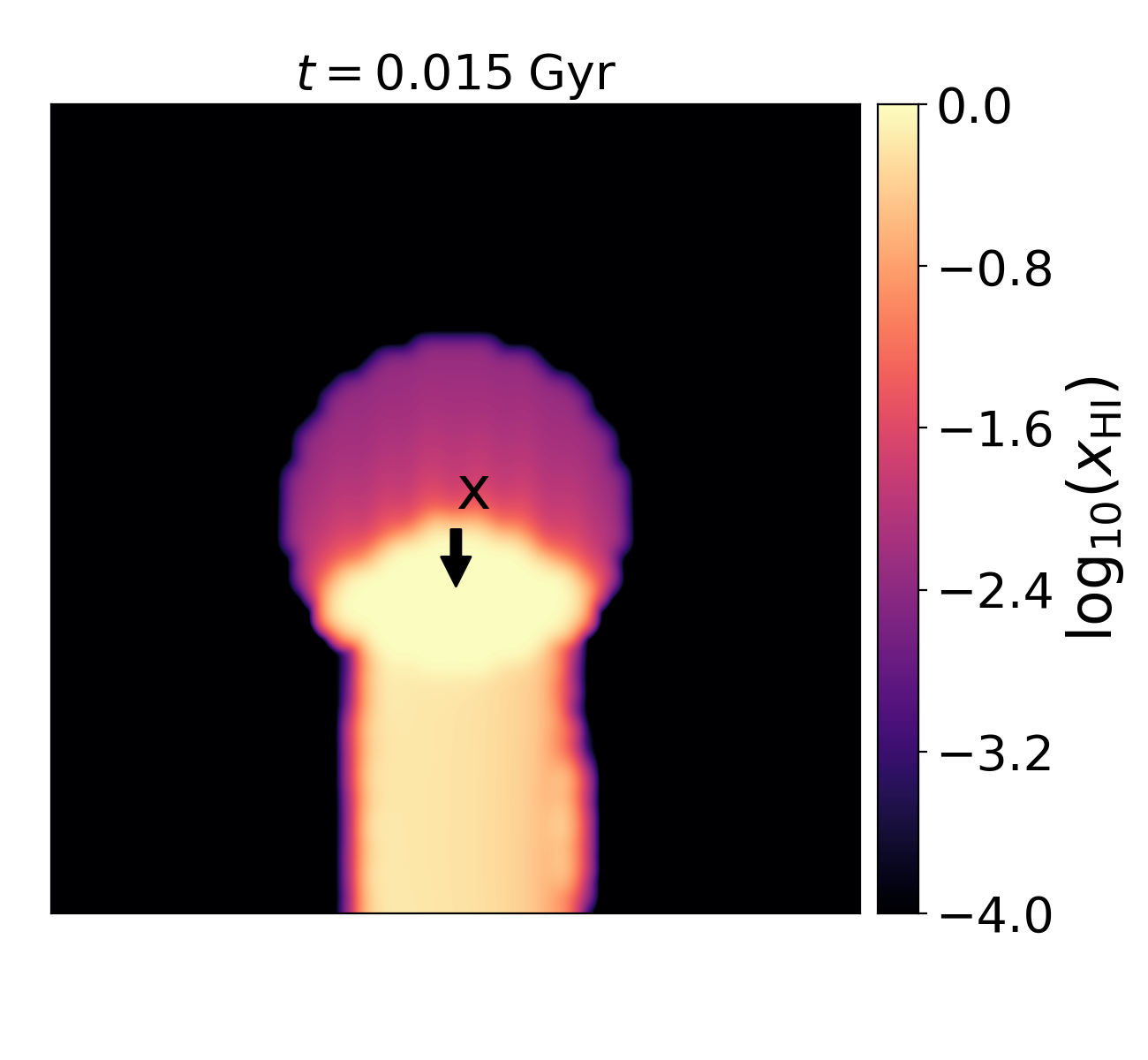}
 \includegraphics[width=0.48\textwidth]{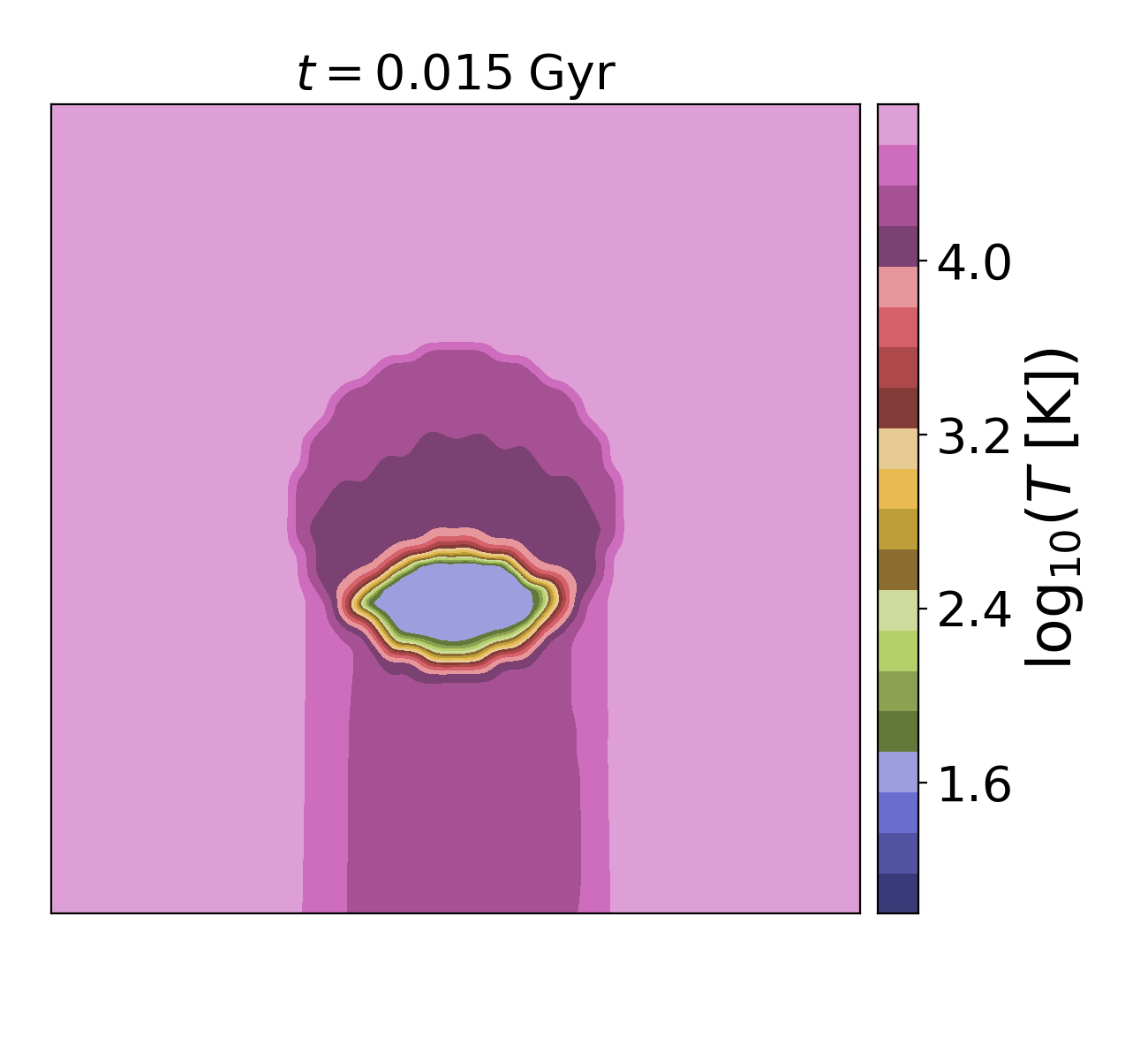}
\caption{\protect\cite{Ilie06RTcom} Test 3: I-front trapping in a dense clump. {\it Left}: a slice of mass-weighted hydrogen neutral fraction through the midplane of the simulation volume at $t=15\;{\rm Myr}$; {\it Right}: same but for temperature. {\it Upper panels}: for the default optically thin direction, $\hat{\bf n}=\hat{\bf f}$; {\it Lower panels}: for an imposed fixed optically thin direction in the $x$ direction (see \S\ref{sec:opticalthindir}), as shown by the arrows.}
\label{fig:radshadow3d}
\end{figure*}

\begin{figure*}
  \includegraphics[width=0.95\textwidth]{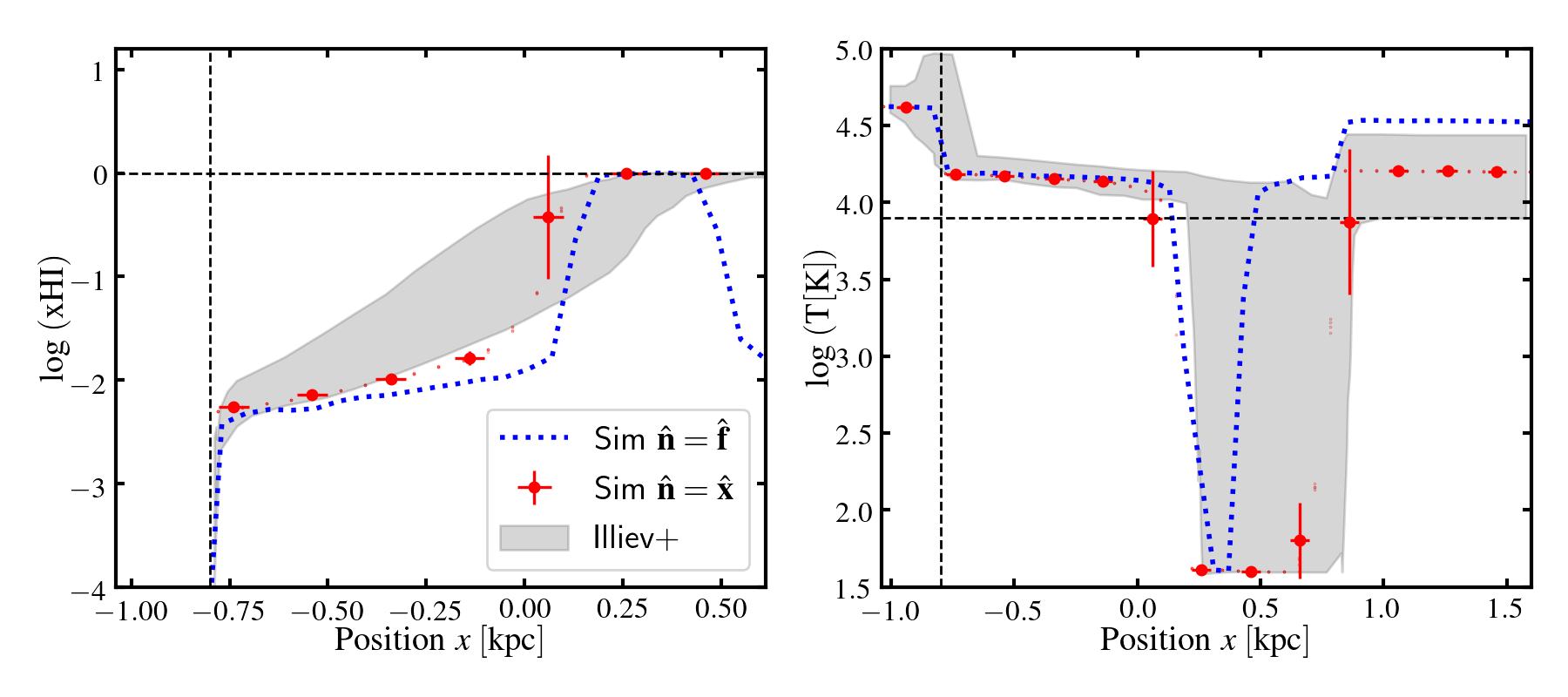}
\caption{\protect\cite{Ilie06RTcom} Test 3: I-front trapping in dense clump at $t=15\;{\rm Myr}$. {\it Left}: the neutral fraction along the axis of symmetry through the centre of the clump, where $x=0$; {\it Right}: the corresponding plot for temperature. The red points represent the values of individual particles (particles with distance < 0.1 kpc from the axis). The vertical errorbars show the mean and standard deviation and the horizontal errorbars show the smoothing length. The blue dashed lines show the result if we do not fix the optically thin direction. The shaded region encompasses the upper and lower bounds of the \protect\cite{Ilie06RTcom} results. Vertical dashed lines show the initial clump front position, whereas horizontal dashed lines show the initial neutral fraction and temperature. We only show the results for the case when the direction of the flux is imposed to be the optically thin direction (\S\ref{sec:opticalthindir}). }
\label{fig:radshadow3dxH0temp}
\end{figure*}

\begin{figure*}
  \includegraphics[width=0.48\textwidth]{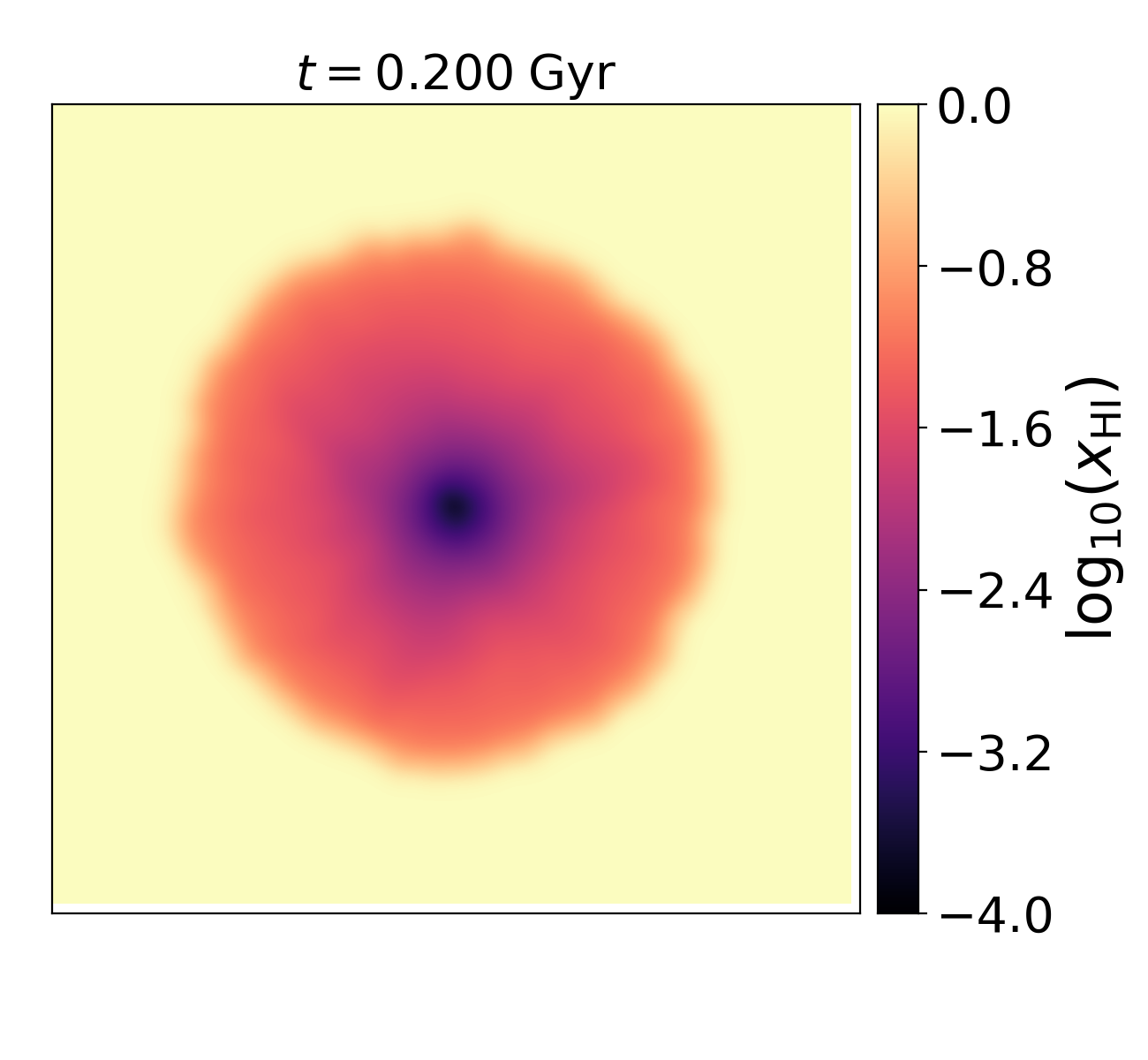}
  \includegraphics[width=0.48\textwidth]{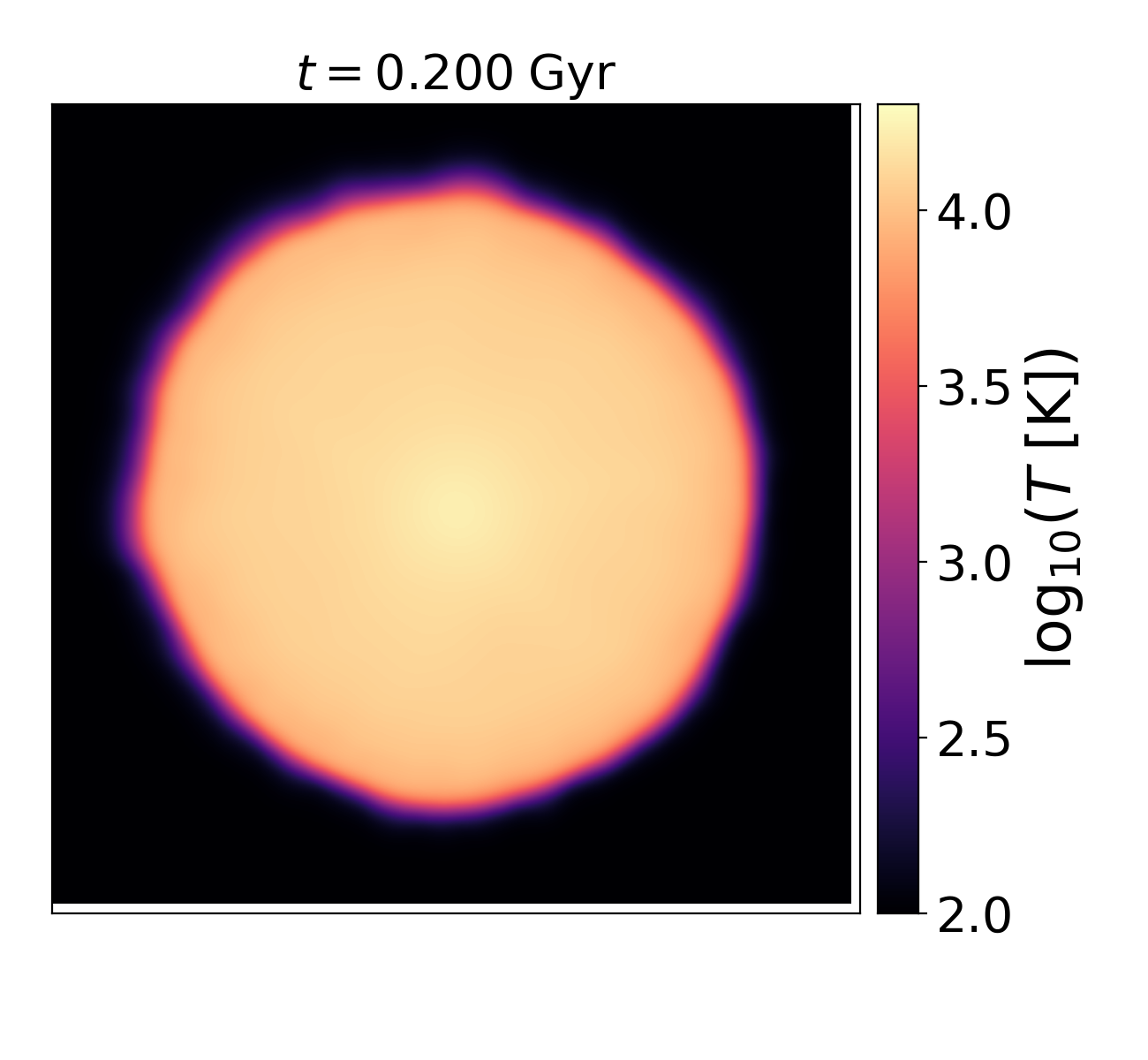}
  \includegraphics[width=0.48\textwidth]{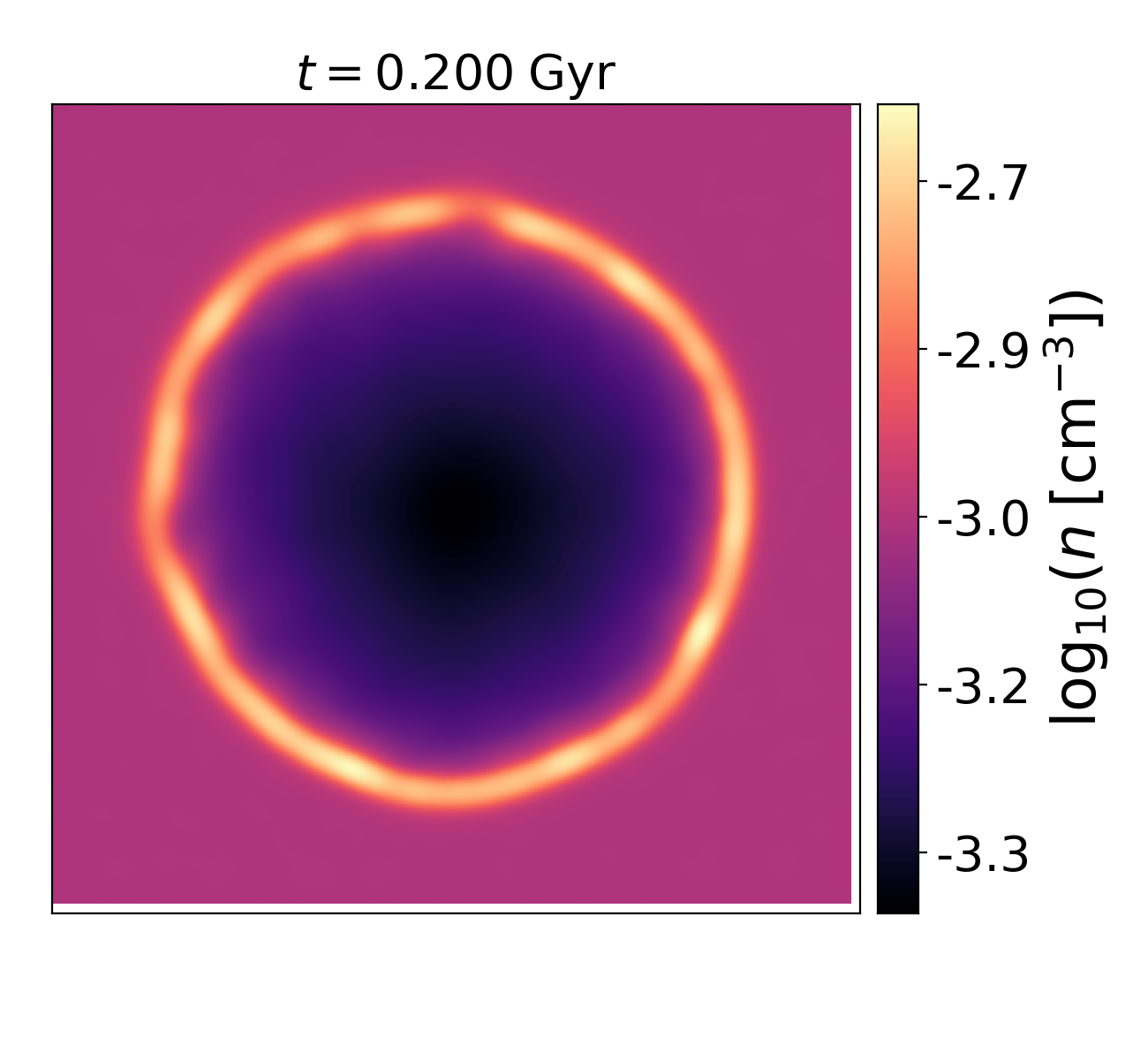}
  \includegraphics[width=0.48\textwidth]{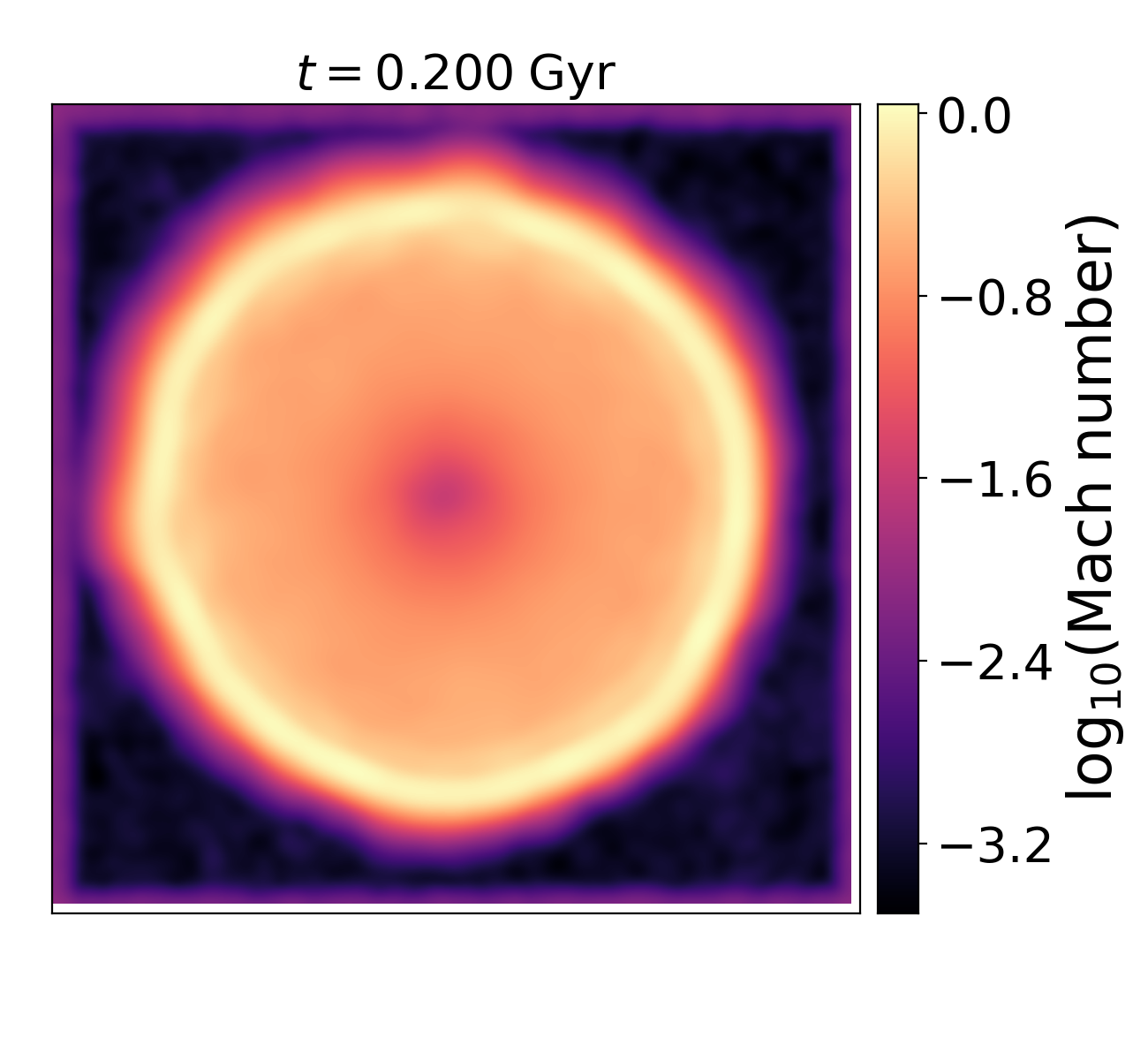}
\caption{Slices of neutral fraction ({\it upper left}), temperature ({\it upper right}), gas density ({\it lower left}), and Mach number ({\it lower right}) through 
an HII region expanding  in an initially uniform medium (\protect\citealt{Ilie09RTcom} Test 5) at $t=200\;{\rm Myr}$ through the centre of the HII region. Hydrodynamics is turned on and we inject radiation over two smoothing lengths. }
\label{fig:rhdtest}
\end{figure*}

\begin{figure*}
  \includegraphics[width=0.95\textwidth]{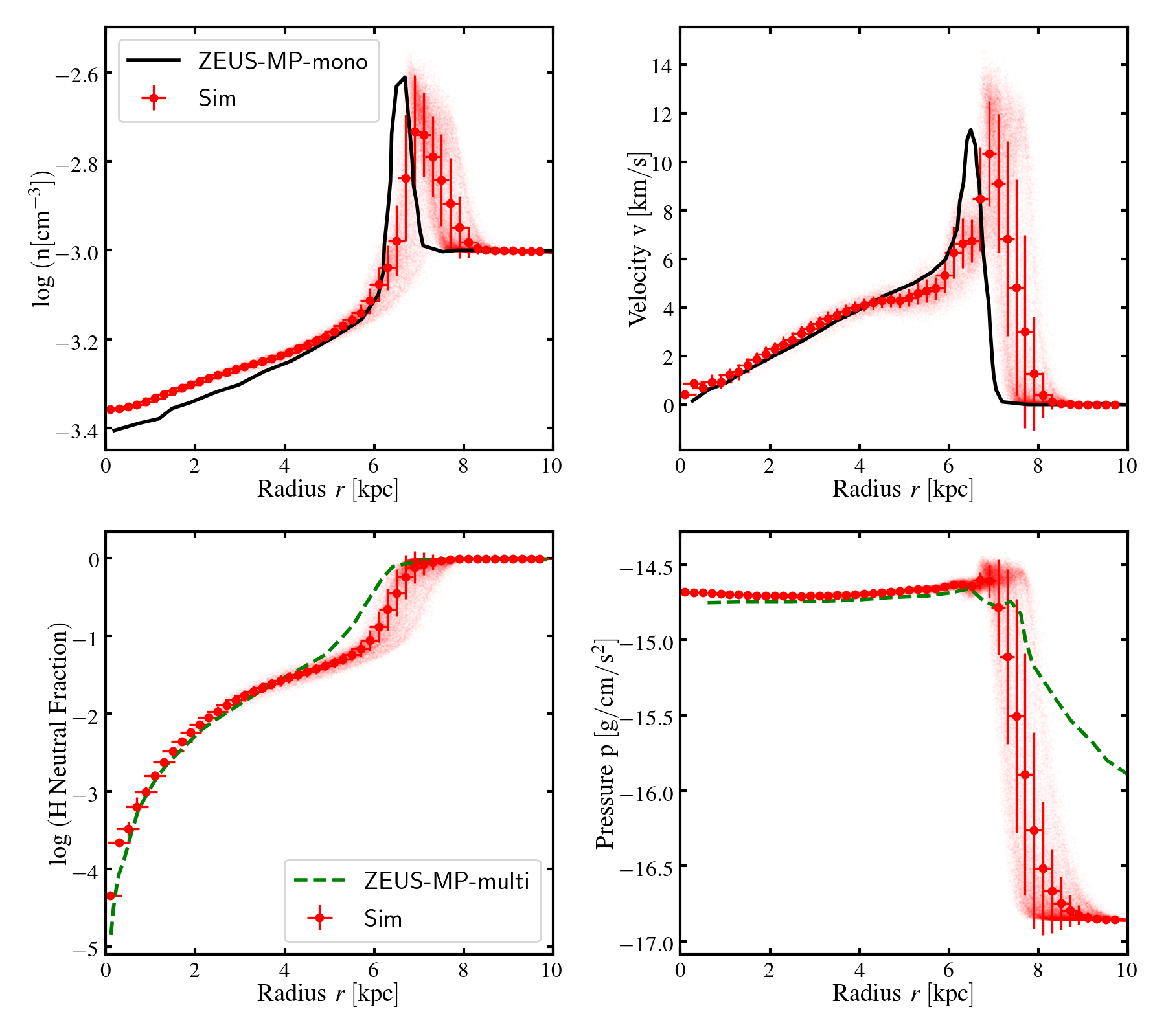}
\caption{Gas density ({\it upper left}), gas velocity ({\it upper right}), neutral fraction ({\it lower left}), and pressure ({\it lower right}) as a function of radius of the HII region expansion in an initially uniform medium (\protect\citealt{Ilie09RTcom} Test 5) at $t=200\;{\rm Myr}$. The red points represent the values of individual particles. The vertical error bars show the mean and standard deviation. The horizontal error bars show the smoothing length. The lines are the results from the {\small ZEUS-MP} code \protect\citep{Whal06ZEUSMP} in \protect\cite{Ilie09RTcom}. Black solid represents monochromatic light, i.e. a single frequency bin, similar to our implementation, whereas  green dashed lines represent multi-frequency transfer.  Injecting radiation over two smoothing lengths results causes the ionization front to propagates slightly faster than seen in {\small ZEUS-MP}.}
\label{fig:rhdtestr}
\end{figure*}

\begin{figure*}
  \includegraphics[width=0.48\textwidth]{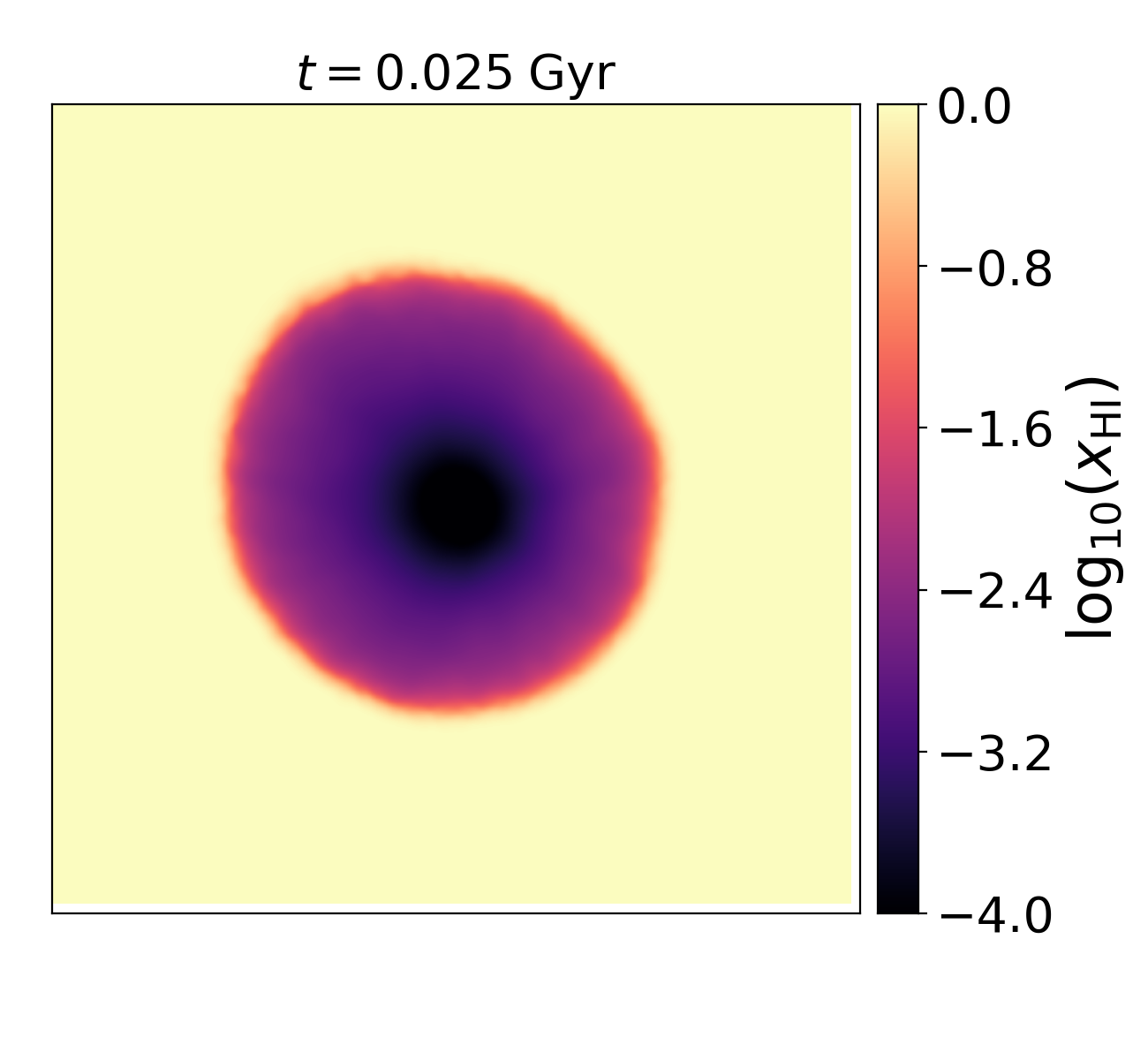}
  \includegraphics[width=0.48\textwidth]{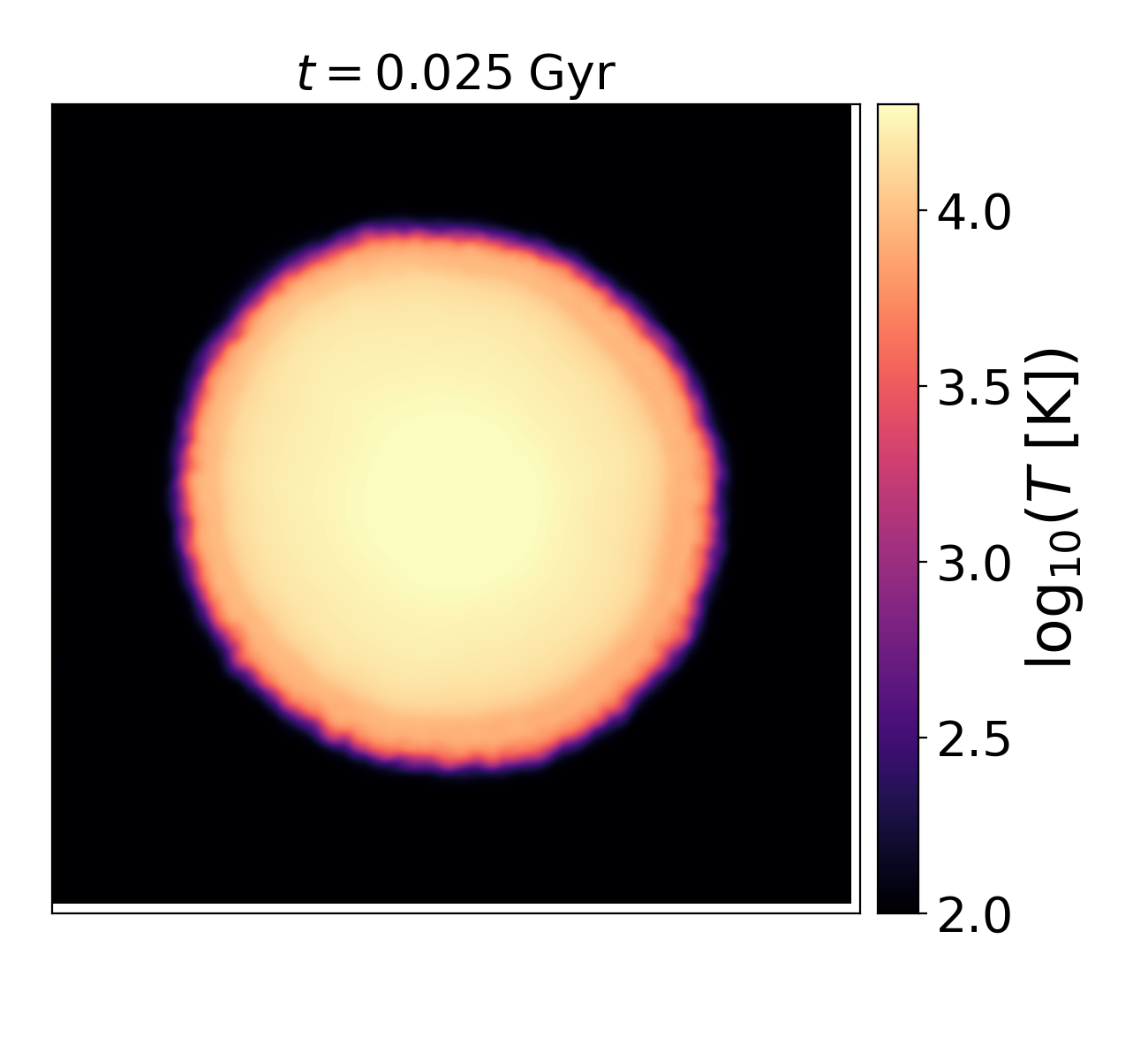}
  \includegraphics[width=0.48\textwidth]{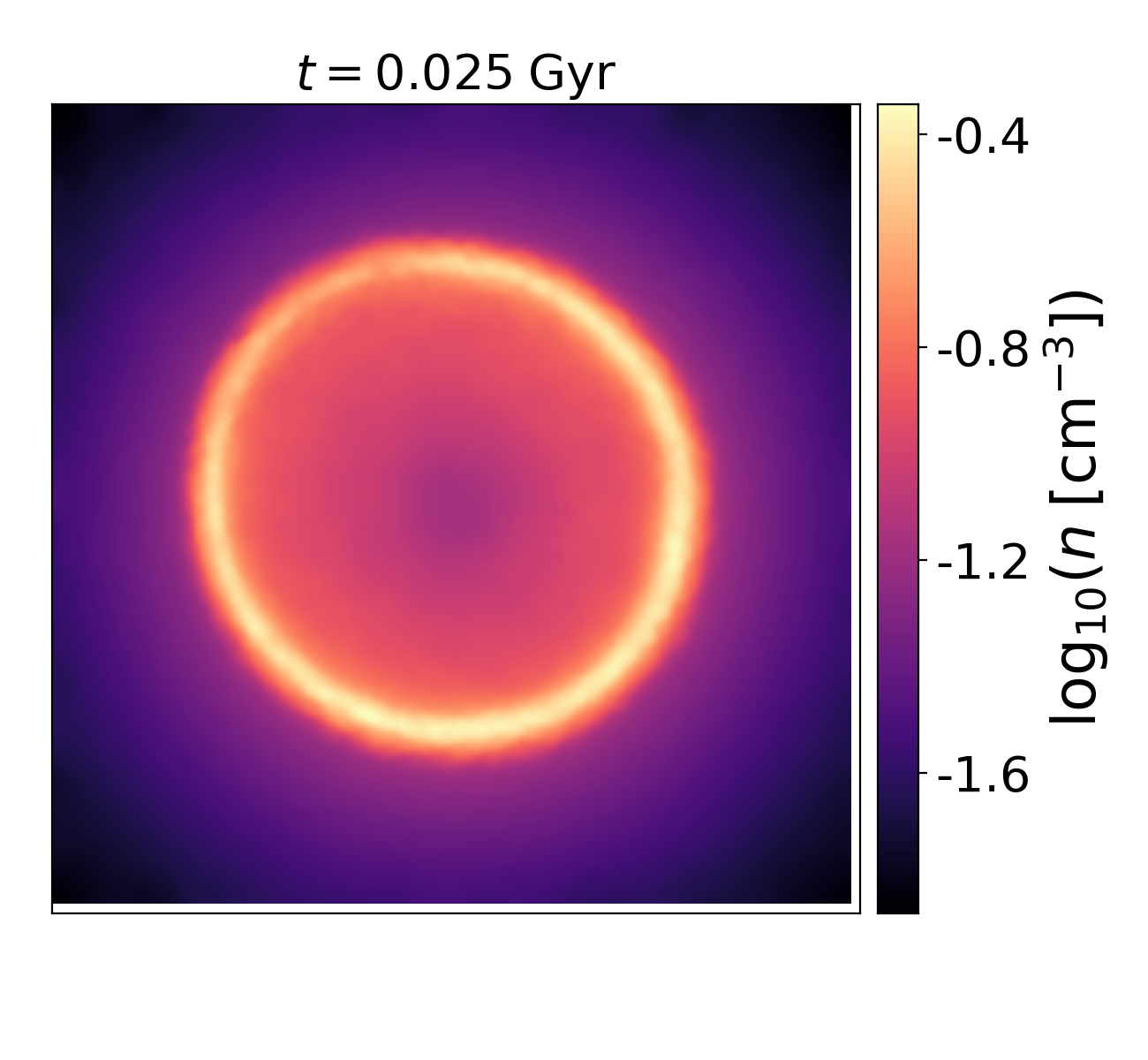}
  \includegraphics[width=0.48\textwidth]{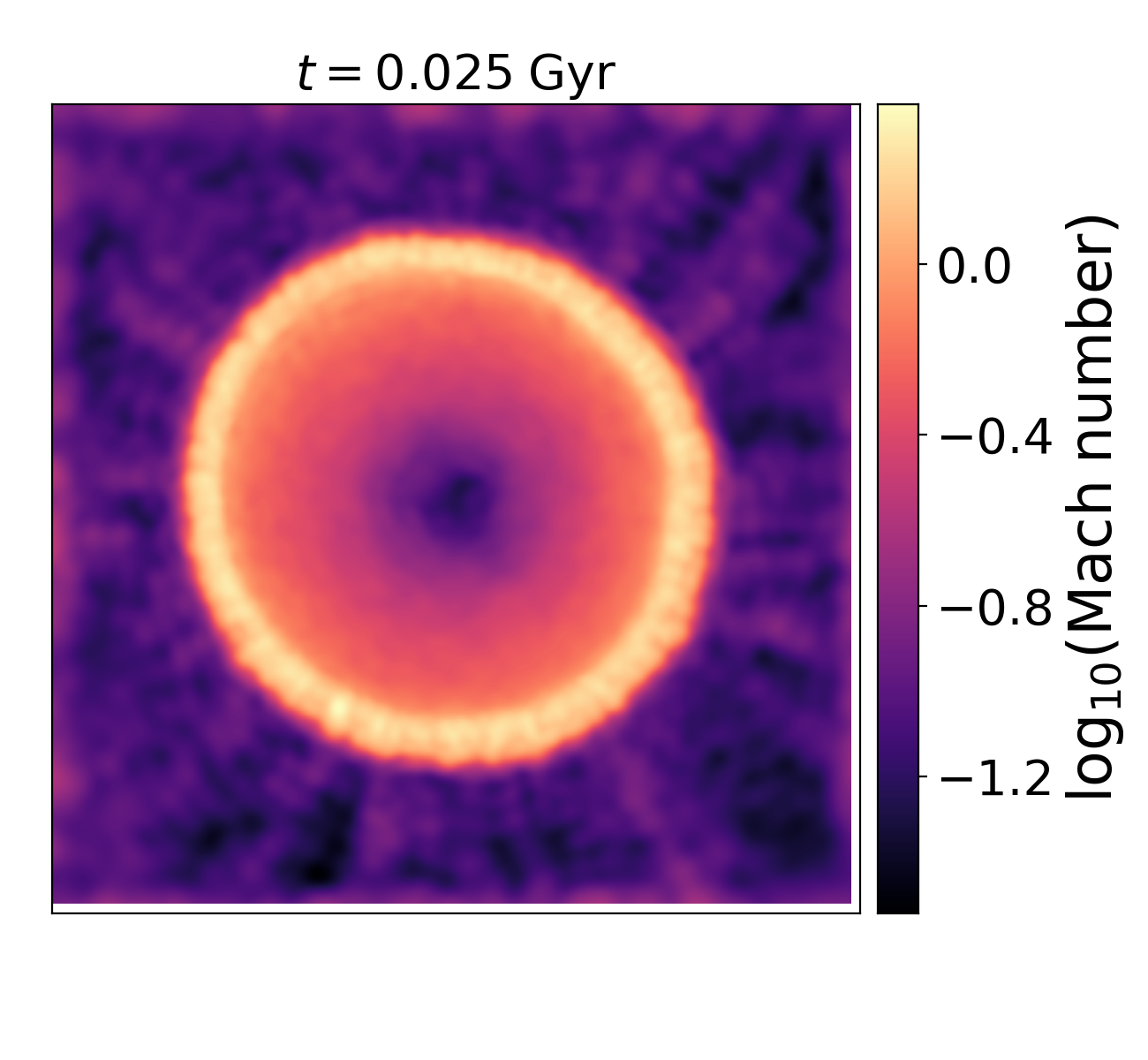}
\caption{Slices of neutral fraction ({\it upper left}), temperature ({\it upper right}), gas density ({\it lower left}), and Mach number ({\it lower right}) through 
an HII region expanding  in a medium
with initial density profile $\propto r^{-2}$  (\protect\citealt{Ilie09RTcom} Test 6) at $t=25\;{\rm Myr}$ through the centre of the HII region. Hydrodynamics is turned on and we inject radiation over two smoothing lengths. }
\label{fig:stromgrenspheredensitygradient}
\end{figure*}

\begin{figure*}
  \includegraphics[width=0.95\textwidth]{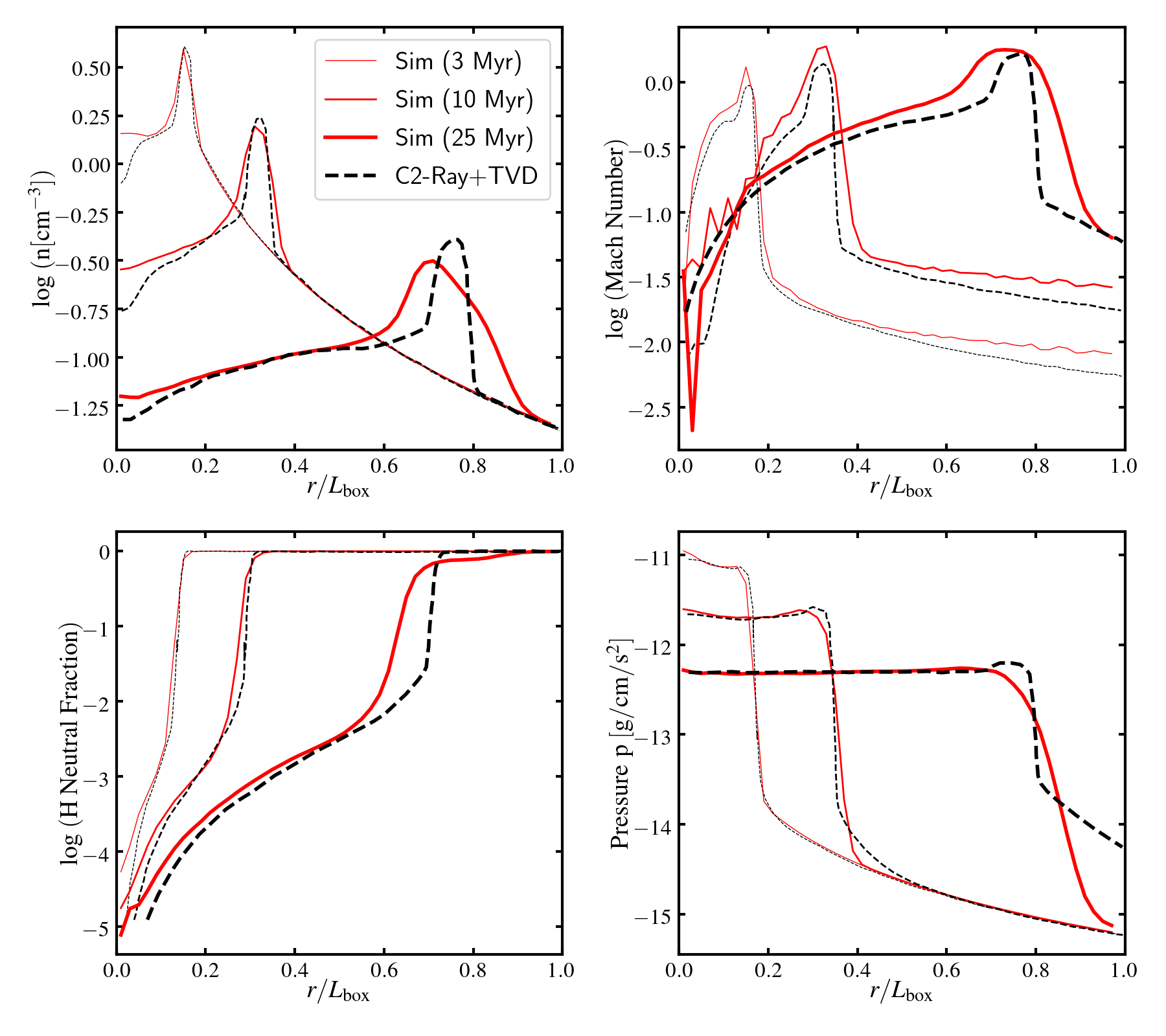}
\caption{Gas density ({\it upper left}), Mach number ({\it upper right}), neutral fraction ({\it lower left}), and pressure ({\it lower right}) as a function of radius of the HII region expanding in an initially $1/r^2$ density profile (\protect\citealt{Ilie09RTcom} Test 6) at times $t=3,10,25\;{\rm Myr}$ (thicker lines at later times). The computational volume in the original \protect\citealt{Ilie09RTcom} test has a linear extent of $L_{\rm box}=0.8~{\rm kpc}$. The {\em red solid lines} represent the mean values from our simulations; the {\em black lines} are the results from the {\small C2-Ray+TVD} code, which uses multi-frequency RT transfer. }
\label{fig:stromgrenspheredensitygradientr}
\end{figure*}

\subsubsection{Propagation in one dimension}
The setup propagates a finite radiation packet in an optically thin medium. We test whether the radiation front travels at the correct speed, $c=1$, without excessive smoothing of the front and without generating artificial oscillations in the radiation density, $E(x)$, behind the radiation packet. Initially, the radiation energy density and flux are uniform and non-zero only for $x<0$, with the radiation flux is pointing in the $+x$ direction initially. Fig. \ref{fig:radfront1d_tm} shows the initial condition and the configuration at $t=5$, when the radiation front has propagated 100 times the mean inter-particle spacing. While there is small broadening of the radiation front caused by the artificial dissipation, numerical oscillations are suppressed significantly and the scheme is stable. The front propagates at the correct speed ($c = 1$) and the radiation energy density $E$ remains constant inside the radiation package, unaffected by the artificial dissipation. The lower panel demonstrates the excellent energy conservation of our scheme in this test problem.

\subsubsection{Propagation in two dimensions}
We repeat the previous test but now in two-dimensions: a rectangular radiation package propagates in empty space in 2D. The constant-mass SPH particle distribution is glass-like, with $256\times 1024$ particles filling the computational volume of horizontal extent $\Delta x=0.5$ and vertical extent $\Delta y=2.0$.

This tests the extent to which radiation leaks out of the package artificially, either perpendicular or parallel to the propagation direction, as a consequence of the artificial dissipation. The propagation direction of the radiation on individual particles is not imposed, but computed from $\hat{\bf n}=\hat{\bf f}$.  Results are shown in Fig.~\ref{fig:radstream2d_defaultdiss} where the radiation package moves upwards from the initial state at time $t=0$ (left panel) to time $t=0.5$ (right panel).

The top panel of the figure demonstrates the ability of the implementation to maintain the direction of the radiation, with little artificial leakage of radiation perpendicular to the beam. At $t=1$, the packet has propagated upwards over 256 mean particle spacings. Two small \lq tails\rq\ of radiation trail the package, where radiation leaked out of the beam. 

In the rightmost panel, labelled `var. $\rho$', we test the radiation propagation in the presence of a particle density gradient. Constant-mass SPH particles are distributed according to a cored power-law density profile $\rho(y)$:
\begin{align}
\rho(y) = 
\left\{\begin{matrix}
\rho_c, & 0.5<y< 1.5\\ 
\rho_c[0.5/(y-1)]^2, & {\rm elsewhere.}
\end{matrix}\right.
\end{align}
Here, $y$ is the direction of propagation of the radiation. The particle density inside the core $\rho_c$ is the same as that in the uniform density test. The radiation distribution is similar to the uniform density case, except that the beam broadens slightly once it enters the low density region at the top of the panel, where the spatial resolution is lower. The total radiation energy
increases by $\sim 1$\% at t=1, mainly because of the numerical oscillations at the beam front. Because we enforce that radiation energy density remains positive everywhere, clipping negative radiation densities increase the total radiation energy. 

Smoothing perpendicular to the beam is quantified in more in detail in the lower panel, which is a cut through the middle of the beam at time $t=1$. This profile has approximately Gaussian-shaped edges, as expected from artificial diffusion (\S\ref{sec:artdiss}); the diffusion coefficient is proportional to $h\,c$. The dependence on the smoothing length, $h$, means that the beam can propagate further without distortion at higher resolution. Due to the finite resolution and, in general, non-uniform underlying SPH particle distribution, it is not possible to completely eliminate radiation leakage perpendicular to the propagation direction without causing instabilities; higher-order shock-capturing schemes \cite[e.g.][]{Liu94WENO} might suppress such artificial leakage more efficiently.

The next test is that of radiation propagating isotropically away from a source in two dimensions, see Fig.~\ref{fig:radfrontsph2d_tm}. The figure confirms that the radiation front preserves rotational symmetry as it propagates out at the speed of light. The radiation energy density is smooth in the radial shell, with no appreciable noise even though the underlying particle distribution is non-uniform. The energy density is small behind the shell. The absence of significant artificial \lq left over\rq\ radiation results from the artificial dissipation switch.

\subsection{Radiation tests without hydrodynamics: constant temperature}
\label{sec:statictest}
\subsubsection{Static Stromgren Sphere with Constant temperature}
The first test is Test~1 in \cite{Ilie06RTcom}. This tests the radiative transfer scheme and the thermo-chemistry solver against an analytical solution: uniform density, neutral gas is photoionized by a source that emits ionizing photons at a constant rate. We keep the density and temperature of the gas constant, {\em i.e.} the gas is not allowed to move, heat or cool. The ionization front propagates into the gas cloud until it reaches its Str\"omgren radius. The analytical solution (assuming grey opacity) is derived and summarised in Appendix \ref{sec:anaStromgren}. 

The numerical parameters are taken to be identical to those used
by \cite{Ilie06RTcom} to allow for a direct comparison: 
the gas cloud consists of pure hydrogen gas with density $n_{\rm H}=10^{-3}{\rm cm^{-3}}$, the collisional ionization coefficient $\beta=3.1\times10^{-16}{\rm cm^3s^{-1}}$ and the recombination coefficient is $\alpha_{\rm B}=2.59\times10^{-13}{\rm cm^3s^{-1}}$ (with the on-the-spot approximation); the photoionization cross section is $\sigma_{\rm \gamma HI}=8.13\times 10^{-18}{\rm cm^2}$. The source emits ionizing radiation at a constant rate of $\dot N_\gamma=5\times10^{48} {\rm photons~s}^{-1}$. The computational volume has linear extent of 20~kpc; the SPH particle distribution is glass-like with approximately $32^3$ particles. In this problem we also test the RSL approximation, using ${\tilde c}=c/10$.

We first compare several implementations of the injection of radiation energy by the source (\S\ref{sec:injection}) and of the optically thin closure relation (\S\ref{sec:opticalthindir}); results are shown in Fig.~\ref{fig:stromgren3dhs}. Without
requiring that the optically thin direction be radial, the ionization front is not spherical when radiation is injected over one smoothing length
(left panel), a consequence of the fact that the SPH particles are not exactly uniformly distributed around the source. This can be remedied by either injecting radiation into gas particles up to two smoothing lengths away from the source (middle panel) or by requiring that the radiation should move radially away from the source, {\em i.e.} ${\hat{\bf n}}={\hat{\bf r}}$ (right panel).

For the actual test, we inject radiation in all gas particles within two smoothing lengths from the source but without imposing a propagation direction, as in the middle panel of Fig.~\ref{fig:stromgren3dhs}. The simulation results at time $t=500~{\rm Myr}$ are compared to the analytical solution derived in Appendix \ref{sec:anaStromgren} in Fig.~ \ref{fig:stromgren3d_tm}. By this time, the system is in a steady-state where ionizations balance recombinations and the ionization front is at the location of the Str\"omgren radius. We can compare the neutral fraction in the simulation to the exact analytical solution.

The mean value of the neutral fraction as a function of radius, $x(r)=n_{\rm HI}(r)/n_{\rm H}$, follows the analytical result closely with relatively small scatter. There are small systematic deviations from the analytical solution near $r=0$, where radiation is injected, and at $r\ge 6$, where the analytical value of neutral fraction drops faster than the simulated value. The latter is due to radiation \lq leaking\rq\ beyond the Str\"omgren radius in the simulation due to the artificial dissipation. The overall performance of the scheme is relative good: (1) the neutral fraction is approximately spherically symmetric; (2) the scheme is photon-conserving and therefore the location of the Str\"omgren radius agrees well with the analytical solution; (3) the scheme is accurate both in the optically thin region near the source as well as in the optically thick region outside the Str\"omgren radius, as well as in the intermediate region. We note that some cone-based \citep[e.g.][]{Pawl08TRAPHIC} or short-characteristic \citep[e.g.][]{Finl09VET} RT schemes could produce artificial \lq ray-like\rq\ features in the neutral fraction, depending on angular resolution; no such features appear in the present scheme\footnote{Deviations from spherical symmetry near the Str\"omgren radius are apparent in Fig. \ref{fig:stromgren3dhs} due to low spatial resolution and irregular particle distribution, but these are significantly less severe than the ``ray effect'' in the cone-based or short-characteristic methods.}.

The effect of using the RSL approximation on the time evolution of the I-front is illustrated in Fig.~\ref{fig:stromgren3devo}. We use a similar setup as above, but to capture the early phase of the expansion of the ionization front (1) we inject radiation over only one smoothing length but impose that the radiation propagates radially outwards, {\em i.e.} $\hat{\bf n}=\hat{\bf r}$; (2) we use higher resolution, $64^3$ particles.

The traditional analytical solution for the time-dependent location of the ionization front, $r_I(t)$, of Eq.~(\ref{eq:anars}) assumes that the front is infinitely thin and that the down-stream gas is fully ionized. In reality, the down-stream gas is not completely ionized and the analytical solution of Eq.~(\ref{eq:anars}) is only approximately correct\footnote{The mean free path of ionizing photons is $\lambda=n_{\rm H}\sigma_{\gamma}\approx 0.04~{\rm kpc}$ and not resolved in the simulation setup.}. Because of this, we compare our simulation results to another simulation code, namely {\small C2-ray} \citep{Mell06c2ray} (see also \citealt{Ilie06RTcom}) and, follow them by defining the position of the I-front as the radius at which $x_{\rm HI}=0.5$.

Fig. \ref{fig:stromgren3devo} demonstrates that our results converge for $\tilde c \to c$, and even when $\tilde c=c/100$ are close to those obtained with {\small C2-ray}. Using $\tilde c=c/1000$, we notice deviations of a few tens of percent at times less than a recombination time, and much smaller than 10 per cent later on. This matches our expectation discussed in \S\ref{sec:RSL} \cite[see also][]{Rosd13ramsert}. The scheme works well at low resolution even when using non-uniform particle distributions. This is important because in typical applications ({\em e.g.} reionization simulations or simulations of the interstellar medium) the gas distribution around the ionizing sources is often at best marginally resolved. 

\subsection{Radiation tests without hydrodynamics: variable temperature}
\label{sec:vttest}

\subsubsection{Static Str\"omgren Sphere with Thermodynamics}
We repeat the previous test, but now allow photo-heating of the ionized gas, testing the interaction between the radiative transfer and the photo-chemistry solver. The parameters of the test are identical to the previous case; the photo-heating and cooling processes are detailed in Appendix \ref{sec:heatcoolparam} 
(the test makes the on-the-spot approximation). We use the optically thin value for
the photo-heating energy rate per ionization, $\epsilon_{\gamma}$ (see Appendix \ref{sec:heatcoolparam}). The underlying particle distribution is glass-like with 32 particles on a side; the injection radius is three smoothing lengths (see \S\ref{sec:injection}); and $\tilde c=0.01c$. Fig.~\ref{fig:stromvartemp} summarises our simulation results. 

We compare our results to results for the same setup published by \citet{Ilie06RTcom}, since this test has no known analytical solution. Unfortunately the comparison is not straightforward because some codes in that paper use different values for the thermo-chemistry coefficients (see Fig. 2 in \citealt{Ilie06RTcom}), and some codes use multi-frequency RT to account for spectral hardening. Spectral hardening leads to pre-heating of gas upstream from the ionization front. To make the comparison appropriate, we
also compare to {\small TT1D} (TestTraphic1D), which is a 1D radiative transfer code developed by \cite{Pawl08TRAPHIC,Pawl11multifRT} for testing the {\small TRAPHIC} code. We will compare to their `{\small TT1D thin}' result, which uses the same assumptions as ours, {\em i.e.} one frequency bin (grey approximation) and photoheating in the optically thin limit. Fig. \ref{fig:stromvartemp} demonstrates that our scheme produces a roughly spherical morphology at this resolution with the mean value at a given radius matching those obtained by {\small TT1Dthin}. In addition, our result falls within the grey-band defined by the range of simulation results published by \cite{Ilie06RTcom}.

\subsubsection{Ionization Front trapping with a Dense Clump}
Next, we illustrate our scheme's ability to trap an ionization front and cast a shadow, by
repeating \lq Test 3\rq\ in \citealt{Ilie06RTcom}. The setup is as follows: a cubic volume of linear extent $4.0~{\rm kpc}$ is filled with hydrogen gas of density $n_{\rm out}=2\times10^{-4}{\rm cm}^{-3}$ and temperature $T_{\rm out}=8000\;{\rm K}$. A spherical cloud with radius $0.8~{\rm kpc}$, hydrogen density $n_{\rm clump}=0.04\;{\rm cm}^{-3}$ and temperature $T_{\rm clump}=40\;{\rm K}$ is placed at the centre of the volume. The system is irradiated from the top with a black-body spectrum with temperature $T_{\rm BB}=10^5{\rm K}$, injecting a constant photon flux of $10^6\,{\rm photons}\,{\rm s}^{-1}{\rm cm}^{-2}$ from the upper computational boundary. We impose absorbing boundary conditions at the lower computational boundary and periodic boundary conditions at all other computational boundaries.

The SPH particle distribution is glass-like with approximately $64^3$ particles (which is lower than the tests published in \cite{Ilie06RTcom} but enough to demonstrate the performance of our scheme). We model the over-density of the gas in the clump by increasing the hydrogen fraction of the SPH particles in the clump. We use the RSL approximation, setting ${\tilde c}=c/10$.

We show the simulation results in Fig.~\ref{fig:radshadow3d}, with (lower panels) and without (upper panels) imposing the optically thin propagation direction for the radiation, ${\hat{\bf n}}={\hat{\bf x}}$, ( \S\ref{sec:opticalthindir}). As expected, the gas in front of the high-density region is highly ionized; inside the high-density region but upstream from the ionization front it is ionised to a level $n_{\rm HI}/n_{\rm H}\sim 10^{-2}$; the ionization front is trapped inside the high-density region at $x\sim 0.1~{\rm kpc}$, and finally the gas is mostly neutral behind the ionization front in the shadow behind the high-density region.

We follow the reasoning of \S\ref{sec:singlegasparcel} to estimate the temperature
immediately after the ionization front has passed,
\begin{align}
    T_1&=\frac{2-x_0}{2-x_1}T_0+\frac{2}{3}\frac{x_0-x_1}{2-x_1}\frac{\epsilon_\gamma}{k_{\rm B}}\,,
\end{align}
where $x_0$ and $T_0$ are the initial neutral fraction and temperature, $x_1$ and $T_1$ are the neutral fraction and temperature when the ionization front has passed, and $\epsilon_\gamma\approx 6.33~{\rm eV}$ is the photoheating per ionization. In front of the high-density region, we take $x_0=1$, $T_0=8000~{\rm K}$ and $x_1\approx 0$ to find $T_1=10^{4.45}~{\rm K}$; upstream from the ionization front inside the high-density region, we take $x_0=1$, $T_0=40~{\rm K}$, $x_1=10^{-2}$ to find $T_1=10^{4.38}~{\rm K}$. Once the front has passed, the gas will continue to be photoheated while also cooling radiatively. In front of the high-density region, the recombination time, $\tau_r\equiv 1/(\alpha_B\,n_{\rm H})\approx 600~{\rm Myr}$, is long compared to the simulation time, and the gas is still heating slowly to its new equilibrium temperature of $10^{4.64}~{\rm K}$.
Inside the high-density region, $\tau_r\approx 3~{\rm Myr}$ is short compared to the simulation time and the gas should be in thermal equilibrium. Scaling the flux so that the neutral fraction is $10^{-2.2}$, close to what is found in the simulation at the front of the high-density region, yields an equilibrium temperature of $\approx 10^{4.18}~{\rm K}$,  consistent with what we find in the simulation (Fig.~\ref{fig:radshadow3dxH0temp}).

When the propagation direction is imposed, the over-dense gas traps the I-front at time $t=15\,{\rm Myr}$, and the run of the neutral fraction and temperature from the top of the volume in the $x$-direction follows our analytical estimates. It also falls within the grey-region, defined by the the locus of the simulation results
for the different codes for Test 3 in \cite{Ilie06RTcom}, as shown in Fig.~\ref{fig:radshadow3dxH0temp}. The 
shadow is relatively sharp, with a small level of ionization at its boundary due to the numerical/artificial diffusion. As expected, the gas is also cooler in the shadow with the temperature there agreeing with the results of
some of the codes in \cite{Ilie06RTcom}. As in the previous test case, our results are not directly comparable to some of the codes in \cite{Ilie06RTcom}, in particular codes that perform multi-frequency RT capture the pre-heating ahead of the ionization front due to the smaller optical depth of higher-energy photons.

When the direction of propagation is not imposed (top panels in Fig.~\ref{fig:radshadow3d} and Fig.~\ref{fig:radshadow3dxH0temp}), the ionization front is still trapped in the high-density region and there is still a fraction of self-shielded gas. However, the numerical and/or artificial diffusion 
wipes-out the shadow. Some high-density gas down-stream from the ionization front also gets ionized as a consequence of artificial diffusion.

While fixing the optically-thin direction is less relevant for simulations with many sources, this approach is useful for a few radiation sources or the propagation of radiation fronts in a few directions, which we are currently investigating.

\subsection{Radiation tests with hydrodynamics and variable temperature}
\label{sec:rhdtest}
In this section we present tests with heating, cooling, and hydrodynamics with radiation\footnote{We ignore the $\nabla{\bf v}:\mathbb{P}$ and radiation pressure terms (see \S\ref{sec:two-moment}) in these test problems.}.

\subsubsection{Str\"omgren Sphere with hydrodynamics}
The first test is a repeat of the HII region, but now allowing the gas to expand as it is heated; this is Test 5 in \cite{Ilie09RTcom}. The problem setup and conditions are exactly identical to the variable temperature Str\"omgren sphere test in \S\ref{sec:vttest}, but we simulate also the hydrodynamics response of the gas. We use the default {\small SPHENIX} SPH formulation for hydrodynamics (\S\ref{sec:SPHform}) with $64^3$ particles in the computational volume.

We do {\it not} fix the optically thin direction here, but instead inject radiation over two smoothing lengths. We find that the ionized region is nevertheless almost spherically symmetric (as it should be). However, the location of the front is slightly further out compared to the profile computed with {\small ZEUS-MP}. This is because this way of injecting radiation does not quite guarantee that the flux drops correctly with distance in the injection region. Figure~ \ref{fig:rhdtest} shows that the gas profiles are approximately spherically symmetric 
and there is no evidence of any numerical instabilities. 

In Fig.~\ref{fig:rhdtestr} we compare our simulation results (red points show values for individual SPH particles, blue points show the mean and scatter of values in radial bins) to those of {\small ZEUS-MP} single-frequency bin result (black solid lines), as published by \cite{Ilie09RTcom}. For the gas neutral fraction and pressure, the single-frequency bin result is not available, so we compare to the multi-frequency {\small ZEUS-MP} result instead. 

As the gas is ionized and heated, the surrounding gas is swept-up in a dense shell. The location of the shell agrees well with that found by {\small ZEUS-MP}, as do gas neutral fraction, density and velocity profiles. In our simulation, the shell is somewhat wider than that found by {\small ZEUS-MP}, partly because of our much lower resolution and possibly due to the applied artificial viscosity in the hydrodynamics solver.   

Our pressure profile agrees with {\small ZEUS-MP-multi} until $r=8\;{\rm kpc}$, where {\small ZEUS-MP-multi} shows a slower falloff in pressure. This is because {\small ZEUS-MP-multi} includes also {\it spectral hardening}, where high frequency photons penetrate further into the neutral medium. Our current single-bin method cannot model this effect.

\subsubsection{Str\"omgren Sphere in a $1/r^2$ density profile with hydrodynamics}
\label{sec:stromgrendensitygradient}

Finally, we examine how our code performs in the case when the density is not uniform  (\citealt{Ilie09RTcom} Test 6). This allows us to validate our hierarchical time-stepping scheme (\S\ref{sec:code}) and test the propagation of radiation down a density slope. 

We initialise a density profile $n_{\rm H}(r)$ as a function of radius $r$ from the center, given by
\begin{align}
n_{\rm H}(r)=
\begin{cases}
n_{\rm core}, & \text{ if } r<r_{\rm core}, \\ 
n_{\rm core}(r_{\rm core}/r)^2, & \text{ if } r> r_{\rm core},
\end{cases}
\end{align}
by distributing $\sim 64^3$ equal-mass SPH particles in a $(2~{\rm kpc})^3$ box\footnote{We use a slightly larger box size than \cite{Ilie09RTcom} (box size $=1.6~{\rm kpc}$) since we adopt periodic boundary condition rather than trans-missive BCs and we want to avoid edge effects.}. We take $n_{\rm core}=3.2\,{\rm cm}^{-3}$ and $r_{\rm core}=91.5\,{\rm pc}$. We also set up a star particle at the center of the box which emits $10^5{\rm K}$ black body spectrum photons at a rate of $\dot{N}_\gamma=10^{50}~{\rm photons~s^{-1}}$. We choose to inject radiation over one smoothing length and set $\tilde{c}=0.01c$ in this test. 

Initially, the R-type ionization front moves quickly on a time-scale $\ll 1 {\rm Myr}$ stalling at the initial Str\"omgren radius at $r_{\rm S}\sim 70 {\rm pc}$.  Then the gas itself starts to expand as it is heated and the ionization front expands slowly towards the edge of the computational volume. The ionization front accelerates once it exits the core region, due to the lower gas density at a larger radius. 

In Fig.~\ref{fig:stromgrenspheredensitygradient} 
we plot slices through the neutral gas fraction, temperature, density, and Mach number at time $t=25~{\rm Myr}$. The profiles are approximately spherically symmetric in the ionized region, although there are some  deviations from spherical symmetry caused by the non-uniform particle distribution.
There is also some noise visible in the Mach number slice outside the ionization front, because it is difficult to set up a completely static SPH density field when the density distribution has a steep gradient.

In Fig.~\ref{fig:stromgrenspheredensitygradientr}, we compare spherically-averaged radial profiles to the results obtained with the 
{\small C2-Ray+TVD} code \citep{Mell06c2ray, Trac04TVD}. Our results agree reasonably well on the location and speed of the ionization front, as well as on the run of density, mach number, ionised fraction and pressure. The biggest differences occur in the outer parts that have not been reached yet by the radiation. These differences are a consequence of how the initial conditions are set up: in SPH, it is difficult to set up 
the initial conditions very accurately. 
Furthermore, {\small C2-Ray+TVD} has higher pressure downstream from the ionization front,
because, unlike in our implementation, it performs multi-frequency RT, so that gas downstream from the ionization front undergoes preheating.

\section{Discussion}
\label{sec:discussions}

In the previous sections, we have demonstrated that our
RT implementation accurately propagates radiation in the optically thin limit, preserving the direction of propagation and advancing the radiation front at the correct speed (\S\ref{sec:method} and \S\ref{sec:purepro}).
In three dimensions, it accurately reproduces the initial expansion of an ionization front around a source, and its asymptotic slow down to the Str\"omgren radius. The implementation also handles ionization front trapping (\S\ref{sec:results}), reproducing results accurately also when the 
hydrodynamics of the gas is accounted for, even at moderate numerical resolution (\S\ref{sec:rhdtest}). 

Importantly, the method has favourable computational scaling, which is proportional only to the number of gas particles, and independent of the number of sources.
By implementing a \lq reduced speed of light\rq\ method, the time-step associated with radiation propagation can be dramatically increased
(see \S\ref{sec:RSL}). The thermo-chemistry uses sub-cycling in order to decouple the short ionization timescales from the much longer radiation propagation timescale (see \S\ref{sec:thermochemisty}). 

The RT implementation inherits the full spatial and temporal adaptivity of the underlying gas dynamics scheme (e.g. \S\ref{sec:rhdtest}). The method as described can be combined with any SPH code, without any need for extra structures (e.g. grids, rays, photon packets, or angular discretization).

However, the method also has limitations, e.g. those associated with the closure relation and numerical noise. In this section, we discuss these limitations in more detail.

\subsection{Limitations of the Two-Moment M1 method}
\subsubsection{Approximations in the Moment Closure}
\label{sec:radcoll}
\begin{figure*}
\includegraphics[width=0.95\textwidth,left]{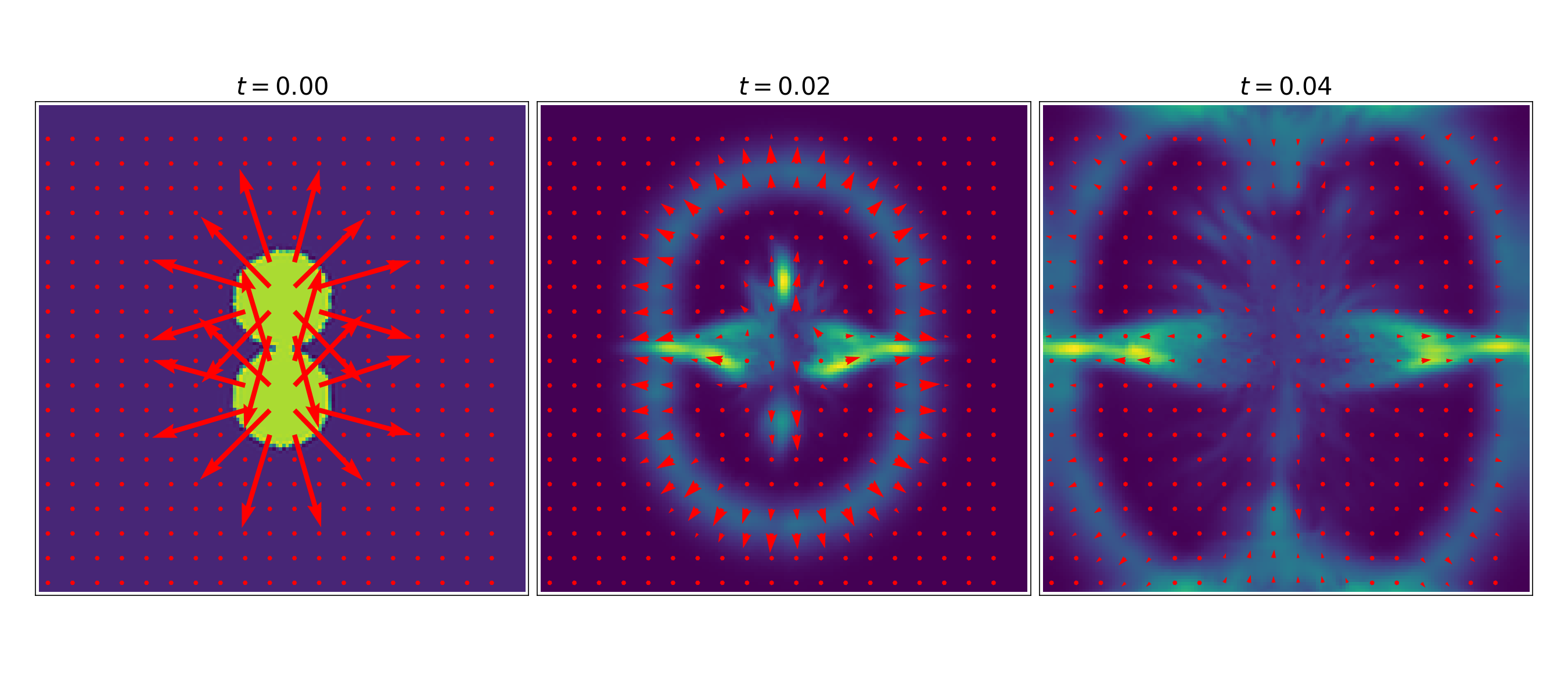}
\caption{Radiation collision in two dimensions: spherical radiation fronts emanate from two sources into
an optically thin medium. The setup is the same as Fig.~\ref{fig:radfrontsph2d_tm}, but with two spherical fronts instead of one. The panels from left to right show the initial conditions and the state at later times of $t=0.02$ and $0.04$ when the spherical fronts overlap. The simulation uses the modified M1 closure relation (Eqs.~\ref{eq:eddclose} and \ref{eq:codeest}). Colors represent the radiation energy density, red arrows show the radiation fluxes. 
Whereas the shells should pass through each other, this closure relation results in a collision of the two fronts, resulting in the formation of horizontal beams of light.}
\label{fig:rad2frontsph2d_tm}
\end{figure*}

\begin{figure}
 \includegraphics[width=0.5\textwidth]{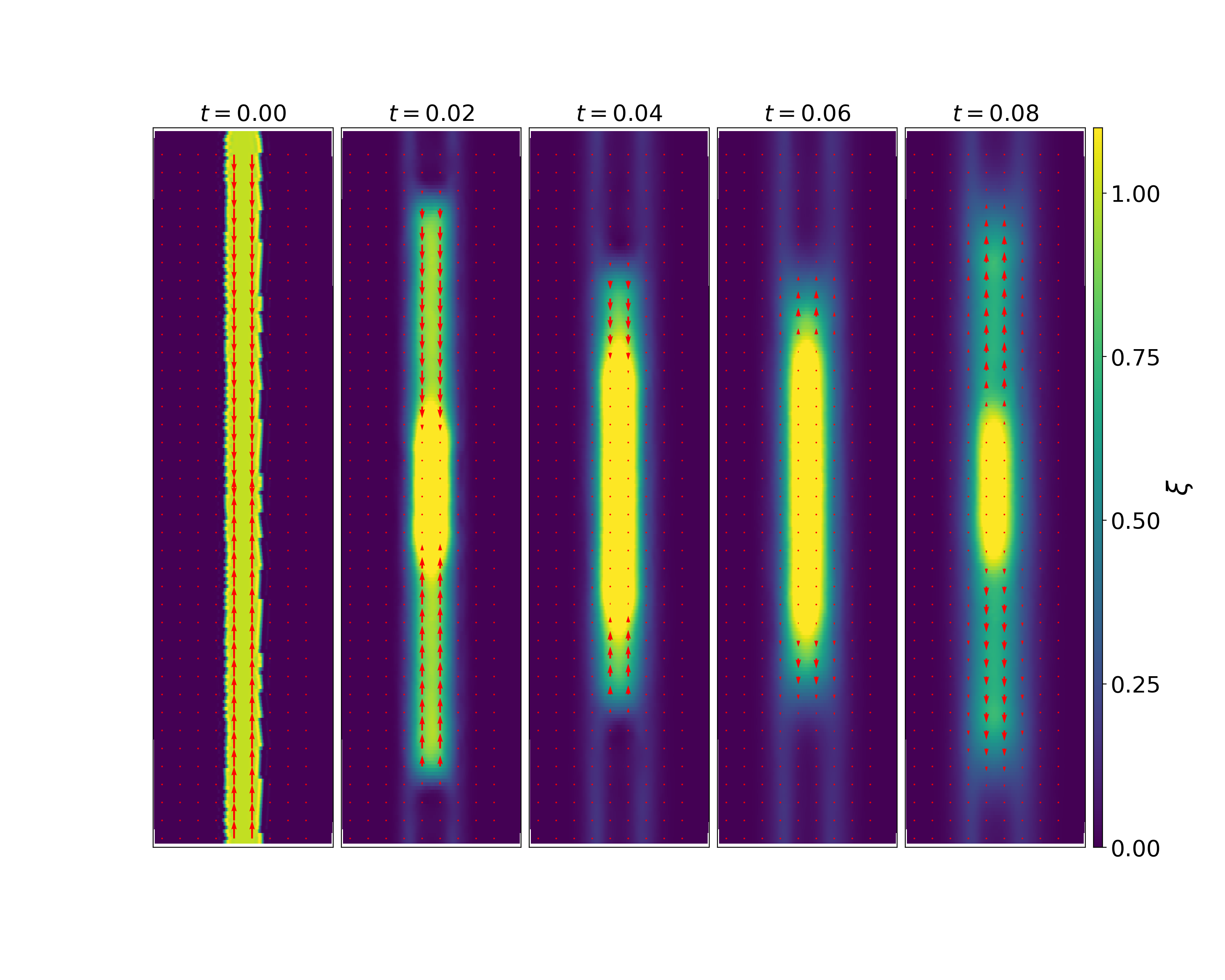}
\caption{A short beam of radiation propagating to the top, colliding with a short beam of radiation propagating to the bottom, simulated with the default two-moment method (see \S\ref{sec:testartdiss}) in an optically thin medium (with the same condition as in Fig.\ref{fig:radstream2d_difdiss}, except there is an extra radiation beam pointing downward). Colors represent the radiation energy density, whereas the red arrows show the radiation fluxes. This shows our modified M1 closure (Eq.\ref{eq:codeest}) can handle the head-on beam collision problems.}
\label{fig:radstreamheadon2d_tm}
\end{figure}

As discussed in \S\ref{sec:method}, the two-moment method results from truncating 
an infinite hierarchy of moment equations by postulating a closure relation for the Eddington tensor.
The choice of closure relation affects the accuracy of the method.

The \lq M1 closure relation\rq\ is not exact in the regime intermediate between optically thin and optically thick, even in the case of a single source. As shown by \citet{Leve84}, this is because 
in such a situation the closure relation cannot be uniquely determined by the first two moments.
Secondly, this closure relation cannot handle situations where particles receive radiation from 
two or more directions, even in the optically thin regime (see \S\ref{sec:momentclosure} and \S\ref{sec:radcoll}). In such cases, beams of radiation \lq collide\rq\ with each other rather than simply pass through one another as they should do. The reason for this is twofold. Firstly, the M1 closure relation of Eq.~(\ref{eq:feddcon}) erroneously implies that the radiation is optically thick when two beams collide
(see also Fig.~1 in \citealt{Rosd13ramsert}). Secondly, the form of the Eddington tensor in Eq.~(\ref{eq:eddclose}) implicitly assumes that radiation is moving in a single direction (single stream, plus an additional isotropic component).

We illustrate the \lq collision of radiation\rq\ in Fig.~\ref{fig:rad2frontsph2d_tm}. The setup is as follows: two sources emit a burst of radiation isotropically into an optically thin medium. This results in two spherical shells of radiation, propagating away from each source. When these shells overlap, they {\em should} simply pass through each other. When this situation is simulated with the M1 closure relation, the shells \lq collide\rq\  instead, erroneously producing two spikes of radiation. Such collisions could significantly distort the radiation morphology {\em e.g.} when sources are associated with multiple star clusters or multiple galaxies.

There are several ways to improve the method to reduce the impact of such collisions. Firstly, we introduced the modified M1 closure relation in Eq.~(\ref{eq:codeest}). This new closure relation does not lead to 
radiation collision in one dimension, as shown in Fig.~\ref{fig:radstreamheadon2d_tm}. The reason is that
this modified closure relation correctly identifies that the radiation is optically thin even where the two beams collide (unlike the original M1 close relation).

Unfortunately, the modified closure scheme still cannot handle the collision of optically thin beams, which is not head-on. Hence it does not resolve the problem illustrated in Fig.~\ref{fig:rad2frontsph2d_tm}. A possible way forward would be to resort to higher order methods, {\em e.g.} \cite{Vika13QMOMRT} and \cite{Leve96keclosure}. However, a stable and efficient high-order method has not been discussed in the astrophysics literature, as far as we are aware. 
Another avenue might be to calculate the closure relation itself more accurately, for example
using short characteristic \citep{Finl09VET,Jian12VET} or using a Monte Carlo scheme \citep{Fouc18VET}. 
Unfortunately, both these schemes are computationally more expensive in the case of multiple bright sources.

\subsubsection{Numerical/Artificial Noise}
Another limitation of the moment method, which is not restricted to the two-moment or M1 methods, is numerical noise. Such noise can destroy the coherence of the radiation even in the optically thin regime (see {\em e.g.} Fig.~ \ref{fig:radstream2d_difdiss}), or cause radiation to propagate into a shadow (see {\em e.g.} Fig. \ref{fig:radshadow3d}).

There are two major sources of noise. Firstly, the discretization of the density field into a disordered set of SPH particles, which is especially problematic when the symmetry of the particle distribution 
differs from that of the radiation field and/or when the numerical resolution is low. In addition,
SPH suffers from \lq zeroth-order errors\rq\ that are also particularly severe when the particle distribution is irregular (see \S\ref{sec:Introduction}, \citealt{Dehnen12}); our SPH formulation is designed
to minimize these (see \S\ref{sec:SPHform}). Higher-order SPH schemes could reduce such SPH noise further ({\em e.g.} \citealt{Vila99bettergrad,Gabu11WPMHD,Ross15improvedSPH,GIZMO,Fron17CRKSPH}).

The second major source of noise is the discontinuity-capturing
artificial dissipation, which can introduce unphysical diffusion and
damping. We have tried to limit the severity of such artefacts by
introducing {\it anisotropic} viscosity as well as a {\it
  higher-order} artificial diffusion scheme. Even so, our artificial
diffusion is not completely anisotropic, so that a small amount of
radiation still diffuses artificially perpendicular to the propagation
direction of a beam of radiation (see
Figs.~\ref{fig:radstream2d_defaultdiss} and \ref{fig:radshadow3d}). We
suggest that higher order artificial anisotropic diffusion, similar to
that used in the HLL Riemann solver \citep{Hart83HLL}, might reduce
these artefacts.

Our method can further reduce unphysical diffusion in cases where it is possible to impose the direction of 
propagation of the radiation.

\subsubsection{Computational Cost}
Another limitation of the M1 method is that the computational cost may be high in some physically interesting situations, {\em e.g.} when capturing the final stages of reionization when the reduced speed of light
needs to be close to $c$ to capture the speed of ionization fronts correctly (e.g.\citealt{Baus15Hreion}). Possible improvements include using a \lq variable speed of light approximation\rq\ 
(see \S\ref{sec:RSL}, \citealt{Katz17reionVSL}) or implementing the radiation transfer on graphics processing units \citep{Ocvi16CoDa}. Subcycling the radiative transfer module can further improve the performance of the overall code \citep{Rosd13ramsert,Kann19AREPORT}. 

However, we emphasise that the computational cost of our method scales with the number of gas particles, $N_{\rm gas}$, more favourably than ray-tracing methods (which additionally scale with the number of sources)
and {\small OTVET} (which scales with the $N\log N$ scaling of its Poisson solver).

\subsection{A comparison with other M1 codes}
The two-moment M1 method is a popular scheme in the field of galaxy formation simulations and is implemented in {\em e.g.} {\small ATON} \citep{Aube08M1}, {\small RAMSES-RT} \citep{Rosd13ramsert}, and {\small AREPO-RT} \citep{Kann19AREPORT}. The first two codes are Eulerian schemes, whereas {\small AREPO-RT} is a moving mesh scheme \citep{Spri10AREPO}. 

Our implementation in the {\sc swift} code takes advantages of the Lagrangian nature of SPH, which is important,
particularly when gas flows at high speeds. This is a great advantage compared to Eulerian codes, especially at relatively low resolution \citep{Robe10gridcodeerror} and high redshift \citep{Pont20gridnoise}. While Eulerian mesh codes can gain adaptivity through the adaptive mesh refinement, the refinement and de-refinement are not trivial and can be noisy.

{\small AREPO-RT} and our code can follow fluid motions and are highly adaptive, so both of them are advantageous in galaxy formation simulations, where the large dynamic range and high speeds of the gas are both numerically challenging. 

We think that the main advantages of our scheme are its computational efficiency and its parallelizability. The moving mesh code requires mesh reconstruction, which can be computationally expensive; such reconstruction is not necessary in SPH. SPH codes can also be parallelized efficiently through task-based parallelism, {\em e.g.} {\small ChaNGa} \citep{Jetl08ChaNGa} and {\small SWIFT} \citep{Scha16SWIFT}, the code in which our method is implemented.

\section{Summary}
\label{sec:conclusions}
We have developed a numerical radiation hydrodynamics scheme based on the two-moment method and using and improved closure relation. The two-moment method is based on the first two moments of the radiation moment hierarchy, and the hierarchy is closed with an Eddington closure relation. The M1 Eddington tensor closure is the simplest anisotropic closure that maximizes the entropy (see \S\ref{sec:method}), our modified closure relation improves the stability of the method, in particular in the optically thin regime. The numerical scheme is implemented in the smoothed particle hydrodynamics (SPH) code {\small SWIFT} \citep{Scha16SWIFT}. The interaction of radiation with a pure hydrogen gas is implemented in non-equilibrium, using sub-cycling to improve computational speed.

Key aspects of the method and its implementation include:
\begin{itemize}
\item the first stable and accurate SPH implementation of the two moment method (Eqs. \ref{eq:sphxi} and \ref{eq:SPHdfrad});
\item improvements to the M1 closure relation (Eq. \ref{eq:codeest}), 
making the method (1) less prone to noise in the optically thin regime (enabling the implementation of the two moment method in SPH) and (2) correct in the optically thin limit;
\item anisotropic artificial viscosity and high-order artificial diffusion schemes  (\S\ref{sec:artdiss}) to capture discontinuities in the radiation. These are essential to propagate radiation accurately in the optically thin regime;
\item an efficient non-equilibrium thermo-chemistry solver with sub-cycling (\S\ref{sec:thermochemisty});
\item implementation in {\small SWIFT}, a task-based parallel SPH galaxy simulation code \citep{Scha16SWIFT}. 
\end{itemize}

The accuracy and stability of the scheme is demonstrated in \S\ref{sec:results}. The scheme:
\begin{itemize}
\item preserves the directions and speed of propagation in the optically thin regime;

%\item {\color{red} is able to cast shadows if the optically thin direction is imposed;}

\item can simulate radiation hydrodynamics in dynamical multi-scale problems;

\item has accuracy comparable to other radiative transfer codes in the cosmological code comparison papers \citep{Ilie06RTcom,Ilie09RTcom}.
\end{itemize}
Note that the method yields robust results for spherically symmetric problems, e.g. Str\"omgren sphere, even without imposing that radiation propagates radially, provided that the injection region is sufficiently well sampled; see \S\ref{sec:injection} \& \ref{sec:statictest}.

The main advantage of our scheme (and of moment methods in general) as compared to other radiative transfer schemes, is that it is computationally efficient in large-scale simulations with numerous sources, since this cost is independent of the number of sources (which has been demonstrated in many other studies; see \S\ref{sec:Introduction}). Our scheme is also highly adaptive in both space and time, since the radiation field is directly sampled by individual SPH particles.

Our scheme can be also implemented in other SPH codes without requiring substantial structural changes. This is because the two-moment equations resemble the set of hydrodynamic equations themselves, e.g. the radiation energy density equation 
resembles the gas density equation, and the radiation flux equation 
resembles the momentum equation of hydrodynamics. The artificial dissipation 
terms for the propagation of radiation are also very similar to the
artificial viscosity and diffusion terms in SPH.

This method was developed to enable tracking of the propagation of ionizing radiation
through the intergalactic medium or interstellar medium in galaxy simulations.
It should be sufficiently accurate to correctly model the propagation of ionization fronts and cast shadows behind optically thick absorbers, while at the same time being sufficiently efficient to be able to handle thousands of sources without overly slowing down the calculations. 

The M1 method has also been applied to
study the effect of diffuse radiation or account for multiple scattering events, for example, when studying the transport of IR radiation in the interstellar medium, where multiply-scattered IR photons are thought to be important in transferring momentum to gas (e.g. \citealt{Skin13M1,Rosd15M1IR,Kann19AREPORT}). We intend to extend our code to be able to study these problems in the future.

We are aiming to release this code to the community in the near future, after we have completed the following two improvements. First, we are implementing multi-frequency radiative transfer so that we can include Helium in the calculations. Multi-frequency radiation also enables the modelling of spectral hardening. Secondly,
we are improving the efficiency of the implementation by decoupling the update of the hydrodynamics variables and the radiation variables, i.e. sub-cycling the radiation transport module. Taking advantage of the very different time-scales for the motion of gas and of radiation should lead to a large speed-up in computing time. We will report on these improvements and demonstrate the efficiency of our code elsewhere.

\section*{ACKNOWLEDGEMENTS}
We thank J.~Borrow, M.~Schaller, M.~Ivkovic, L.~Hausammann, and C.~Correa for their assistance with the implementation in {\small SWIFT}, and A.~Richings, Y.~Revaz, and A.~Benítez-Llambay for helpful discussions. We thank Joakim Rosdahl for a thoughtful and constructive referee's report, which led to significant improvements to our paper.

This work was supported by the Science and Technology Facilities Council (STFC) through Consolidated Grants ST/P000541/1 and ST/T000244/1 for Astronomy at Durham. We acknowledge support from the European Research Council through ERC Advanced Investigator grant, DMIDAS [GA 786910] to CSF. This work used the DiRAC@Durham facility managed by the Institute for Computational Cosmology on behalf of the STFC DiRAC HPC Facility (www.dirac.ac.uk). The equipment was funded by BIS capital funding via STFC capital grants ST/K00042X/1, ST/P002293/1, ST/R002371/1 and ST/S002502/1, Durham University and STFC operations grant ST/R000832/1. DiRAC is part of the National e-Infrastructure.

The research in this paper made use of the {\small SWIFT} open source simulation code
(http://www.swiftsim.com, \citealt{Scha18SWIFTascl}) version 0.9.0 This work also made use of  matplotlib \citep{Hunt07matplotlib}, numpy \citep{vand11numpy}, scipy \citep{Jone01scipy}, swiftsimio \citep{Borr20swiftsimio}, and NASA’s Astrophysics Data System. 

\section*{DATA AVAILABILITY}
The data underlying this article will be shared on reasonable request to the corresponding author (TKC).

\bibliographystyle{mn2e}
\bibliography{mn-jour,mybib}

\appendix

\section{Thermo-chemistry rate coefficients}
\label{sec:heatcoolparam}

\begin{figure}
 \includegraphics[width=0.45\textwidth]{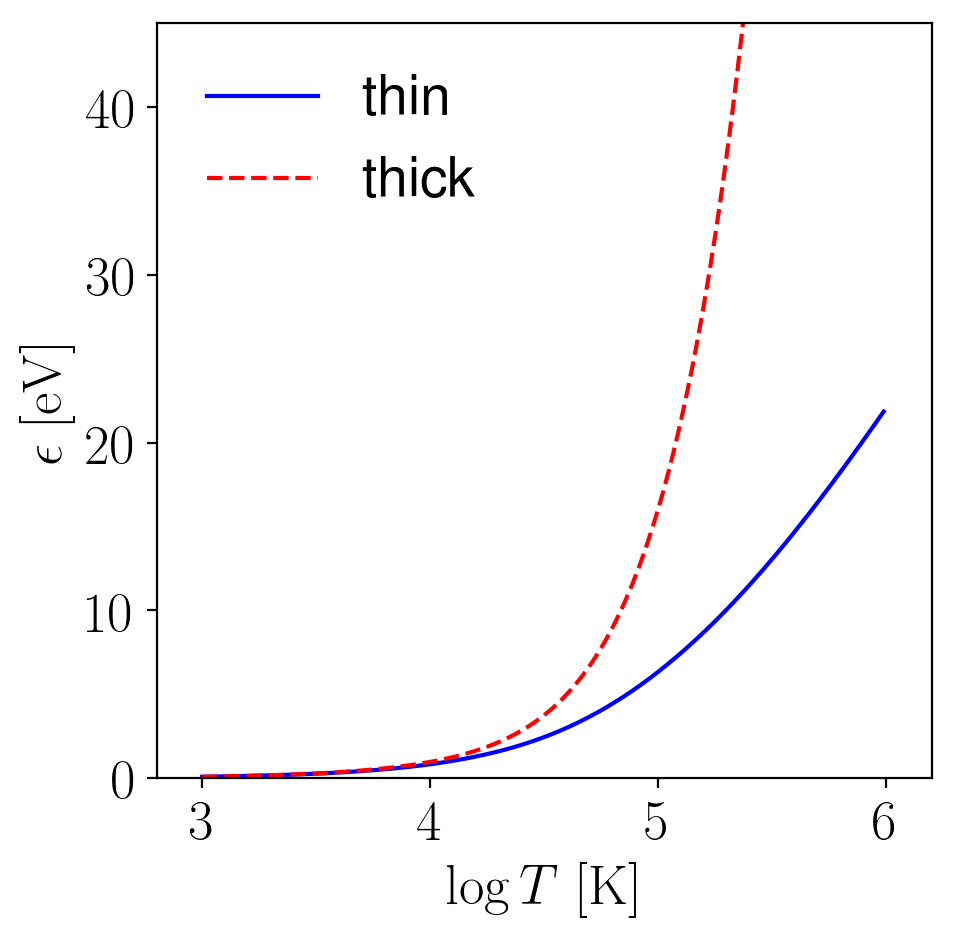}
\caption{Injected energy, $\epsilon$, in eV per photo-ionisation as a function of the temperature, $T$, of the irradiating black-body spectrum. {\em Solid blue line} is the optically thin case, {\em dashed red line} is the optically thick case.}
\label{fig:Heating}
\end{figure}

\begin{figure*}
 \includegraphics[width=0.95\textwidth]{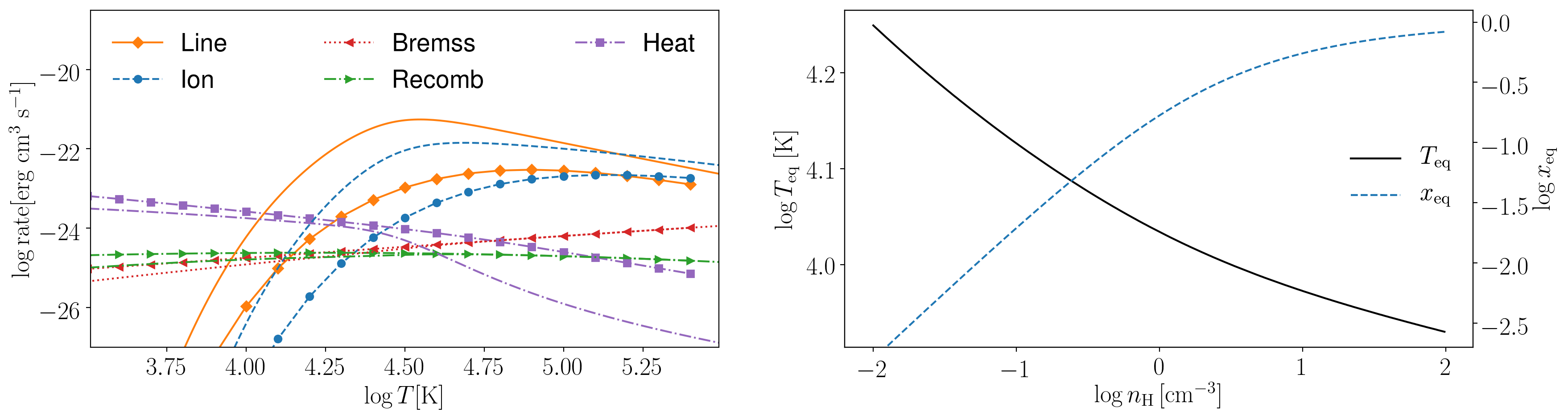}
\caption{{\bf Left panel:} Heating and cooling rates from Table~\ref{table:heatcoolparam} as a function of temperature under thermal equilibrium: collisional line cooling rate, $x(1-x)\Gamma_{\rm line,e{\rm HI}}(T)$ ({\em orange solid}),
collisional ionization cooling rate, $x(1-x)\Gamma_{\rm ion,e{\rm HI}}(T)$ ({\em blue dashed}),
Bremsstrahlung cooling rate, $(1-x)^2\Gamma_{\rm ff,e{\rm HII}}(T)$ ({\em red dotted}), 
recombination cooling rate $(1-x)^2\Gamma_{B,e{\rm HII}}(T)$ ({\em green long-dashed}) and photo-heating rate,
$\epsilon_\gamma x\,\Gamma_{\gamma,{\rm HI}}/n_{\rm H}$ ({\em purple dot-dashed}). Each rate is shown twice: for a hydrogen number density of $n_{\rm H}=1~{\rm cm}^{-3}$ (lines with symbols), and for $n_{\rm H}=10^{-2}{\rm cm}^{-3}$ (lines without symbols). The assumed photo-ionisation rate is $\Gamma=10^{-12}{\rm s}^{-1}$ and the energy injected per photo-ionization is $\epsilon_\gamma=6.33~{\rm eV}$, appropriate for a black-body spectrum of temperature $10^5~{\rm K}$ {\bf Right panel: } corresponding equilibrium temperature ($T_{\rm eq}$, solid black line) and neutral fraction ($x_{\rm eq}$, blue dashed line).}
\label{fig:Rates}
\end{figure*}

\begin{table*}
\centering
\begin{tabular}{llll}
\hline
\hline
&Recombination rate \citep{Hui97EOSIGM} by fitting \cite{Ferl92}\\
$\alpha_{\rm A}$  
&= $1.269\times 10^{-13}{\rm cm^3s^{-1}}\lambda^{1.503}[1.0+(\lambda/0.522)^{0.470}]^{-1.923}$   \\
$\alpha_{\rm B}$  
&= $2.753\times 10^{-14}{\rm cm^3s^{-1}}\lambda^{1.5}[1.0+(\lambda/2.740)^{0.407}]^{-2.242}$   \\
\hline
 &Collisional ionization rate \cite{Theu98P3MSPH} modified from \cite{Cen92hydro}   \\
 $\beta$ 
 &= $1.17\times 10^{-10}{\rm cm^3s^{-1}}T^{1/2}\exp(-157809.1/T)(1+T_5^{1/2})^{-1}$     \\
\hline
 &Collisional ionization cooling rates  \cite{Theu98P3MSPH} modified from \cite{Cen92hydro}  \\
$\Gamma_{\rm ion,e{\rm HI}}$ &$=2.54\times 10^{-21}{\rm erg\;cm^3\;s^{-1}}T^{1/2}\exp(-157809.1/T)(1+T^{1/2}_5)^{-1}$\\
\hline
 &Collisional excitation cooling rates   \cite{Theu98P3MSPH} modified from \cite{Cen92hydro} \\
 $\Gamma_{\rm line,e{\rm HI}}$& $=7.5\times 10^{-19}{\rm erg\;cm^3\;s^{-1}}\exp(-118348/T)(1+T^{1/2}_5)^{-1}$\\
 \hline
&Recombination cooling rates taken from \cite{Hui97EOSIGM} (fitted from \cite{Ferl92}) \\
$\Gamma_{A,e{\rm HII}}$ & $=1.778\times 10^{-29}{\rm erg\;cm^3\;s^{-1}K^{-1}}T\lambda^{1.965}[1.0+(\lambda/0.541)^{0.502}]^{-2.697}$\\
$\Gamma_{B,e{\rm HII}}$ & $=3.435\times 10^{-30}{\rm erg\; cm^3s^{-1}K^{-1}}T\lambda^{1.970}[1.0+(\lambda/2.250)^{0.376}]^{-3.720}$\\
 \hline
 &Bremsstrahlung cooling rate   \cite{Theu98P3MSPH} modified from \cite{Cen92hydro} and \cite{Spit78ISMbook} \\
  $\Gamma_{\rm ff,e{\rm HII}}$& $= 1.42\times 10^{-27}{\rm erg\;cm^3\;s^{-1}}T^{1/2}\{1.1+0.34\exp(-[5.5-\log(T)]^2/3)\}$\\
 \hline
 \hline
\end{tabular}
\caption{Coefficients for heating and cooling of hydrogen. $T_n=T/(10^n{\rm K})$ and temperature $T$ is in K. $\lambda = 315614 {\rm K}/T$.  If the on-the-spot approximation is applied, we used the case B recombination cooling (in Eq.~\ref{eq:dedt}). Otherwise, the case A recombination cooling is used.}
\label{table:heatcoolparam}
\end{table*}

The hydrogen photo-ionization rate, $\Gamma_{\gamma{\rm HI}}$ in units ${\rm s}^{-1}$, is \cite[e.g.][]{Oste89Nebulae}
\begin{align}
\Gamma_{\gamma,{\rm HI}}=\int^\infty_{\nu_{{\rm HI}}}\frac{4\pi J_\nu}{2\pi\hbar\nu}\sigma_{\gamma}(\nu)\,{\rm d}\nu\,,\label{eq:photoionization}
\end{align}
where $h\nu_{{\rm HI}}\approx 13.6~{\rm eV}$ is the hydrogen binding energy, $J_\nu$ is the angular-averaged specific intensity $I$ (in unit of ${\rm erg\; cm^{-2}s^{-1} Hz^{-1}}$), and $\sigma_{\rm \gamma}$ is the photo-ionization cross-section of hydrogen as a function of frequency, $\nu$, where we used the fit from \cite{Vern96sigmacross}. Defining the frequency-averaged photo-ionization cross-section
\begin{align}
\langle\sigma_\gamma\rangle \equiv \left [\int^\infty_{\nu_{\rm HI}}\frac{4\pi J_\nu}{2\pi\hbar\nu}\sigma_{\gamma }(\nu)\,{\rm d}\nu\,\right ]\left [ \int^\infty_{\nu_{\rm HI}}\frac{4\pi J_\nu}{2\pi\hbar\nu}\,{\rm d}\nu\right ]^{-1},
\label{eq:sigmagamma}
\end{align}
the photo-ionization rate of Eq.~(\ref{eq:photoionization}) is
\begin{align}
\Gamma_{\gamma,{\rm HI}}=\langle\sigma_\gamma\rangle \int^\infty_{\nu_{{\rm HI}}}\frac{4\pi J_\nu}{2\pi\hbar\nu}\,{\rm d}\nu=\langle\sigma_\gamma\rangle\tilde{c}n_\gamma,
\end{align}
where the frequency-averaged photon flux is:
\begin{align}
\tilde{c} n_\gamma \equiv \int^\infty_{\nu_{{\rm HI}}}\frac{4\pi J_\nu}{2\pi\hbar\nu}\,{\rm d}\nu\,.
\end{align}
For reference, the spectrum of a black-body (BB) with temperature $T=10^5~{\rm K}$ has
$\langle \sigma_\gamma\rangle=1.62\times 10^{-18}{\rm cm}^2$. 

The energy injected into the gas per photo-ionization is
\begin{align}
\epsilon_\gamma &= \left [\int^\infty_{\nu_{{\rm HI}}}\frac{4\pi J_\nu}{\hbar\nu}\,\sigma_{\gamma }(\nu)(\hbar\nu-\hbar\nu_{{\rm HI}})\,{\rm d}\nu \right ]\nonumber\\
&\times \left [ \int^\infty_{\nu_{{\rm HI}}}\frac{4\pi J_\nu}{2\pi \hbar\nu}\,\sigma_{\gamma }(\nu)\,\,{\rm d}\nu\ \right ]^{-1}\,,
\label{eq:photoheatperion}
\end{align}
in the \lq optically thin\rq\ limit where the probability that a photon of frequency $\nu$ is responsible for the ionization is set by the photo-ionization cross-section. 

In the \lq optically thick\rq\ limit, we simply assume that every photon
with $h\nu\geqslant h\nu_{{\rm HI}}$ causes an ionization, and replace $\sigma_{\gamma}\to 1$. This increases the value of $\epsilon_\gamma$ as higher-energy photons contribute relatively more to the ionizations \cite[e.g.][]{Abel99}. This is only an approximation: in the optically thick-limit, hard photons with $h\nu\gg h\nu_{{\rm HI}}$ 
tend to partially ionize gas upstream from the ionization front, which is not describe accurately by simply increasing the value of $\epsilon_\gamma$.

In the special case of a BB spectrum, $J_\nu=B_\nu$, with
\begin{align}
B_\nu=\frac{4\pi\hbar\nu^3}{c^2}\frac{1}{\exp(2\pi\hbar\nu/k_{\rm B}T)-1}\,,
\end{align}
the photo-heating per photo-ionization is
\begin{align}
\frac{\epsilon_\gamma}{kT} &= \frac
{\int_{\zeta_{T}}^\infty \zeta ^3\,(\exp(\zeta )-1)^{-1}\,\sigma_{\gamma{\rm HI}}(\zeta )\,d\zeta }
{\int_{\zeta_{T}}^\infty \zeta^2\,(\exp(\zeta )-1)^{-1}\,\sigma_{\gamma{\rm HI}}(\zeta )\,d\zeta }
-\zeta_{T}\,,  
\end{align}
where $\zeta _{T}\equiv 2\pi \hbar \nu_{{\rm th}}/(k_{\rm B}T)\approx 1.578\times 10^5{\rm K}/T$ is a dimensionless fraction. The value of $\epsilon_\gamma$ as a function of the temperature $T$ of the BB is plotted  in Fig.~\ref{fig:Heating}. For reference, $\epsilon_\gamma\approx 6.33~{\rm eV}$ ($16.0~{\rm eV}$) in the optically-thin (thick) case, when $T=10^5~{\rm K}$.

Table \ref{table:heatcoolparam} lists the interpolation formula for the ionization, recombination, heating, and cooling coefficients for hydrogen, as used in the thermo-chemistry described in section \S\ref{sec:thermochemisty}.
For reference, the cooling rate due to collisional line excitation, collisional ionization,
thermal Bremsstrahlung and recombination radiation (in the on-the-spot approximation), in units of ${\rm erg}~{\rm cm}^{-3}~{\rm s}^{-1}$, are respectively 
\begin{align}
    \left.\rho\frac{du}{dt}\right|_{\rm line} &=-x(1-x)\Gamma_{\rm line,e{\rm HI}}\,n^2_{\rm H}\\
    \left.\rho\frac{du}{dt}\right|_{\rm Ion} &=-x(1-x)\Gamma_{\rm ion,e{\rm HI}}\,n^2_{\rm H}\\
    \left.\rho\frac{du}{dt}\right|_{\rm Bremss} &=-(1-x)^2\Gamma_{\rm ff,e{\rm HII}}\,n^2_{\rm H}\\
    \left.\rho\frac{du}{dt}\right|_{\rm Recomb} &=-(1-x)^2\Gamma_{{\rm B},e{\rm HII}}\,n^2_{\rm H}\,,
\end{align}
whereas the photo-heating rate is
\begin{align}
\left.\rho\frac{du}{dt}\right|_{\rm Heat} &=x\epsilon_\gamma\,\Gamma_{\gamma, {\rm HI}}\,n_{\rm H}\,.
\end{align}
These rates are plotted in Fig.\ref{fig:Rates}, together with the thermal equilibrium values of the temperature and neutral fraction. For a gas temperature of $10^{4}{\rm K}\leqslant T\leqslant 10^5{\rm K}$, line cooling typically dominates the cooling rate, whereas at lower temperatures recombination cooling takes over. For reference, the case-A and case-B recombination coefficients are $\alpha_{\rm A}=4.29\times 10^{-13}{\rm cm^3s^{-1}}$ and $\alpha_{\rm B}=2.59\times 10^{-13}{\rm cm^3s^{-1}}$ respectively, and the collisional ionization coefficient $\beta=1.25\times 10^{-17}{\rm cm^3s^{-1}}$ at a gas temperature of $T=10^4{\rm K}$.

\section{Thermo-chemistry sub-cycling}
\label{sec:thermochemsub}
\begin{figure}
 \includegraphics[width=0.45\textwidth]{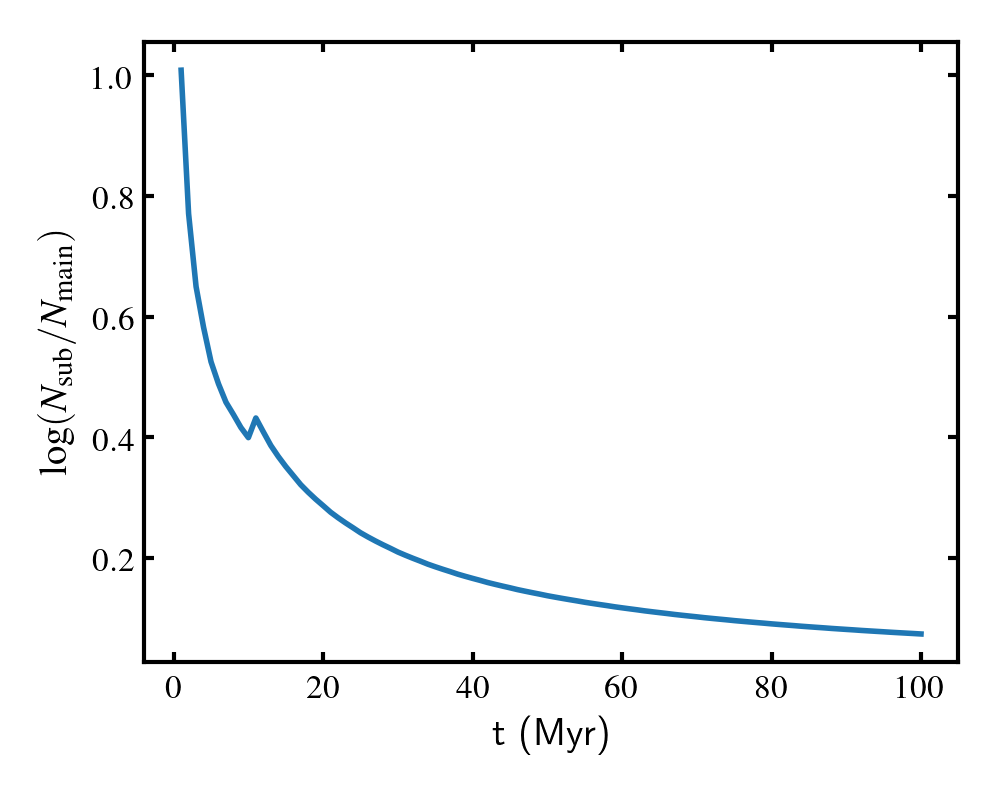}
\caption{The number of sub-cycles per global time-step in the single particle photo-ionization heating test, described in \S\ref{sec:singlegasparcel}, as a function of time. The global time-step $\Delta t_{\rm main}=0.1\;{\rm Myr}$. This ratio can be up-to a factor of ten initially, reducing to tens of percents after the gas reaches equilibrium.}
\label{fig:Nsubsingleparcel}
\end{figure}

\begin{figure}
 \includegraphics[width=0.45\textwidth]{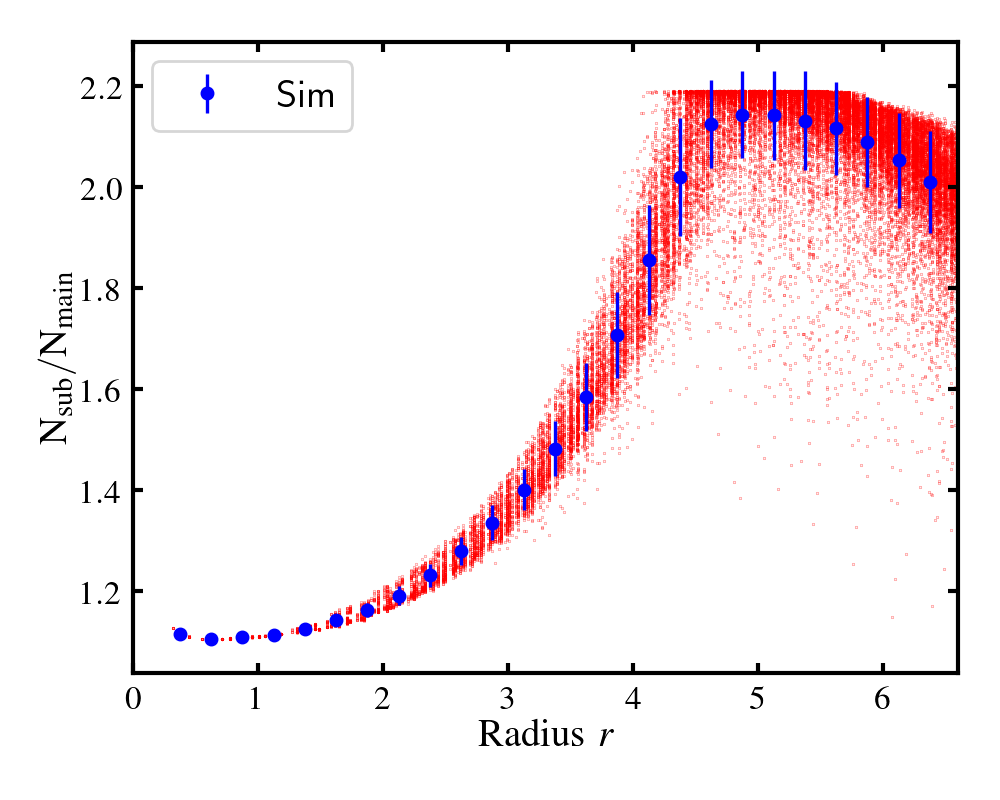}
\caption{The number of sub-cycles per global time-step for the variable temperature Str\"omgren sphere test with static gas particles described in  \S\ref{sec:statictest}.
This ratio is plotted for every SPH particle ({\em red dots}) as a function of its distance to the centre, at time $t=100~{\rm Myr}$; {\em blue points with error bars} show the mean and variance in radial bins. The speed-up due to sub-cycling is around a factor of two, mainly near the location of the ionization front.}
\label{fig:Nsubstromvartemp}
\end{figure}

Here we analyze the computational efficiency of using sub-cycling in the thermo-chemistry solver, as described in \S\ref{sec:thermochemisty}. Note that during sub-cycling, we only need to solve a few equations (at most quadratic) for each active gas particle, making the computational cost small compared to calculating hydrodynamical or gravitational forces, or performing the radiative transfer, all of which require loops over neighbouring particles
and potentially communication between compute nodes. Therefore, the number of sub-cycling steps per global time-step is a measure of the efficiency of sub-cycling.

We plot this ratio for the single gas particle test (see \S\ref{sec:singlegasparcel}) in Fig.~\ref{fig:Nsubsingleparcel}, and for the static 3D Str\"omgren test (see \S\ref{sec:statictest}) in Fig.~\ref{fig:Nsubstromvartemp}. 
The number of sub-cycling steps per global time-step is typically highest in
{\em nearly} neutral regimes, where the ratio can be up-to an order of magnitude. 
This is because the gas in this regime is far from equilibrium, and the sub-cyle time-step
is limited by the photo-ionization time scale, $\tau_i=1/\Gamma_{\gamma, {\rm HI}}$, which can be {\em much} shorter than the global time-step set by the {\sc cfl} condition.
Once the gas is highly ionized,  sub-cyle time-step is usually set by the recombination 
time-scale, which is typically much longer than $\tau_i$. In addition, the gas may be in ionization equilibrium or even thermal equilibrium, so that the chemistry time-step is long.
In the kind of astrophysical application that we have in mind, for example reionization simulations, sub-cycling is essential since otherwise the short time-step required in gas being over run by an ionization front will grind the code to a halt.
Finally, our sub-cycling scheme parallelizes well, since it does not require any communication.

\section{Analytic solution of the Str\"omgren sphere}
\label{sec:anaStromgren}
In the classical Str\"omgren sphere problem \citep{Stro39stromgrensphere}, a source emitting ionizing photons at a constant rate $\dot{N}_\gamma$ is embedded in a spherical cloud, initially filled with completely neutral hydrogen atoms with density $n_{\rm HI}=n_{\rm H}$. As the source switches on, an ionization front expands around the source, and the gas inside the ionization front, radius $R_I$, will be mostly ionized, $x\equiv n_{\rm HI}/n_{\rm H}\ll 1$, and outside $R_I$ will be mostly neutral.

The equations describing the evolution of the neutral fraction and photon density of such an idealized system are
\begin{align}
\frac{\partial n_{{\rm HI}}}{\partial t}&=-n_{{\rm HI}}c\sigma_\gamma n_\gamma +n_en_{{\rm HII}}\alpha_{A}-n_en_{\rm HI}\beta\,,
\label{eq:stromdnHI}
\end{align}
\begin{align}
\frac{\partial n_\gamma}{\partial t}&=-n_{{\rm HI}}c\sigma_{\gamma} n_\gamma + n_en_{{\rm HII}}(\alpha_{\rm A}-\alpha_{\rm B}) + S_\gamma,\nonumber\\
&=\frac{\partial n_{{\rm HI}}}{\partial t} -\alpha_{\rm B} n_en_{{\rm HII}} + S_\gamma.
\label{eq:simngamma}
\end{align}

As a first approximation to describe such a system, we assume that the gas inside $R_I$ is fully ionized, $x=0$, and outside $R_I$ is fully neutral, $x=1$, and that the ionization front is infinitely sharp. We further neglect collisional ionization, setting $\beta=0$. In this case, $n_{{\rm HI}}(r)=n_{{\rm HI}} \Theta (R_I-r)$, where $\Theta (x)$ is the step function. Integrating each term of Eq.~(\ref{eq:simngamma}) over the volume centered at the source, we find
\begin{eqnarray}
\int \frac{\partial n_{{\rm HI}}}{\partial t}{\rm d}V &=& \int \frac{\partial}{\partial t}\left [ n_{{\rm HI}} \Theta (R_I-r)\right ]{\rm d}V\nonumber\\
&=&n_{{\rm HI}} \int  \delta(R_I-r)\frac{\partial R_I}{\partial t}{\rm d}V\nonumber\\ 
&=& -4\pi R_I^2 n_{{\rm HI}} \frac{\partial R_I}{\partial t}\nonumber\\
\int \alpha_{\rm B} n_en_{{\rm HII}}{\rm d}V &\approx & \alpha_{\rm B} n^2_{\rm H}\frac{4\pi}{3}R^3\nonumber\\
\int S_\gamma{\rm d}V &=& \dot{N}_\gamma\,.
\end{eqnarray}
Combined, these yield the well-known equation,
\begin{align}
\dot{N}_\gamma=4\pi R_I^2 n_{{\rm H}}\frac{\partial R_I}{\partial t}+\alpha_{\rm B} n^2_{{\rm H}}\frac{4\pi}{3}R_I^3\,,
\label{eq:strogremdt}
\end{align}
with solution
\begin{align}
R_I(t)=R_S[1-\exp(-t/\tau_r)]^{1/3}\,.
\label{eq:anars}
\end{align}
Here, the recombination time $\tau_r=(\alpha_{\rm B} n_{\rm H})^{-1}$ and the Str\"omgren radius
\begin{align}
R_S \equiv \left(\frac{3\dot{N}_\gamma}{4\pi \alpha_{\rm B} n_{{\rm H}}^2}\right)^\frac{1}{3}.
\label{eq:Stromgrenradius}
\end{align}
For times $t\gg \tau_r$, and greater than the ionization time scale $\tau_i$,
\begin{align}
    \tau_i\equiv \frac{(4\pi/3)n_{\rm H}}{R^3_s}{\dot N_\gamma}\,,
\end{align}
the ionization front reaches an equilibrium location where ionizations balance recombinations.

To derive the profile of the hydrogen neutral fraction in equilibrium analytically, we work with the time independent equation
in the on-the-spot approximation, $\alpha_{\rm A}\to\alpha_{\rm B}$,
\begin{align}
\frac{\partial n_\gamma }{\partial t}=-\nabla\cdot{\bf f}_\gamma -n_{\rm HI}c\sigma_\gamma n_\gamma=0\,,
\end{align}
In spherical symmetry (and neglecting the scattered radiation), ${\bf f}_\gamma = n_\gamma c\hat{\bf r}$ so that
\begin{align}
\nabla\cdot (cn_\gamma \hat{\bf r} ) + n_{\rm HI}c\sigma_\gamma n_\gamma = 0.
\label{eq:stromgrendivF}
\end{align}
Writing this is spherical coordinates yields
\begin{align}
\frac{1}{r^2}\frac{\partial }{\partial r}\left ( r^2 n_\gamma \right ) + n_{\rm HI}\sigma_\gamma n_\gamma = 0\,,
\end{align}
with formal solution
\begin{align}
n_\gamma = \frac{\dot N_\gamma}{4\pi c\,r^2}\exp\left(-\int^r_0n_{\rm HI}(r')\sigma_\gamma {\rm d}r'\right)\,.
\label{eq:stromgrenngamma}
\end{align}
This derivation implicitly assumes that the mean free-path of photons is much less than $R_S$, which should be a good approximation for typical HII regions.

The steady-state neutral hydrogen profile follows by balancing ionizations and recombinations, {\em i.e.} substituting the previous relation into Eq.~(\ref{eq:stromdnHI}) and setting $\partial n_{\rm HI}/\partial t=0$,
\begin{align}
\frac{x\sigma_\gamma\dot{N}_\gamma}{4\pi r^2}\exp\left(-\tau(r)\right) &=(1-x)^2n_{\rm H}\alpha_{\rm B}
-x(1-x)\beta\,n_{\rm H}^2\,,
\label{eq:intsolxH0}
\end{align}
where the optical depth is given by
\begin{align}
\tau(r)=n_{\rm H}\sigma_\gamma \int^r_0\,x(r'){\rm d}r'\,.
\end{align}

The nature of the solution is brought out better by casting these equations in dimensionless form, 
\begin{align}
    \frac{\tau_S x}{q^2}\,\exp(-\tau)&=3(1-x)^2-3\frac{\beta}{\alpha_{\rm B}}x(1-x)\nonumber\\
    \tau&=\tau_S\int_0^q\,x(q')dq'\,,
\label{eq:qtau}    
\end{align}
where the dimensionless radius $q\equiv r/R_s$, and the \lq Str\"omgren optical depth\rq\ $\tau_S\equiv n_{\rm H}\sigma_\gamma\,R_S$. This is an integral equation for the neutral fraction, $x(q)$. We follow \cite{Alta13revrt} to convert this into an easier to integrate differential equation: take the logarithm of both sides and differentiate with respect to $q$, which yield
\begin{align}
    \left[\frac{1}{x}+\frac{2(1-x)-(1-2x)\beta/\alpha_{\rm B}}{(1-x)^2-x(1-x)\beta/\alpha_{\rm B}} \right]\frac{dx}{dq}=\tau_S\,x+\frac{2}{q}\,,
    \label{eq:anabeta}
\end{align}
with boundary condition for $q\to 0$
\begin{align}
    x\to \frac{3}{\tau_S}\,q^2\,.
\end{align}
Provided that collisional ionizations can be neglected,
the differential equation simplifies to
\begin{align}
    \frac{dx}{dq}=\frac{x(1-x)}{1+x}\,\left(\tau_S\,x+\frac{2}{q}\right)\,,
    \label{eq:Stromgren} 
\end{align}
which shows that there is a one-parameter family of solutions that are characterised by the value of $\tau_S$.
The numerical integration of Eq.~(\ref{eq:anabeta})  with its associated boundary conditions is plotted as the line labelled \lq Analytic\rq\ in Fig.~\ref{fig:stromgren3d_tm}.

\section{Moment derivations}
\label{sect:appendix-moments}
This short Appendix aims to elucidate the analogy between
taking moments of the Boltzmann equation to derive the fluid equations, and taking moments of the RT equation to derive the moment equations for radiation. The collisional Boltzmann equation is
\begin{equation}
    \frac{\partial}{\partial t}f+{\bf v}\cdot\frac{\partial}{\partial{\bf x}}f + {\bf a}\cdot\frac{\partial}{\partial {\bf v}}f=\left(\frac{Df}{Dt} \right)_{\rm coll}\,,
    \label{eq:BE}
\end{equation}
where the right hand side (R.H.S.) is the collision term, and the distribution function $f$ is a function of position, ${\bf x}$, velocity ${\bf v}$, and time, $t$. We will suppress this dependency to avoid clutter. Moments of the equation
are derived by multiplying Eq.~(\ref{eq:BE}) with some function $Q({\bf v})$ and integrating over velocities, 
\begin{eqnarray}
\frac{\partial}{\partial t} \int Q({\bf v})\,f\,d{\bf v}&+&\frac{\partial}{\partial{\bf x}}\cdot\,\int {\bf v}Q({\bf v})f\,d{\bf v} \nonumber\\
&+& {\bf a}\cdot\int  \left\{\frac{\partial}{\partial {\bf v}} Q({\bf v})f - f\frac{\partial}{\partial {\bf v}}Q({\bf v})\right\}\,d{\bf v}\nonumber\\
&=&\int\, Q({\bf v})\,\left(\frac{Df}{Dt} \right)_{\rm coll}\,d{\bf v}\,.
\label{eq:BE2}
\end{eqnarray}
The first term in curly brackets is the flux $Q\,f$ evaluated at the integration limits of the velocity.
Provided we integrate over all velocities, we can assume that $f$ goes to zero sufficiently fast that this term vanishes. Writing the velocity as ${\bf v}={\bf V}+{\bf w}$, where ${\bf V}$ is the mean and ${\bf w}$ is the random component of ${\bf v}$, the density, momentum, pressure and viscous stress tensor, are defined as 
\begin{eqnarray}
\rho &\equiv & m\int f\,d{\bf v}\nonumber\\
P &\equiv & \frac{1}{3}\int w^2\,f\,d{\bf v}\nonumber\\
\Pi^{ij} &\equiv & P\delta^{ij}- \int w^i\,w^j f\,d{\bf v}\,.
\end{eqnarray}
The fluid equations then follow by realising that integrals over collision term on the R.H.S. of Eq.~(\ref{eq:BE2}) vanish for functions $Q$ that are conserved during collisions. This is the case for $Q=m$ (the particle's mass), and $Q=m{\rm v}$ (the particle's momentum), which then yield the first two moment equations (the continuity and Euler equations), 
\begin{eqnarray}
\frac{\partial}{\partial t}\rho &+& \frac{\partial}{\partial{\bf x}}\cdot \rho{\bf V}=0\nonumber\\
\frac{\partial}{\partial t}\rho{\bf V}^i&+&\frac{\partial}{\partial{\bf x}^j}
\left( \rho{\bf V}^j+P\,\delta^{ij}-\Pi^{ij}\right) - \rho\,{\bf a}^i=0\,.
\end{eqnarray}
Contrast this derivation with taking moments of the radiative transfer equation \cite[e.g][]{Leve81FLD}\footnote{Here we only briefly illustrate the concept, so we suppress the acceleration term of Eq.~(\ref{eq:RT1}) for simplicity. This missing term is included in \S\ref{sec:two-moment} which is based on the derivation by \cite{Buch83rtff}. \cite{Gnedin97} \cite[see also e.g.][]{Petk09OTVET, Cantalupo11} did include the rate of change of frequency in Eq.~(\ref{eq:RT1}) but only to include cosmological redshifting of photons while neglecting changes of radiation energy density from gas velocity.},
\begin{equation}
    \frac{1}{c}\frac{\partial}{\partial t}I+{\bf n}\cdot \frac{\partial}{\partial{\bf x}}I=\left(\frac{DI}{Dt}\right)_{SS}\,,
    \label{eq:RT1}
\end{equation}
where the specific intensity $I$ is a function of position, ${\bf x}$, direction, ${\theta, \phi}$ in spherical coordinates, frequency, $\nu$, and time, $t$. The Cartesian coordinates of the unit vector in direction ${\bf n}$ are ${\bf n} = (\sin(\theta)\cos(\phi), \sin(\theta)\sin(\phi), \cos(\theta))$. The R.H.S. now represents photon sources and sinks. 

We proceed as before, by multiplying with some function $Q(\theta, \phi)$, which can be a scalar or a tensor, and integrating the RT equation over solid angle $d\Omega$, but not over frequency. We then take $Q=1$ and $Q={\bf n}$ to yield the first two moment equations,
\begin{eqnarray}
\frac{\partial}{\partial t} E + \frac{\partial}{\partial x^i} F^i &=&\int \left(\frac{DI}{Dt}\right)_{SS}\,d\Omega\nonumber\\
\frac{1}{c^2}\frac{\partial}{\partial t} F^i + \frac{\partial}{\partial x^j} {\mathbb P}^{ij} &=&\int
{\bf n}^i\,\left(\frac{DI}{Dt}\right)_{SS}\,d\Omega\nonumber\\
E &=& \frac{1}{c}\int I\,d\Omega=3P\nonumber\\
{\mathbb P}^{ij} &\equiv & P\,\delta^{ij} - \Pi^{ij}\nonumber\\
{\bf F}^i &=& \int {\bf n}^i\,I\,d\Omega\nonumber\\
\Pi^{ij} &=& P\,\delta^{ij} - \int {\bf n}^i\,{\bf n}^j\,Id\Omega\,.
\label{eq:RT2}
\end{eqnarray}
Here, $E$ is the energy density, $P$ the radiation pressure, ${\bf F}$ the flux, and the trace-less tensor $\Pi$ is the radiation equivalent of the viscous stress tensor. These are the moment equations of Eq.(\ref{eq:durad}) and Eq.~(\ref{eq:dfrad}) in the fluid frame, ${\bf v}=0$.

In the special case where the sources plus sinks term have the form of isotropic absorption, $\left({DI}/{Dt}\right)_{SS}=-\kappa\,I$, where $\kappa$ is the isotropic absorption coefficient, the sink terms in Eq.~(\ref{eq:RT2}) are $-\kappa E$ and $-\kappa{\bf F}$ for the first and second equation, respectively. Provided that the term $(1/c^2)\partial{\bf F}/\partial t$ can be neglected, the moment equations then combine to
\begin{align}
\frac{\partial}{\partial t} E = \nabla\cdot\left[ \frac{1}{\kappa} \nabla {\mathbb P}\right] - \kappa E\, ,
\end{align}
which is a diffusion equation. In the isotropic case, $\Pi^{ij}=0$, and a Gaussian package of the form $E({\bf x}, t)=\left(2\pi\sigma^2\right)^{-3/2}\exp[-x^2/(2\sigma^2)]\exp(-\kappa t)$ is a solution, 
spreading out as $\sigma^2(t)=\sigma^2(t=0)+2t/(3\kappa)$ while dimming, $\propto \exp(-\kappa t)$.
\label{lastpage}
\bsp
\end{document}